\documentclass[10pt,journal,compsoc]{IEEEtran}
\ifCLASSOPTIONcompsoc
  \usepackage[nocompress]{cite}
\else
  \usepackage{cite}
\fi
\ifCLASSINFOpdf
   \usepackage[pdftex]{graphicx}
  \graphicspath{{figs/}{figures/}{pictures/}{images/}{./}}
   \DeclareGraphicsExtensions{.pdf,.jpeg,.png}
\else
\fi
\usepackage{amsmath}
\usepackage{amsthm}
\usepackage{amsbsy}
\usepackage{algorithm}
\usepackage{algpseudocode}
\algnewcommand\algorithmicin{\textbf{Input}}

\usepackage{array}
\newcommand{\PreserveBackslash}[1]{\let\temp=\\#1\let\\=\temp}
\newcolumntype{C}[1]{>{\PreserveBackslash\centering}p{#1}}
\newcolumntype{R}[1]{>{\PreserveBackslash\raggedleft}p{#1}}
\newcolumntype{L}[1]{>{\PreserveBackslash\raggedright}p{#1}}

\ifCLASSOPTIONcompsoc
  \usepackage[caption=false,font=footnotesize,labelfont=sf,textfont=sf]{subfig}
\else
  \usepackage[caption=false,font=footnotesize]{subfig}
\fi
\usepackage{bm}
\mathchardef\mhyphen="2D % Define a "math hyphen"

\usepackage[switch]{lineno}
\usepackage{xcolor}
\usepackage{color,soul}
\usepackage{array}
\usepackage{textcomp}
\usepackage{stfloats}
\usepackage{url}
\usepackage{verbatim}
\usepackage{graphicx}
\usepackage{cite}
\usepackage{comment}
\usepackage{upgreek}
\usepackage{float}
\usepackage{makecell}
\usepackage{rotating}
\usepackage{capt-of}
\usepackage{pdflscape}
\usepackage{subcaption}
\usepackage{enumitem}
\usepackage{comment}

\begin{document}
%
% paper title
% Titles are generally capitalized except for words such as a, an, and, as,
% at, but, by, for, in, nor, of, on, or, the, to and up, which are usually
% not capitalized unless they are the first or last word of the title.
% Linebreaks \\ can be used within to get better formatting as desired.
% Do not put math or special symbols in the title.
\title{Fast and Faithful Edge Bundling using Spectral Sparsification}
%
%
% author names and IEEE memberships
% note positions of commas and nonbreaking spaces ( ~ ) LaTeX will not break
% a structure at a ~ so this keeps an author's name from being broken across
% two lines.
% use \thanks{} to gain access to the first footnote area
% a separate \thanks must be used for each paragraph as LaTeX2e's \thanks
% was not built to handle multiple paragraphs
%
%
%\IEEEcompsocitemizethanks is a special \thanks that produces the bulleted
% lists the Computer Society journals use for "first footnote" author
% affiliations. Use \IEEEcompsocthanksitem which works much like \item
% for each affiliation group. When not in compsoc mode,
% \IEEEcompsocitemizethanks becomes like \thanks and
% \IEEEcompsocthanksitem becomes a line break with idention. This
% facilitates dual compilation, although admittedly the differences in the
% desired content of \author between the different types of papers makes a
% one-size-fits-all approach a daunting prospect. For instance, compsoc 
% journal papers have the author affiliations above the "Manuscript
% received ..."  text while in non-compsoc journals this is reversed. Sigh.

\author{Xingjue Jiang, Seok-Hee Hong, Amyra Meidiana, and Xianyuan Zeng% <-this % stops a space
\IEEEcompsocitemizethanks{\IEEEcompsocthanksitem X. Jiang, S.-H. Hong, A. Meidiana, and X. Zeng are with the University of Sydney.\protect\\
% note need leading \protect in front of \\ to get a newline within \thanks as
% \\ is fragile and will error, could use \hfil\break instead.
E-mail: xjia9238@uni.sydney.edu.au, \{seokhee.hong, amyra.meidiana\} @sydney.edu.au, xzen6984@uni.sydney.edu.au
}%<-this % stops an unwanted space
%\thanks{Manuscript received April 19, 2005; revised August 26, 2015.}
}

% note the % following the last \IEEEmembership and also \thanks - 
% these prevent an unwanted space from occurring between the last author name
% and the end of the author line. i.e., if you had this:
% 
% \author{....lastname \thanks{...} \thanks{...} }
%                     ^------------^------------^----Do not want these spaces!
%
% a space would be appended to the last name and could cause every name on that
% line to be shifted left slightly. This is one of those "LaTeX things". For
% instance, "\textbf{A} \textbf{B}" will typeset as "A B" not "AB". To get
% "AB" then you have to do: "\textbf{A}\textbf{B}"
% \thanks is no different in this regard, so shield the last } of each \thanks
% that ends a line with a % and do not let a space in before the next \thanks.
% Spaces after \IEEEmembership other than the last one are OK (and needed) as
% you are supposed to have spaces between the names. For what it is worth,
% this is a minor point as most people would not even notice if the said evil
% space somehow managed to creep in.

% The paper headers
%\markboth{IEEE Transactions on Visualization and Computer Graphics}%
\markboth{Hong et al: Fast and Faithful Edge Bundling using Spectral Sparsification}
{Hong et al: Fast and Faithful Edge Bundling using Spectral Sparsification}
% The only time the second header will appear is for the odd numbered pages
% after the title page when using the twoside option.
% 
% *** Note that you probably will NOT want to include the author's ***
% *** name in the headers of peer review papers.                   ***
% You can use \ifCLASSOPTIONpeerreview for conditional compilation here if
% you desire.

% The publisher's ID mark at the bottom of the page is less important with
% Computer Society journal papers as those publications place the marks
% outside of the main text columns and, therefore, unlike regular IEEE
% journals, the available text space is not reduced by their presence.
% If you want to put a publisher's ID mark on the page you can do it like
% this:
%\IEEEpubid{0000--0000/00\$00.00~\copyright~2015 IEEE}
% or like this to get the Computer Society new two part style.
%\IEEEpubid{\makebox[\columnwidth]{\hfill 0000--0000/00/\$00.00~\copyright~2015 IEEE}%
%\hspace{\columnsep}\makebox[\columnwidth]{Published by the IEEE Computer Society\hfill}}
% Remember, if you use this you must call \IEEEpubidadjcol in the second
% column for its text to clear the IEEEpubid mark (Computer Society jorunal
% papers don't need this extra clearance.)

% use for special paper notices
%\IEEEspecialpapernotice{(Invited Paper)}

% for Computer Society papers, we must declare the abstract and index terms
% PRIOR to the title within the \IEEEtitleabstractindextext IEEEtran
% command as these need to go into the title area created by \maketitle.
% As a general rule, do not put math, special symbols or citations
% in the abstract or keywords.

\IEEEtitleabstractindextext{%
\begin{abstract}
Edge bundling reduces the visual complexity of drawings of large and complex graphs by clustering ``compatible'' edges. However, it often introduces distortion by bundling ``unrelated'' edges, resulting in misleading, ambiguous drawings.  
Moreover, existing edge bundling methods often have high computational complexity, leading to scalability issues.

This paper presents new edge bundling methods and faithfulness metrics for edge bundling using  \emph{spectral sparsification}, which sparsifies a graph $G$ into a subgraph $G'$ with $O(n \log n)$ edges, based on the  {\em effective resistance} values of edges, preserving the spectrum of $G$, closely related to important structural properties of $G$, such as connectivity, clustering, and the commute distance. 

We first present a new edge bundling method  \emph{SEB} (Spectral Edge Bundling), introducing {\em effective resistance}-based compatibility for spectral-faithful bundling, improving distortion and ambiguity. 
Then, we present a general framework \emph{FEB} (Fast Edge Bundling) utilizing spectral sparsification to improve the efficiency of existing bundling methods while maintaining a similar quality.
We also present \emph{FBQ} (Fast Bundling Quality) framework for \emph{proxy bundle faithfulness} metrics, for measuring how FEB faithfully preserves the ground truth structure in the original edge bundling, with two variants, $FBQ_{JS}$ (utilizing Jaccard Similarity) and  $FBQ_{SQ}$ (utilizing sampling quality metrics).

Extensive experiments using various  
real-world and synthetic graphs demonstrate the {\em effectiveness} of SEB for edge bundling, outperforming state-of-art bundling methods on quality metrics, with 46\% and 17\% average improvement in distortion and ambiguity respectively for SEB2.
Furthermore, experiments successfully demonstrate the {\em efficiency} of the FEB framework, with 61\% runtime improvement over the original edge bundling methods without sparsification, while maintaining a similar quality,  with 74\% similarity based on $FBQ_{SQ}$. 
\end{abstract}

% Note that keywords are not normally used for peerreview papers.
\begin{IEEEkeywords}
Graph drawing, edge bundling
\end{IEEEkeywords}}

% make the title area
\maketitle

% To allow for easy dual compilation without having to reenter the
% abstract/keywords data, the \IEEEtitleabstractindextext text will
% not be used in maketitle, but will appear (i.e., to be "transported")
% here as \IEEEdisplaynontitleabstractindextext when the compsoc 
% or transmag modes are not selected <OR> if conference mode is selected 
% - because all conference papers position the abstract like regular
% papers do.
\IEEEdisplaynontitleabstractindextext
% \IEEEdisplaynontitleabstractindextext has no effect when using
% compsoc or transmag under a non-conference mode.

% For peer review papers, you can put extra information on the cover
% page as needed:
% \ifCLASSOPTIONpeerreview
% \begin{center} \bfseries EDICS Category: 3-BBND \end{center}
% \fi
%
% For peerreview papers, this IEEEtran command inserts a page break and
% creates the second title. It will be ignored for other modes.
\IEEEpeerreviewmaketitle

\IEEEraisesectionheading{\section{Introduction}\label{sec:introduction}}
% Computer Society journal (but not conference!) papers do something unusual
% with the very first section heading (almost always called "Introduction").
% They place it ABOVE the main text! IEEEtran.cls does not automatically do
% this for you, but you can achieve this effect with the provided
% \IEEEraisesectionheading{} command. Note the need to keep any \label that
% is to refer to the section immediately after \section in the above as
% \IEEEraisesectionheading puts \section within a raised box.

% The very first letter is a 2 line initial drop letter followed
% by the rest of the first word in caps (small caps for compsoc).
% 
% form to use if the first word consists of a single letter:
% \IEEEPARstart{A}{demo} file is ....
% 
% form to use if you need the single drop letter followed by
% normal text (unknown if ever used by the IEEE):
% \IEEEPARstart{A}{}demo file is ....
% 
% Some journals put the first two words in caps:
% \IEEEPARstart{T}{his demo} file is ....
% 
% Here we have the typical use of a "T" for an initial drop letter
% and "HIS" in caps to complete the first word.
%\IEEEPARstart{T}{his} demo file is intended to serve as a ``starter file''
%for IEEE Computer Society journal papers produced under \LaTeX\ using
%IEEEtran.cls version 1.8b and later.
% You must have at least 2 lines in the paragraph with the drop letter
% (should never be an issue)

\IEEEPARstart{E}{dge} bundling reduces the visual complexity in drawings of large and complex graphs and improves \emph{readability} by clustering compatible edges together into ``bundles''~\cite{Lhuillier2017StateOT}. 
Specifically, edges are bundled together if they are ``compatible'', mostly based on the geometry of edges in earlier edge bundling methods such as GBEB (Geometry-Based Edge Bundling)~\cite{gbeb} and FDEB (Force-Directed Edge Bundling)~\cite{forceDirBund}.

However, bundling edges solely based on geometric properties may often introduce distortion by bundling structurally ``unrelated'' edges, resulting in misleading, ambiguous drawings,
and failing to accurately represent the ground truth structure of the graph. 
In addition to readability, edge bundling should be \emph{faithful}, such that the bundled drawing faithfully displays the ground truth structure of the graph.%~\cite{faithfulness}.  

Recent \emph{structure-based} bundling, therefore, incorporates structural properties of the graph to avoid ambiguity, improving the faithfulness of the edge bundling methods. 
Examples include Confluent drawings~\cite{dickerson2004confluent}, TGI-EB (Topology-Geometry-Importance Edge Bundling)~\cite{nguyen2012tgi} using graph topology and centrality, Clustered Edge Routing~\cite{bouts2015clustered} which uses well-separated pair decomposition, and EPB (Edge-Path Bundling)~\cite{edge-path} grouping edges based on paths in the graph to avoid non-existing connections. 

Moreover, edge bundling methods are often computationally intensive, with high runtime complexity; for example, FDEB runs in $O(m^2n)$ time and EPB runs in $O(m^2 \log n)$ time, where $m$ is the number of edges and $n$ is the number of vertices of a graph $G$. 

Existing fast bundling methods, such as FFTEB (Fast Fourier Transform Edge Bundling)~\cite{ffteb}, whose runtime depends on the size of the bitmap image, rather than the size of the graph, still retain readability issues, and SEPB (Spanner EPB)~\cite{sepb}, which runs faster than EPB in practice, still has $O(m^2 \log n)$ worst-case runtime. 
Therefore, there is a strong demand for fast edge bundling methods to address the scalability issues for bundling drawings of large and complex graphs.

In this paper, we present new edge bundling approaches incorporating \emph{SS (Spectral Sparsification)} for improving both the effectiveness and efficiency of edge bundling methods. 
Spectral sparsification~\cite{spielman2007spectral} sparsifies the edges of a dense graph $G$ to a subgraph $G'$ with $O(n \log n)$ edges, based on the {\em effective resistance} values of edges, computed as the voltage drop along edges when modelling the graph as an electrical network~\cite{spielman-srivastava}. 
Note that $SS$ preserves the spectrum of $G$, which is closely related to important structural properties, such as connectivity, clustering, and commute distance~\cite{von2007tutorial,chung1997spectral}.

Recent work has demonstrated both the effectiveness and efficiency of spectral sparsification ($SS$) for sampling and drawing large and complex graphs. 
Specifically, $SS$ outperforms popular random sampling methods on various sampling quality metrics~\cite{eades2018drawing,meidiana2019topology,hu2019spectral,Hu3}.
Moreover, $SS$ has been successfully integrated with the most popular graph drawing algorithms, force-directed and stress minimization methods, to improve the force (resp., stress) computation from quadratic to sublinear, while maintaining good quality drawings~\cite{meidiana2020sublinear-tvcg,meidiana2021stress}. 
Therefore, it is a natural approach to leverage spectral sparsification for improving the effectiveness and efficiency of edge bundling of drawings of large and complex graphs.

In this paper, we first present the \emph{FEB (Fast Edge Bundling)} framework, which utilizes  spectral sparsification~\cite{spielman2011spectral} 
to improve the runtime efficiency of existing edge bundling methods while maintaining a similar quality to the original bundling methods without sparsification.
We then present \emph{SEB (Spectral Edge Bundling)}, introducing {\em effective resistance-based compatibility}, 
%with geometry-based compatibility 
to improve the spectral faithfulness of the bundled drawing, including ambiguity and distortion. 
%We also present \emph{SBQ (Spectral Bundling Quality)}, a faithfulness metric for measuring how faithfully the spectrum of a graph is preserved in a bundled drawing.

Moreover, we also present the \emph{FBQ (Fast Bundling Quality)} metrics for measuring \emph{proxy bundling faithfulness}, i.e., how the FEB framework faithfully preserves the ground truth structure in the original edge bundling, based on the Jaccard Similarity and sampling quality metrics. 
Specifically, our main contributions can be summarized as follows:

\begin{enumerate}
    \item We present a fast edge bundling framework FEB based on {\em spectral sparsification}, where we compute an edge bundling on a smaller subgraph to reduce runtime while maintaining a similar quality bundling. 
    Extensive experiments demonstrate the efficiency and effectiveness of FEB, running 61\% faster than the original edge bundling without sparsification while maintaining similar quality using visual comparison.
    
   \item We present a new edge bundling method SEB based on \emph{spectral sparsification}, by introducing {\em effective resistance-based compatibility} with two variations, SEB1 and SEB2. 
   Extensive experiments using various data sets demonstrate the effectiveness of SEB, outperforming state-of-art bundling methods on quality metrics: 46\% average improvement on ambiguity and 17\% average improvement on distortion for SEB2.
   %and with 5\% better SBQ. YYY XXXXX 

  %\item We present SBQ, a faithfulness metric for edge bundling, measuring how faithfully a bund`led drawing represents the ground truth spectrum of the graph. Experiments show that SBQ can effectively measure the spectral faithfulness of bundled drawings.
    %e.g., higher scores for drawings which do not exaggerate artificially introduced cycles and paths.

    \item We present the FBQ framework, to measure how faithfully the FEB framework preserves the ground truth structures of the original bundling without sparsification. 
    Specifically, we present $FBQ_{JS}$, using \emph{JS (Jaccard similarity)}, and $FBQ_{SQ}$, based on the sampling quality metrics.
    % We also present the FBQ metric framework for measuring how well FEB maintains the quality compared to the original bundling methods.
    Experiments show the effectiveness of FBQ metrics for measuring the faithfulness of the FEB framework, maintaining a similar quality bundling, with 74\% similarity based on $FBQ_{SQ}$.
    %, with on average 70\% similarity to the original bundling. 
                
\end{enumerate}

\section{Related Work}
\label{sec:litrev}

\subsection{Spectral Sparsification }

A {\em spectral sparsification} ($SS$) is a subgraph $G'$ of a large dense graph $G$ with \(O(n \log n)\) edges~\cite{spielman2011spectral} preserving the \emph{spectrum} of $G$, which is closely related to important structural properties including connectivity, clustering, and commute distance~\cite{von2007tutorial,chung1997spectral}.

$SS$ stochastically selects edges based on their \textit{effective resistance} values, equivalent to the probability of the edge being included in a random spanning tree of the graph~\cite{spielman-srivastava}. Modeling the graph as an electrical network, the effective resistance of an edge is measured as the voltage drop along the edge and can be computed in near-linear time~\cite{spielman-srivastava}.

Recent work has demonstrated both the effectiveness and efficiency of $SS$ for sampling and drawing large and complex graphs. 
Specifically, $SS$ has been shown to outperform random sampling methods on various sampling quality 
metrics~\cite{eades2018drawing,hu2019spectral}. 
Moreover, $SS$ has been integrated with graph topology, such as biconnected component decomposition using BC (Block-Cut vertex) trees~\cite{Hu3} or triconnected component decomposition using SPQR trees~\cite{meidiana2019topology} to reduce the runtime while improving the sampling quality.

More recently, $SS$ has also been successfully used for state-of-the-art fast graph drawing algorithms for large and complex graphs while maintaining good quality. 
For example, the {\em SublinearForce} framework~\cite{meidiana2020sublinear-tvcg} successfully utilizes $SS$ to significantly improve the runtime of the force computation of the most popular force-directed algorithm, from quadratic to sublinear time, also resulting in better quality drawings. 

Similarly, $SS$ has also been used for the fastest stress-based algorithms {\em SublinearSM} and {\em SublinearSGD}~\cite {meidiana2021stress}, significantly improving the runtime of stress computation, from quadratic to sublinear time, while maintaining similar quality drawings.

\subsection{Edge Bundling Methods}

\emph{Edge bundling} is a method for reducing the visual complexity of drawings of complex graphs~\cite{Lhuillier2017StateOT}. 
For each pair of edges $e_i, e_j \in E$ in a drawing $D$ of a graph $G$, $e_i$ and $e_j$ are bundled together if their ``similarity'' exceeds a certain threshold. 
When two edges $e_i$ and $e_j$ are bundled, the lines representing the edges are distorted towards each other, making them appear as a single ``bundle'' rather than two separate lines.

The similarity functions for edge bundling can generally be divided into four categories: {\em structural, attribute, geometric}, and {\em image-based}~\cite{Lhuillier2017StateOT}. 
As our new SEB method integrates structural and geometric-based similarities, in this paper we focus on bundling methods based on structural or geometric-based similarity; see the survey~\cite{Lhuillier2017StateOT} for other edge bundling methods. 
%It takes as input a drawing $D$ of $G = (V,E)$ and computes a \emph{distance function} $\upkappa$ for each pair of edges $e_i, e_j \in E$ such that if $\upkappa (e_i, e_j)$ is lower than a threshold, $e_i$ and $e_j$ are ``bundled'' together.

One of the earlier geometry-based bundling methods, \emph{GBEB (Geometry-Basd Edge Bundling)}~\cite{gbeb}, defines a control mesh on top of the drawing, and then clusters edges based on their proximity to control points. 
However, the quality of the result is highly dependent on the quality of the control mesh, and automated mesh generation methods often fail to consider the graph's topological structure.

\textit{Force-Directed Edge Bundling} (FDEB)~\cite{forceDirBund} adds sub-division points to each edge in the drawing $D$ and an attraction force is computed between a pair of edges using a \emph{compatibility} function $C_G(p,q)$, which decides whether two edges $p, q \in E$ are ``compatible'' enough to be bundled together based on their geometry such as angle, scale, position, and visibility compatibility. 
\textit{Topology-Geometry-Importance Edge Bundling} (TGI-EB)~\cite{nguyen2012tgi} incorporates topology (e.g., clustering topology) and importance (e.g., centrality) compatibility together with the geometric compatibility.

\textit{Edge Path Bundling} (EPB)~\cite{edge-path} is one of the latest edge bundling methods, aiming to reduce the \emph{ambiguity} in bundled drawings. Specifically, EPB prioritizes bundling long edges, where if an edge $e = (u,v)$ is selected for bundling, it is bundled along the shortest path between $u$ and $v$ using the vertices on the shortest path as the control points. 

Edge bundling methods tend to have high runtime complexity, such as the $O(m^2n)$ runtime for FDEB and $O(m^2 \log n)$ runtime for EPB. 
Methods to reduce the runtime has been introduced, such as \emph{Fast Fourier Transform Edge Bundling (FFTEB)}~\cite{ffteb}, which leverages Fast Fourier Transform (FFT) to speed up the computation using kernel density estimation, 
%However, FFTEB does not solve underlying readability issues,
and \emph{Spanner Edge-Path Bundling (SEPB)}~\cite{sepb}, which accelerates EPB using the decomposition into biconnected components.  
% the worst-case runtime of SEPB is still $O(m^2 \log n)$.

\subsection{Edge Bundling Quality Metrics}

A number of metrics have been presented to evaluate the quality of edge-bundled drawings. 
%including both readability~\cite{aestheticMetric,purchase2004evaluating} and faithfulness metrics~\cite{faithfulness}. 
%
\emph{Ink reduction}~\cite{edge-path} is a readability metric measuring how a bundled drawing simplifies the visual clutter, using the ratio between the number of colored pixels in the bundled drawing and the number of colored pixels in the original drawing. 

Specifically, ink reduction is computed as $\operatorname{ink}_J=\frac{\sum_{i=1}^m \sum_{j=1}^n I^B(i, j)}{\sum_{i=1}^m \sum_{j=1}^n J^B(i, j)}$, where $J$ is the grayscale image of the original drawing, $I$ is the grayscale bitmap image of the bundled drawing, and superscript $B$ denotes binarization where a pixel is considered occupied if it exceeds a threshold grey value. A lower value for $\operatorname{ink}_J$ is better, as it denotes less ink used.

\emph{Distortion}~\cite{edge-path} measures the average factor by which edge lengths increase in the bundled drawing, where a lower distortion is desirable as higher distortion leads to adjacencies being harder to read~\cite{huang2009graph}. 
%XXXX A distortion value of 1 denotes a lack of distortion, with higher values denoting higher distortion. 

Specifically, distortion is computed as $\operatorname{dist}(\Gamma)=\frac{1}{|E|} \sum_{(u, v) \in E} \frac{d_{\Gamma}(u, v)}{||u-v||}$, where $d_{\Gamma}(u, v)$ is the length of curve connecting the vertices $u$, $v$ in the bundled drawing, and $||u-v||$ is the Euclidean distance between $u$ and $v$. A lower value for $\operatorname{dist}(\Gamma)$ is better.

\emph{Ambiguity}~\cite{edge-path} is a faithfulness metric measuring the extent of perceivable \emph{false connections} in a bundled drawing, where a low ambiguity is desired. 
When two independent edges $(u,v)$ and $(x,y)$ are bundled together, the adjacency between the end vertices of the edges becomes ambiguous as \emph{false neighbors}. 

The ambiguity metric measures the ratio of false neighbors to the total number of true and false neighbors, as follows:  $\operatorname{amb}^\gamma(\Gamma)=\frac{\sum_{v \in V} \sum_{e=(v, w) \in E}\left|N_{\Gamma}^f(v, e)\right|}{\sum_{v \in V} \sum_{e=(v, w) \in E}\left|N_{\Gamma}(v, e)\right|}$, where $\gamma$ is the distance threshold, and $N_{\Gamma}(v, e)$ is the set of reachable neighbors of $v$ along edge $e$ in the bundled drawing where a neighbor is considered reachable if it is within a hop distance of $\gamma$. 

Variations of the ambiguity metric are defined using different values of $\gamma$, e.g., $Amb^1$ for $\gamma = 1$ and $Amb^2$ for $\gamma = 2$. A lower value for ambiguity is better, denoting more faithfulness.

\subsection{Faithfulness Metrics for Complex Graph Drawing}

{\em Faithfulness} metrics are introduced for evaluating drawings of large and complex graphs, by measuring how faithfully the {\em ground truth} structure of the graph is displayed in a drawing, complementing traditional {\em readability} metrics (or aesthetic criteria), which evaluate graph drawing based on how humans perceive the drawing.  

Various faithfulness metrics have been presented: 

\begin{itemize}
    \item \emph{Stress} measures how proportional the geometric distances between vertices in a drawing are to the shortest path distance between the vertices in the graph~\cite{battista1998graph}.
    \item \emph{Shape-based metrics} measures how faithfully the ``shape'' of the drawing, quantified using the \emph{proximity graph} of the vertex point set of the drawing, represents the ground truth structure of a graph~\cite{eades2017shape,hong2022dgg}.
    \item \emph{Cluster faithfulness}~\cite{meidiana2019quality} measures how faithfully the ground truth clustering of vertices is represented as the geometric clustering in the drawing.
    \item \emph{Automorphism faithfulness}~\cite{meidiana2020symmetry} measures how faithfully the ground truth automorphisms of a graph are represented as the symmetries in the drawing.
    \item \emph{Proxy quality metrics}~\cite{nguyen2017proxy} measure how faithfully the drawing of a sample graph represents the ground truth structure of the whole graph by computing the similarity between a graph $G$ and the ``shape'' of the drawing $D'$ of a sample graph $G' \subset G$.
\end{itemize}

%%%%%%%% explain in prox bundling quality metric

\begin{figure*}[h]
\centering
\includegraphics[width = 0.8\textwidth]{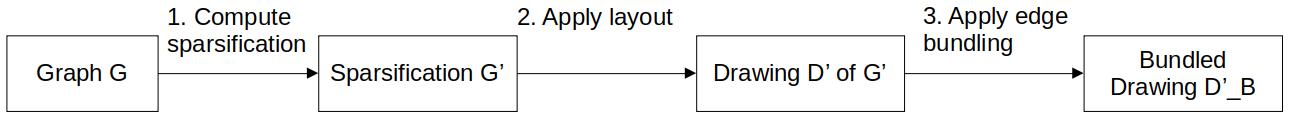}
\caption{FEB Framework.}
\label{tab:Sparsified Bundling}
\end{figure*}

\section{Fast and Faithful Edge Bundling }

\subsection{FEB (Fast Edge Bundling) Framework.}

Graph sampling is a popular approach to reducing the runtime for drawing large and complex graphs $G$, i.e., drawing a smaller sample $G'$ of $G$ is much faster than drawing $G$.
In particular, spectral sparsification ($SS$) has demonstrated effectiveness for sampling (i.e., outperform random sampling on quality metrics) and efficiency for drawing (i.e., fast drawing algorithms with similar quality)~\cite{eades2018drawing,meidiana2019topology,hu2019spectral,Hu3,meidiana2020sublinear-tvcg,meidiana2021stress}.
We, therefore, leverage $SS$ for efficient and effective edge bundling on the drawings of large and complex graphs.

Specifically, we present the \textit{FEB} framework, which computes edge bundling on the drawing $D'$ of a sparsification $G' \subset G$, instead of the drawing $D$ of the whole graph $G$. 
Figure \ref{tab:Sparsified Bundling} shows the FEB framework consisting of the following steps:

\begin{enumerate}
    \item Compute a sparsification $G'$ of $G = (V, E)$.
    \item Compute a drawing $D'$ of $G'$, either using a fixed layout or by applying a graph drawing algorithm.
    \item Compute a bundled drawing $D'_B$ of $D'$ using an edge bundling algorithm. 
\end{enumerate}

More precisely, we present a specific instance of the FEB framework leveraging  $SS$, which computes a sparsification $G'$, preserving important structural properties of $G$, such as connectivity, clustering, and commute distance~\cite{eades2018drawing,meidiana2019topology,hu2019spectral,Hu3,spielman2011spectral}. 
We expect a bundled drawing $D'_{B'}$ of $G'$ using $SS$ can maintain high similarity to the original bundled drawing $D_B$ of $G$. 

Furthermore, as $SS$ reduces the number of edges to $O(n \log n)$ for dense graphs with $O(n^2)$ edges, and successfully improves the runtime of drawings of large complex graphs~\cite{meidiana2020sublinear-tvcg,meidiana2021stress}, 
we expect that $SS$ can significantly improve the runtime of edge bundling algorithms (i.e, 
$D'_{B}$ of $G'$ can be computed much faster than $D_B$ of $G$).

\subsection{FBQ (Fast Bundling Quality) Faithfulness Metrics}

We now present the \emph{FBQ (Fast Bundling Quality)} faithfulness metrics for evaluating the effectiveness of the FEB framework. 
Specifically, we measure the \emph{proxy bundling faithfulness}, i.e., how faithfully the bundled drawing $D'_B$ computed by FEB preserves the structures of the original bundled drawing $D_B$ without sparsification, by extending the concept of proxy faithfulness metric for evaluating drawings of graph samples~\cite{nguyen2017proxy}.% XXXXX REF.

\begin{figure*}
\centering
\includegraphics[width = 0.8\textwidth]{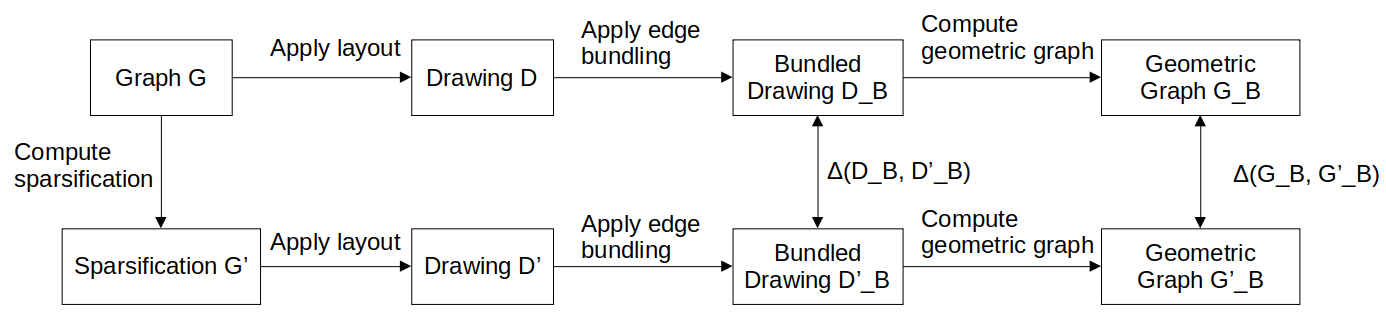}
\caption{FBQ Faithfulness Metrics framework.} 
\label{tab:Sparsified Bundling-experiment}
\end{figure*}

Figure \ref{tab:Sparsified Bundling-experiment} shows the FBQ metrics, where we measure the effectiveness of FEB (i.e., the similarity between $D_B$ and $D'_B$) by computing the similarity between $G_B$ and $G'_B$. 
Specifically, the FBQ metric is computed as follows:

\begin{enumerate}
    \item Compute a sparsification $G'$ of $G$.
    \item Compute drawings $D$ of $G$ and $D'$ of $G'$, either using a fixed layout or by applying a graph drawing algorithm.
    \item Compute bundled drawings $D_B$ of $D$ and $D'_B$ of $D'$ using an edge bundling algorithm. 
    \item Construct geometric graphs  $G_B$ of $D_B$ and $G_B'$ of $D_B'$. 
    \item Compare the similarity between $D_B$ and $D'_B$ (i.e., $\Delta (D_B,D'_B)$)
    by computing the similarity between $G_B$ and $G'_B$ (i.e., $\Delta (G_B,G'_B)$).

\end{enumerate}

\begin{comment}
\begin{figure}
     \centering
     \subfloat[$D$]{
         \centering
         \includegraphics[width=0.19\columnwidth]{metricA.png}
         \label{fig:geo-drawing}
     }
     \subfloat[$D_B$]{
         \centering
         \includegraphics[width=0.19\columnwidth]{metricB.png}
         \label{fig:geo-bundle}
     }
     \subfloat[$G_B$]{
        \centering
        \includegraphics[width=0.19\columnwidth]{metricC.png}
        \label{fig:geo-geograph}
     }
        \caption{Example of computing a geometric graph $G_B$. (A) Edges $e=(s,t)$ and $e'=(u,v)$ in the original drawing $D$. (B) Edges $e$ and $e'$ in the bundled drawing $D_B$. (C) All edges between the vertices $(s, t, u, v)$ in the geometric graph $G_B$.} 
        \label{fig:geograph_construct}
\end{figure}
\end{comment}

In Step 4, we construct a \emph{geometric graph} from the bundled drawing as follows~\cite{edge-path}. 
Consider a bundled drawing $D_B$, where edges $e_1 = (s,t)$ and $e_2 = (u,v)$ are bundled together, with vertices $s$ and $u$ on the same end of the bundle and $t$ and $v$ are on the other end of the bundle. 
Once the edges are bundled, not only the endpoints of each edge $e_1$ and $e_2$ would appear as neighbors, but also the pairs ($s$,$v$) and ($t$,$u$). 
The geometric graph $G_B$ thus contains four edges $(s,t), (u,v), (s,v), (t,u)$,  including the edges between these pair of vertices ``implied'' as neighbors by bundling. %Figure \ref{fig:geograph_construct} shows an example of how to construct a geometric graph $G_B$ from a bundled drawing $D_B$, based on the criteria described in~\cite{edge-path}.
%After applying the edge bundling algorithm on the drawing $D$, %(Figure \ref{fig:geograph_construct}\subref{fig:geo-drawing}), 
%some edges will form bundles. %, such as in Figure \ref{fig:geograph_construct}\subref{fig:geo-bundle}. 
%Here, edge $e=(s,t)$ is bundled with edge $e^{\prime}=(u,v)$ and the two edges touch at point $p$; this intersection makes it difficult to differentiate which pairs of vertices were originally connected by an edge in $G$.
%Thus, in the geometric graph $G_B$, we add all the edges $(s,t),(s,v),(u,t),(u,v)$ (see Figure \ref{fig:geograph_construct}(\subref{fig:geo-geograph})).

For Step 5, we present two variants of the $FBQ$ metric, based on how to  measure the similarity between $G_B$ and $G'_B$ (i.e., $\Delta (G_B, G'_{B})$).
The first variant $FBQ_{JS}$ is based on the \emph{Jaccard Similarity (JS)}~\cite{jaccard}, computed as follows:
    \[
    FBQ_{JS}\left(G_B, G'_{B}\right)=\frac{1}{|V|} \sum_{v \in V} \frac{\left|N_B(v) \cap N_B^{\prime}(v)\right|}{\left|N_B(v) \cup N_B^{\prime}(v)\right|}
    \]

\noindent where $N_B(v)$ (resp., $N_B^{\prime}(v)$) is the set of neighbors of $v$ in $G_B$ (resp., $G'_B$).

The second variant $FBQ_{SQ}$ is based on the \emph{sampling quality metrics}, which computes the similarity between a sample graph and the original graph using important structural graph properties. 

Specifically, the similarity between $G_B$ and $G_{B'}$ is computed using \emph{Kolmogorov-Smirnov (KS) distance}, which measures the similarity between two cumulative distribution functions (CDFs) as follows:

\[
FBQ_{SQ} = KS (SQ(G_B),SQ(G'_{B}))
\]

\noindent
where $SQ(G_B)$ (resp, $SQ(G_{B'})$) is the CDF of a selected graph property computed on $G_B$ (resp., $G_B'$). 

More specifically, we define $FBQ$ based on the graph properties,  popular sampling quality metrics $SQ$, as follows:

\begin{itemize}
\item $FBQ_{SQ(DG)}$: based on the \emph{Degree Correlation Associativity (DEG)}, which measures the likelihood that vertices connect to other vertices with similar degrees~\cite{degcorr}.

\item $FBQ_{SQ(CC)}$: using \emph{Clustering Coefficient (CC)}, which measures the tendency of vertices to be clustered together~\cite{clust}.
\end{itemize}

%    \item[\textit{Degree Correlation Associativity (DEG)}~\cite{degcorr}] measures the likelihood that vertices are connected to other vertices with similar degree (denoted as $FBQ_{SQ(DG)}$)
%    \item[\textit{Clustering Coefficient (CC)}~\cite{clust}] measures the extent of vertices to be clustered together (denoted as $FBQ_{SQ(CC)}$)
%\end{description}

\begin{comment}
\begin{itemize}
    \item \textit{Degree Correlation (DEG)} \cite{degcorr}. Measures the likelihood that vertices are connected to other vertices with a similar degree. 
    \item \textit{Average Neighbour Degree (AND)} \cite{and}. Computes the average degree of a vertex's neighbours.
    \item \textit{} \cite{clust}. Calculates the extent of vertices to be clustered together. 
    \item \textit{Closeness Centrality (CLOSE)} \cite{closeness}. A centrality measure that sums up the length of all shortest paths between the vertex and other vertices in the graph. 
\end{itemize}
\end{comment}

\section{Spectral-Faithful Edge Bundling}

\subsection{Effective Resistance-based Compatibility}

Spectral sparsification ($SS$) has demonstrated the effectiveness and efficiency for sampling (i.e., outperforms random sampling~\cite{eades2018drawing,meidiana2019topology,hu2019spectral,Hu3}) and drawing (i.e., sublinear-time drawing algorithms~\cite{meidiana2020sublinear-tvcg,meidiana2021stress}) of large and complex graphs. 

Therefore, we leverage $SS$ for edge bundling for the {\em effectiveness} (i.e., preserving the spectrum of a graph) and {\em efficiency} (i.e., near-linear runtime), by introducing the  \emph{effective resistance-based compatibility} between edges.

We select effective resistance values to design a new compatibility measure since they are highly correlated to the commute distances in the graph, therefore bundling the edges with similar structural properties together, resulting in more {\em spectrum faithful edge bundling}.

Moreover, computing the effective resistance-based compatibility runs much faster (i.e., near-linear time) than the existing betweenness centrality-based compatibility 
used in the TGI-EB~\cite{nguyen2012tgi},
since computing betweenness centrality takes $O(n^3)$ time for dense graphs, which is not scalable for large and complex graphs.

Specifically, we present two variants for effective resistance-based compatibility $C_{ER}$. 
The first variant uses the absolute difference in effective resistance values, such that edges with only a small difference in effective resistance can be bundled together:
\begin{equation} \label{eq:cer1}
    C_{ER_1}(P,Q) = 1 - | ER(P) - ER(Q) |
\end{equation}

\noindent
where $P, Q \in E$ are edges in a graph $G = (V, E)$ and $ER(P)$ denotes the normalized effective resistance value of edge $P$. 
$C_{ER_1}(P, Q) = 1$ when the values of $ER(P)$ and $ER(Q)$ are the same 
(i.e., higher compatibility values for edges with similar effective resistance values).

The second variant uses the ratio between the effective resistance values to minimize the bias introduced by using absolute difference,
i.e., the same magnitude of difference disproportionately affects the compatibility value when the effective resistance values are low. 
Using ratios limits this effect, so compatibility can be defined equally on the higher and lower end of effective resistance values. 

More specifically, the second variant is defined as:
\begin{equation} \label{eq:cer2}
    C_{ER_2}(P,Q) = \frac{min(ER(P),ER(Q))}{max(ER(P),ER(Q)}
\end{equation}

\subsection{SEB: Spectral Edge Bundling Algorithm}

We now present SEB, a spectral-faithful edge bundling, which integrates the new effective resistance-based compatibility into the force-directed edge bundling (FDEB) model. 

Specifically, FDEB takes a drawing $D$ of a graph $G = (V, E)$, as input and divides each edge $e$ into segments separated by \emph{division points} $p_1, p_2, \ldots, p_d$. 
A force $F_{pi}$ is then exerted on each division point $p_i$, where the magnitude and direction of the force are based on the \emph{geometric compatibility} between the edge $e$ (that $p_i$ lies on) and every other edge in $E$. 

In SEB, for an edge $P \in E$ in a bundled drawing $D_B$ of graph $G = V,E)$, the force $F_{pi}$ exerted on a division point $p_i$ on the projection of $P$ in $D_B$ is defined as:

\begin{equation}
    \begin{aligned}
    F_{pi} = k_p \left ( || p_{i-1} - p_i|| + || p_i - p_{i+1} || \right ) \\
    + \sum_{Q \in E} \frac{C_G(P,Q) C_{ER}(P,Q)}{||p_i - q_i||}
    \end{aligned}
\end{equation}

\noindent
where $p_{i-1}$ (resp., $p_{i+1}$) is the division point directly before (resp., after) $p_i$ on $P$, $q_i$ is the division point on $Q \in E$ with the closest distance to $p_i$, $C_G(P, Q)$ is the geometric compatibility used in FDEB computed using geometric properties of the two edges, and $C_{ER}(P, Q)$ denotes our new effective resistance-based compatibility (either $C_{ER_1}$ or $C_{ER_2}$). 

Note that $C_G(P, Q)$ is computed as a product of several geometric compatibility functions. 
Therefore, we multiply $C_{ER}(P, Q)$ with $C_G(P, Q)$ to obtain a compatibility score that balances the geometry in the drawing (with geometric compatibility) and the ground truth structure of the graph (with effective resistance compatibility).

\section{FEB Experiments}

%\subsection{Experiment Design}

We now present experiments for evaluating the efficiency and effectiveness of the FEB framework (i.e., FEB runs significantly faster than the original bundling methods without sparsification, while maintaining a similar quality). 

Spectral sparsification reduces the number of edges of the graph to be bundled, thereby improving the runtime efficiency, while preserving important structural properties of the graph, enabling the bundled drawing to faithfully represent the original edge bundling. 
%We hypothesize the following:

%\begin{description}
%\item[Hypothesis 1.] \textit{FEB runs faster than direct edge bundling approaches. }
%\item[Hypothesis 2.] \textit{FEB computes bundled drawings with similar quality to direct edge bundling approaches.}
%\end{description}

We use the same data sets and edge bundling methods in Section \ref{sec:exp_comp}, where we denote FEB using an original bundling method X as FX (i.e., FEB with SEB is denoted as FSEB).
For each bundling method, we compute bundled drawings $D_B$ (resp., $D'_B$) from a drawing $D$ (resp., $D'$) of the whole graph $G$ (resp., the sparsification $G'$).

%\subsection{Results}

\subsection{Runtime Comparison}

\begin{figure}[h!]
    \centering
    \subfloat[Runtime]{
        \includegraphics[width=0.22\columnwidth]{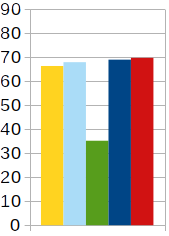}
        \label{fig:runtime-average}
    }
    \subfloat[$FBQ_{JS}$]{
        \includegraphics[width=0.22\columnwidth]{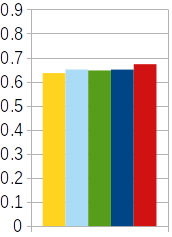}
        \label{fig:feb_all_avg}
    }
    \subfloat[$\scriptstyle FBQ_{SQ(DG)}$]{
        \includegraphics[width=0.22\columnwidth]{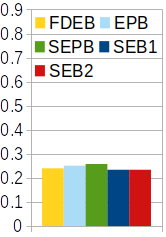}
        \label{fig:feb_kc_deg_avg.png}
    }
    \subfloat[$\scriptstyle FBQ_{SQ(CC)}$]{
        \includegraphics[width=0.22\columnwidth]{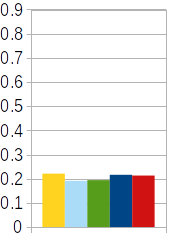}
        \label{fig:feb_kc_cc_avg.png}
    }
    \caption{(a) Runtime improvement: FEB methods obtain 61\% runtime improvement than original bundling methods; (b) $FBQ_{JS}$ (higher=better) 
    for FEB methods, on average 0.68, and (c-d) $FBQ_{SQ}$ (lower=better) for FEB methods, on average 0.26, showing high similarity between $D_B$ and $D'_B$.}
    \label{fig:feb-comb-avg}
\end{figure}

Figure \ref{fig:feb-comb-avg}\subref{fig:runtime-average} 
shows the runtime improvement achieved by FEB, computed as $\frac{t(X)-t(FX)}{t(X)}$, where $t(X)$ is the runtime by the original bundling method X and $t(FX)$ is the runtime by FEB using X.

On average, FEB obtains 61\% runtime improvement over the original bundling methods. 
Note that FEB with the FDEB family (i.e., FFDEB, FSEB1, FSEB2) obtains 68\% runtime improvement on average, where FSEB2 obtains the highest improvement at 70\%.

FEB with the EPB family (i.e., FEPB, FSEPB) obtains 53\% runtime improvement on average, due to SEPB having 35\% improvement since it is designed as a fast variation of EPB.

\begin{comment}
\begin{figure}[h]
    \centering
    \subfloat[Runtime]{
        \includegraphics[width=0.17\textwidth]{runtime-average.png}
        \label{fig:runtime-average}
    }
    \subfloat[$FBQ_{JS}$]{
        \includegraphics[width=0.17\textwidth]{feb_all_avg.png}
        \label{fig:feb_all_avg}
    }
    \caption{(a) Significant  runtime improvement by FEB, on average 61\% faster than the original bundling methods. %XXXXX on average numberXXXX
    (b) Average $FBQ_{JS}$ (higher=better) 
    for FEB methods, on average 68\%,  demonstrating the high similarity between $D_B$ and $D'_B$.
    %XXXX on average number XXXXX. 
    %All FEB methods significantly improve runtime over the original bundling methods; overall, FSEB2 obtains the highest $FBQ_{JS}$ on average. %FSEB2 achieves the highest runtime improvement and is closely followed by FSEB1 being the second place. Then FEPB has the third place for runtime improvement followed by FFDEB and FSEPB.
    }
    \label{fig:feb-comb-avg}
\end{figure}
\end{comment}

%FSEB2 obtains the highest improvement at 69.77\%, with FSEPB obtaining the lowest improvement of 35.2\%, likely due SEPB already being a fast modification of EPB.  %FSEB2 demonstrates the most significant average improvement at 69.77\% over SEB2, closely followed by FSEB1 at 69.01\%, FEPB at 67.97\%, FFDEB at 66.36\% and lastly FSEPB of 35.2\%.
%XXX compare by family XXXX explain why XXX 
%A bar chart for runtime comparison in terms of the data set can be found in figure \ref{fig:runtime-dataset} of Appendix \ref{appendixB}. XXXX CHART XXXXX

\begin{table}[h!]
    \centering
    \caption{$FBQ_{JS}$ results (higher=better) for all FEB methods. FSEB2 performs best on scale-free graphs and GION graphs, and FSEB1 performs best on black-hole graphs, while FEPB performs best on geographic graphs.}
    \footnotesize
    \begin{tabular}{|c|c|c|c|c|c|}
        \hline & \multicolumn{5}{|c|}{ $FBQ_{JS}$ } \\ % & \multirow[b]{2}{*}{ Avg of all EB method } \\
        \hline & FFDEB  & FEPB   & FSEPB  & FSEB1  & FSEB2 \\ % & \\
        \hline
        airlines   & 0.358  & 0.3909 & 0.3761 & 0.3625 & 0.4093 \\ \hline
        airtraffic & 0.4989 & 0.7004 & 0.6831 & 0.4996 & 0.511  \\ \hline
        migration  & 0.3848 & 0.7494 & 0.7305 & 0.4003 & 0.4277 \\
        \Xhline{4\arrayrulewidth} \textit{Average} & 0.424  & \textbf{0.614}  & 0.597  & 0.421  & 0.449\\ % & 0.499 \\
        \hline
        \hline yeastppi   & 0.5958 & 0.4413 & 0.4436 & 0.6533 & 0.6984 \\ \hline
        facebook   & 0.6282 & 0.6816 & 0.6909 & 0.6379 & 0.6664 \\ \hline
        lastfm     & 0.6218 & 0.6085 & 0.6085 & 0.6815 & 0.7397 \\ % & 0.65 \\
        \Xhline{4\arrayrulewidth} \textit{Average} & 0.615  & 0.577  & 0.581  & 0.658  & \textbf{0.702} \\ % & 0.647 \\
        \hline
        \hline 6\_gion    & 0.6952 & 0.6786 & 0.6702 & 0.7004 & 0.7267 \\ \hline
        7\_gion    & 0.6557 & 0.6694 & 0.67   & 0.6598 & 0.67   \\ % & 0.66 \\
        \Xhline{4\arrayrulewidth} \textit{Average} & 0.675  & 0.674  & 0.67   & 0.68   & \textbf{0.698} \\ % & 0.680 \\
        \hline
        \hline G443       & 0.9571 & 0.6371 & 0.6333 & 0.9552 & 0.9528 \\ \hline
        Cycle907   & 0.8718 & 0.8859 & 0.904  & 0.8755 & 0.8733 \\ \hline
        Cycle896   & 0.7229 & 0.7096 & 0.701  & 0.7251 & 0.7224 \\ % & 0.72 \\
        \Xhline{4\arrayrulewidth} \textit{Average} & 0.851  & 0.744  & 0.746 & \textbf{0.852} & 0.850 \\ 
        \hline
    \end{tabular}
    \label{tab:jaccard_metric}
\end{table}

\subsection{Quality Metrics Comparison}

%Table \ref{tab:avg-SBQ-compare} shows the average $\Delta SBQ$ of each FEB variant; to save space, each method name FX is shortened to F(X) in the table. In general, most FEB variants obtain a low magnitude of $\Delta SBQ$, only up to 5.5\%. Compared to the average runtime improvement at over 60\%,  this shows a favorable runtime-quality trade-off.% and supports Hypothesis 2.

Figure \ref{fig:feb-comb-avg}\subref{fig:feb_all_avg} demonstrates the high similarity between the bundled drawings computed by FEB ($D'_B$) and the original bundled drawings ($D_B$), with $FBQ_{JS}$ on average 68\%.   

Table \ref{tab:jaccard_metric} shows the details of $FBQ_{JS}$ for each FEB method, where FSEB2 obtains the highest $FBQ_{JS}$ on average. 
Overall, FEB with SEB methods performs better than FEB with EPB methods on most graphs:  
% except for geographic graphs, 
20\% better on scale-free and black-hole data sets on average.
%, while FEPB methods perform 30\% better on the geographic graphs.

Figures \ref{fig:feb-comb-avg}\subref{fig:feb_kc_deg_avg.png}-\subref{fig:feb_kc_cc_avg.png} show even 
higher similarity between the bundled drawings by FEB ($D'_B$) and the original bundled drawings ($D_B$), with $FBQ_{SQ(DG)}$ and $FBQ_{SQ(CC)}$.
On both metrics, FEB obtains 0.26 (i.e., 74\% similarity to the original bundled drawings) on average, showing that the bundled drawings by FEB faithfully display the structure of the underlying graph to a similar extent as the original bundled drawings. 
FSEB methods obtain slightly better $FBQ_{SQ(DG)}$ over FEB with the EPB family (EPB, SEPB), which obtain slightly better $FBQ_{SQ(CC)}$.

\begin{table*}[h!]
    \centering
    \footnotesize
    \begin{tabular}{|c|c||c|c|}
        \hline $D$  & $D'$  & $D_B$(FDEB) & $D'_B$(FFDEB) \\
        \hline  \includegraphics[width=0.22\textwidth]{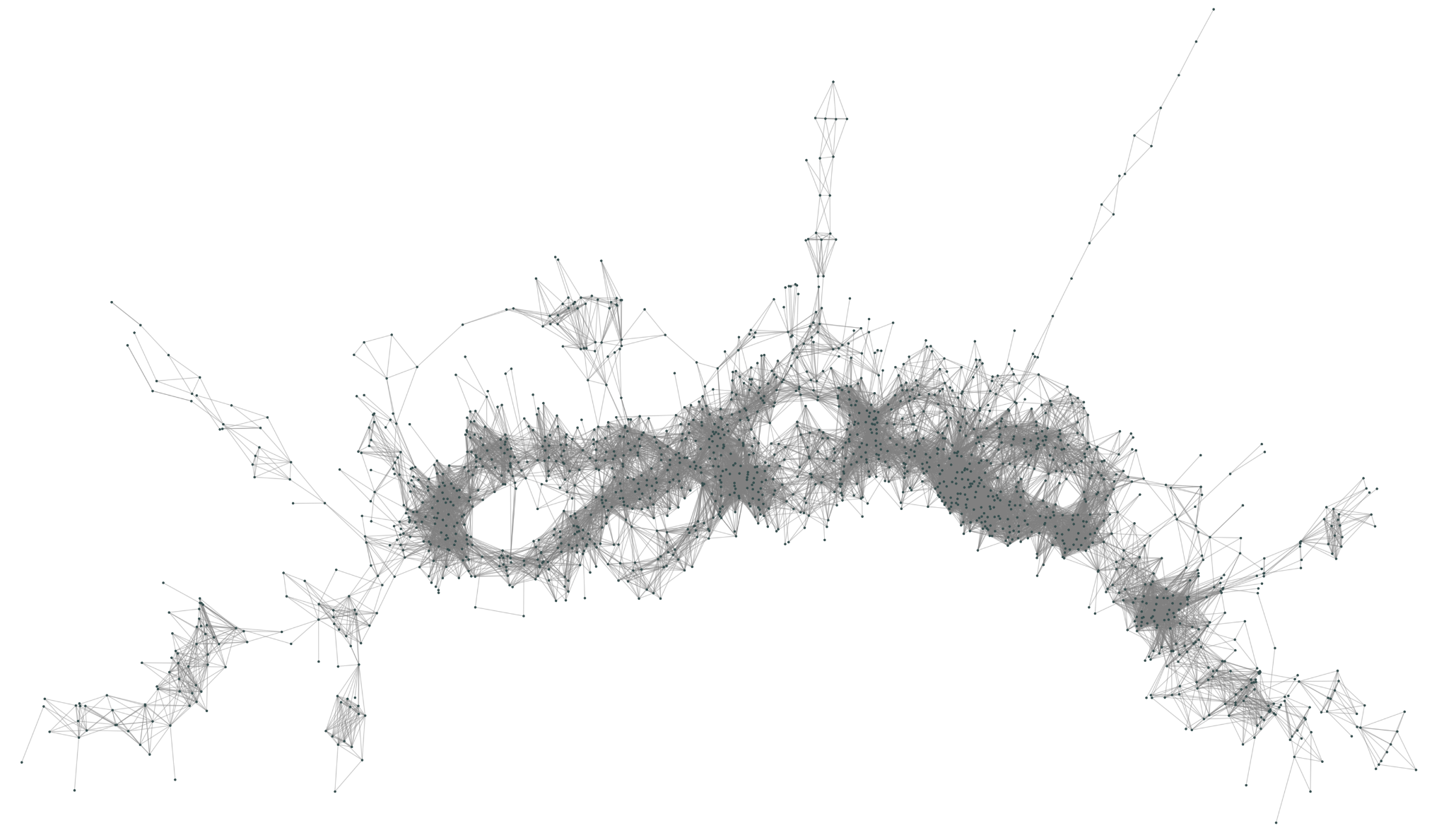}& \includegraphics[width=0.22\textwidth]{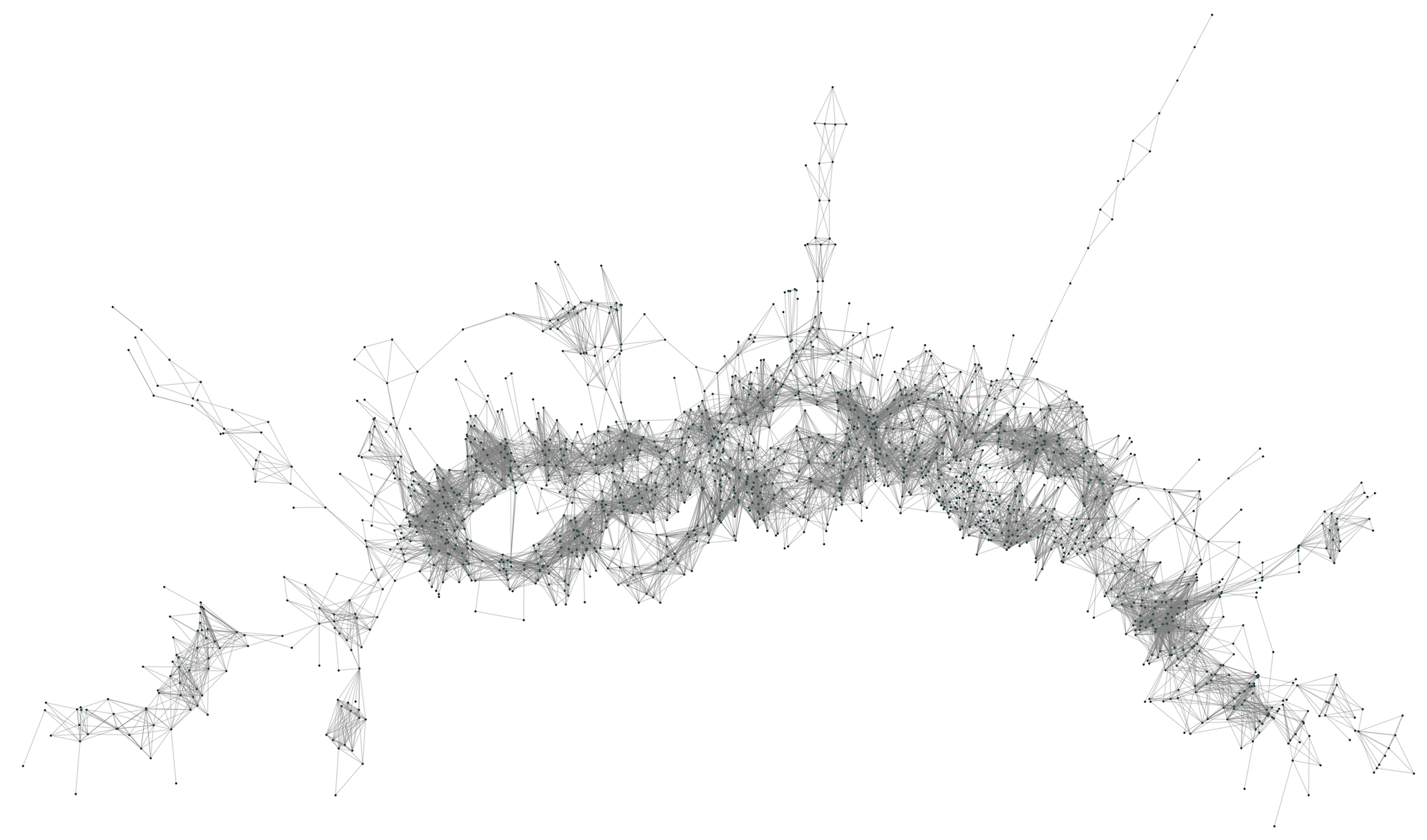} & \includegraphics[width=0.22\textwidth]{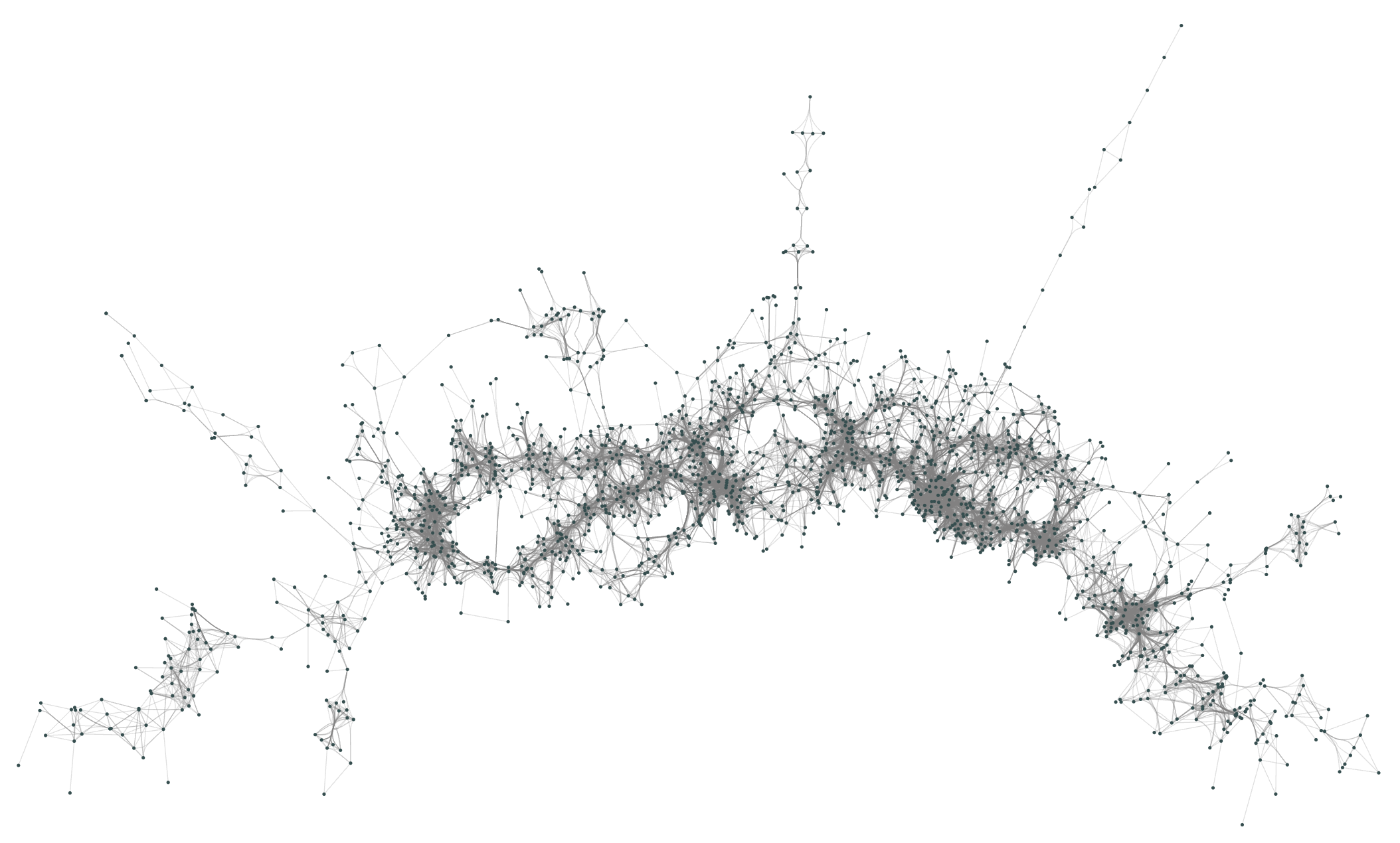} & \includegraphics[width=0.22\textwidth]{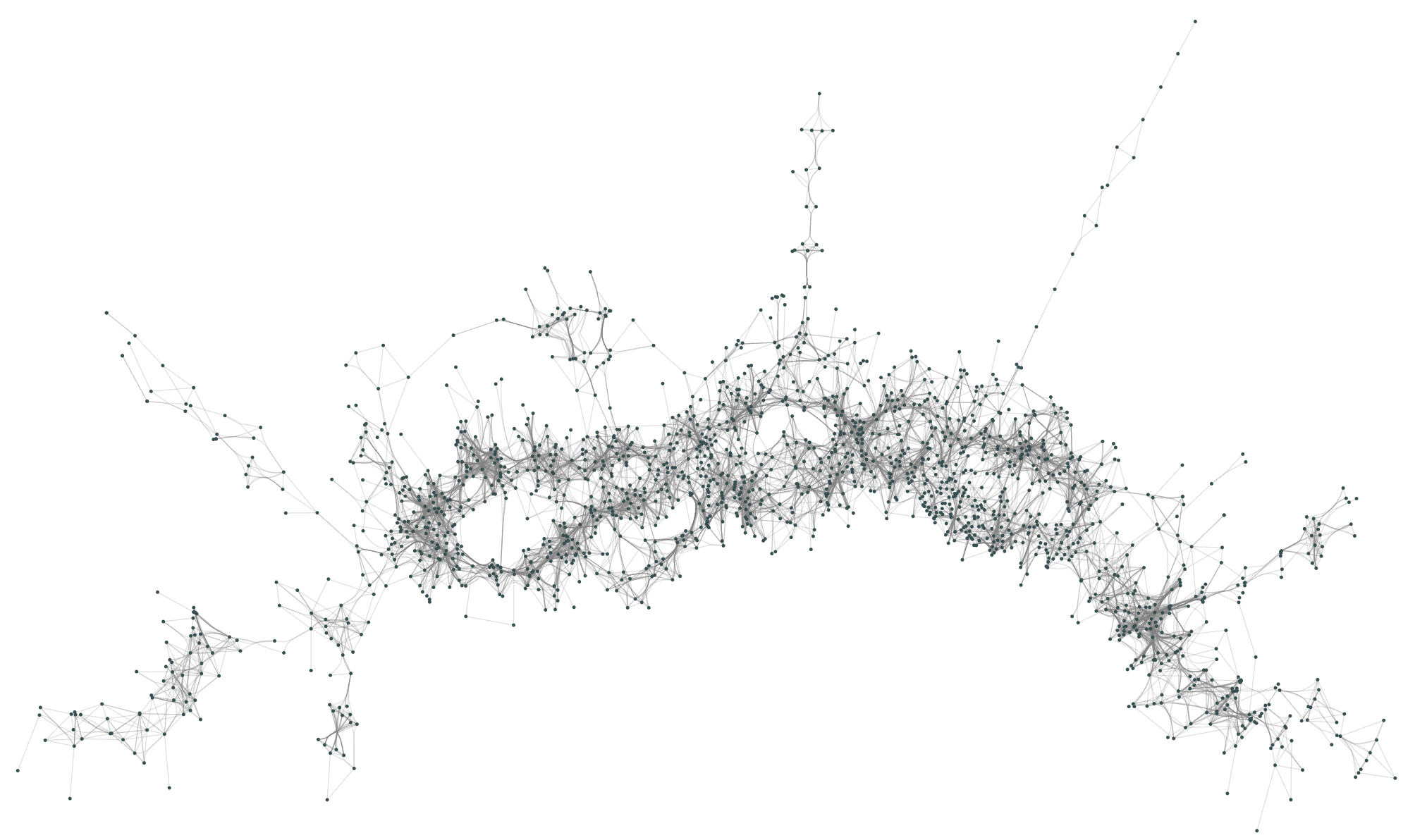} \\
        \hline $D_B$(EPB) & $D'_B$(FEPB)  & $D_B$(SEPB) & $D'_B$(FSEPB) \\
    \hline \includegraphics[width=0.22\textwidth]{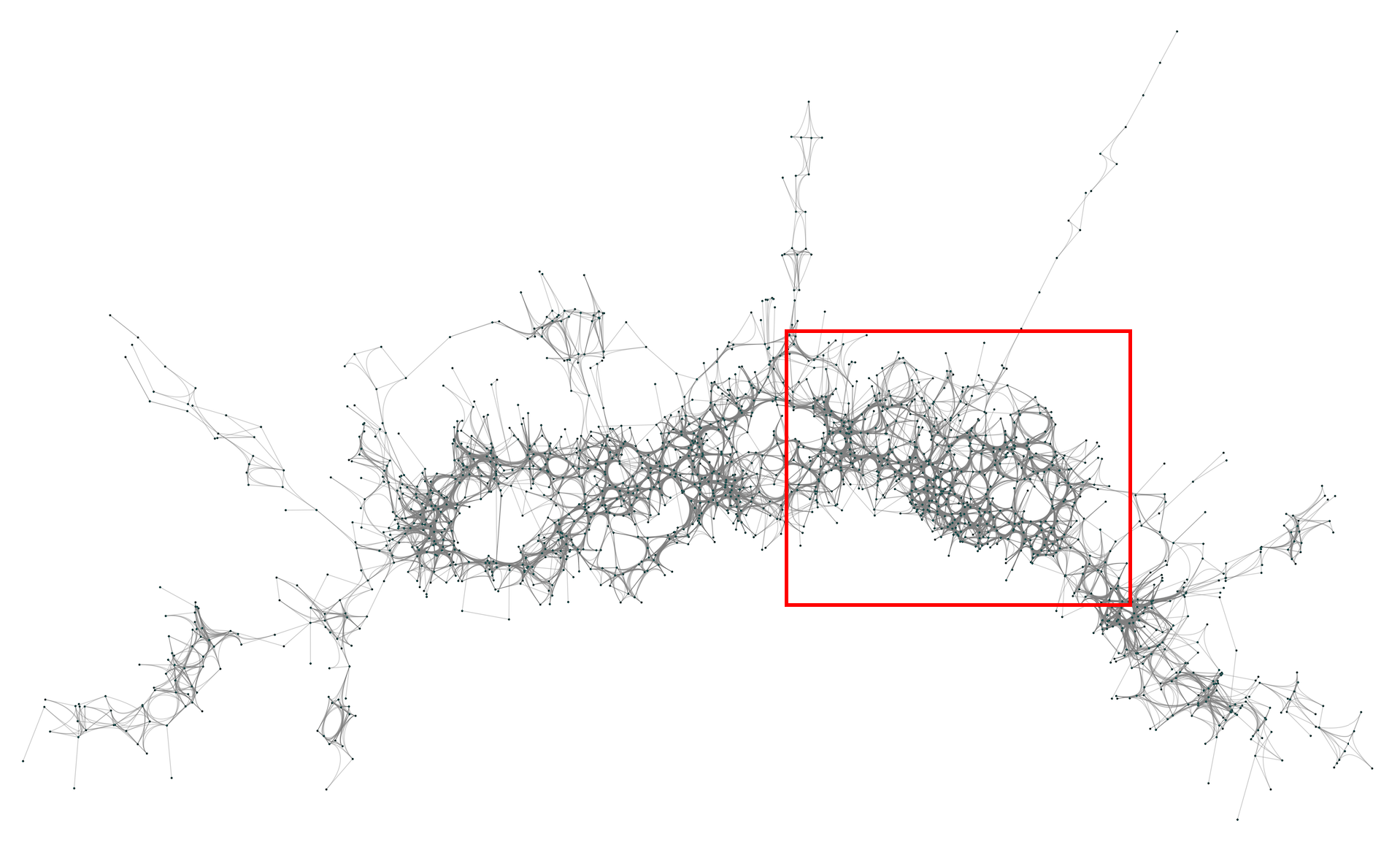} & \includegraphics[width=0.22\textwidth]{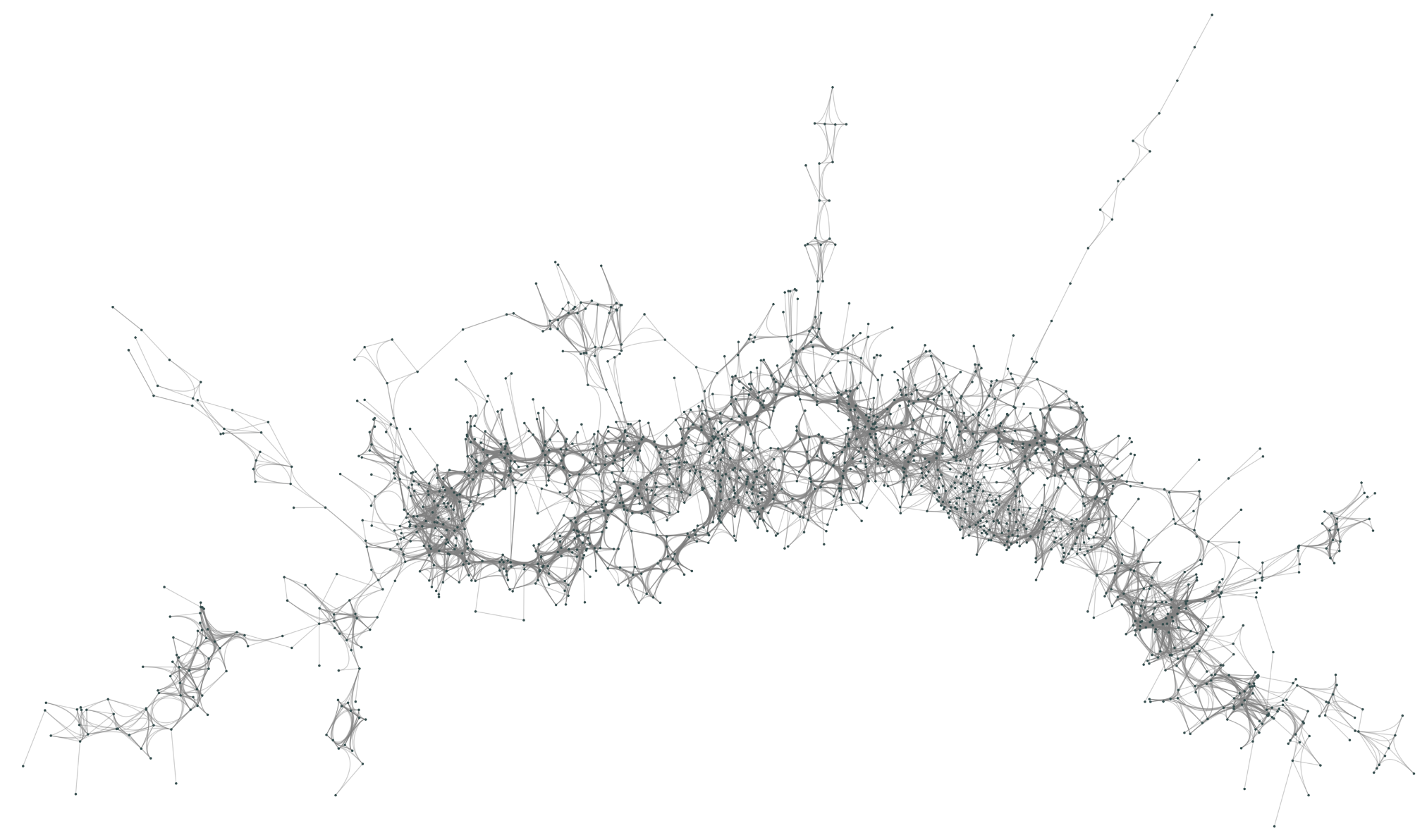} & \includegraphics[width=0.22\textwidth]{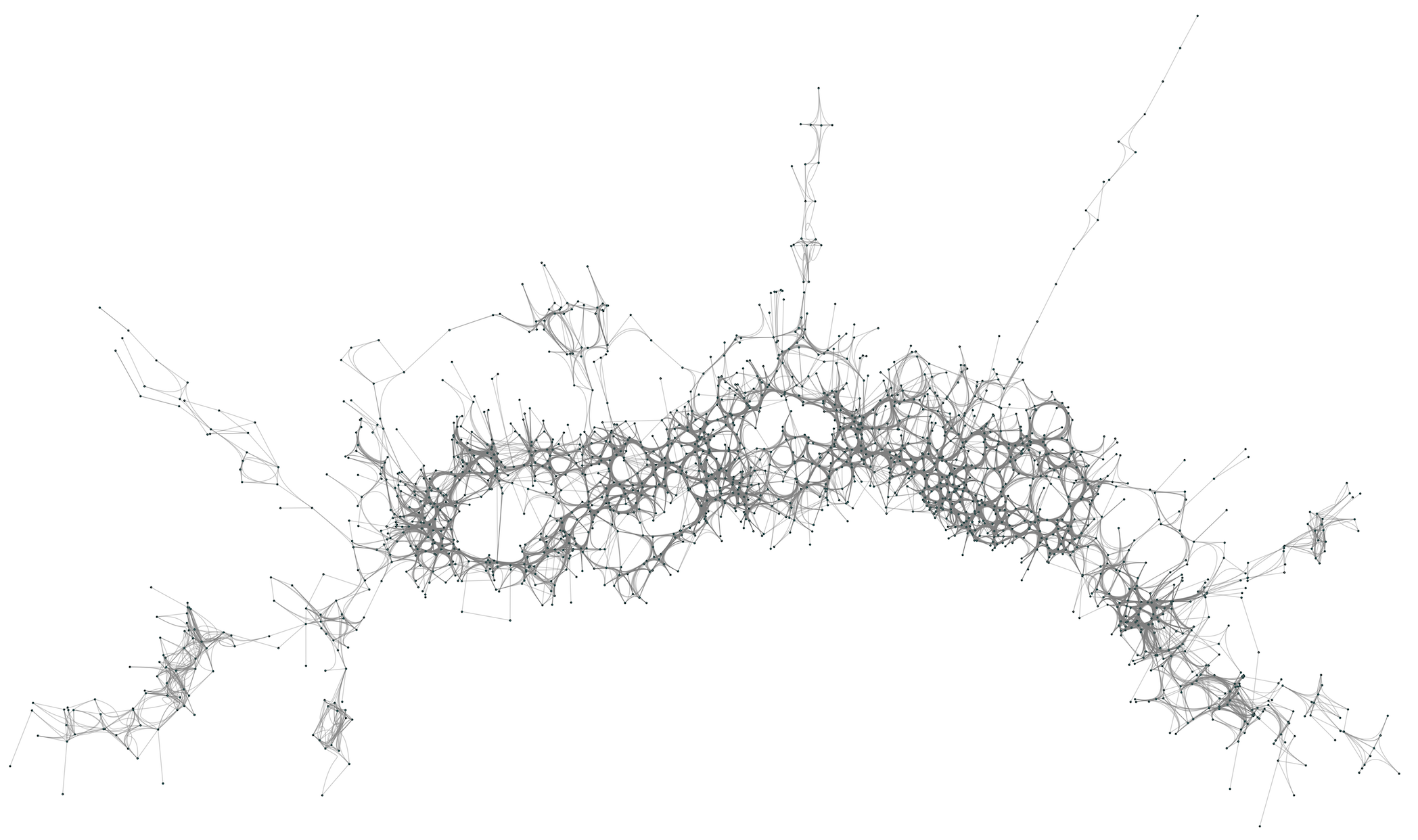} & \includegraphics[width=0.22\textwidth]{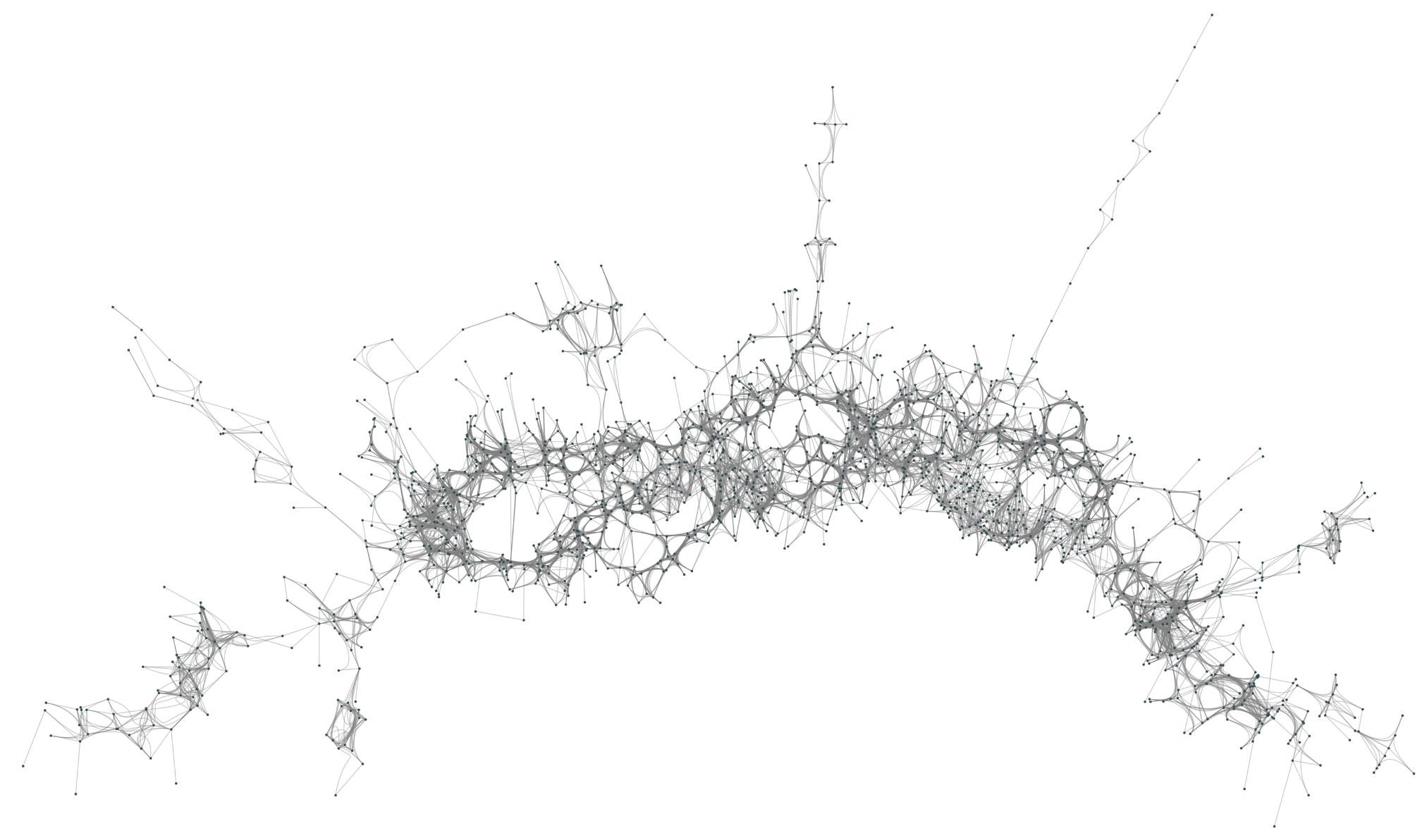}\\
    \hline $D_B$(SEB1) & $D'_B$(FSEB1)  & $D_B$(SEB2) & $D'_B$(FSEB2) \\
    \hline \includegraphics[width=0.22\textwidth]{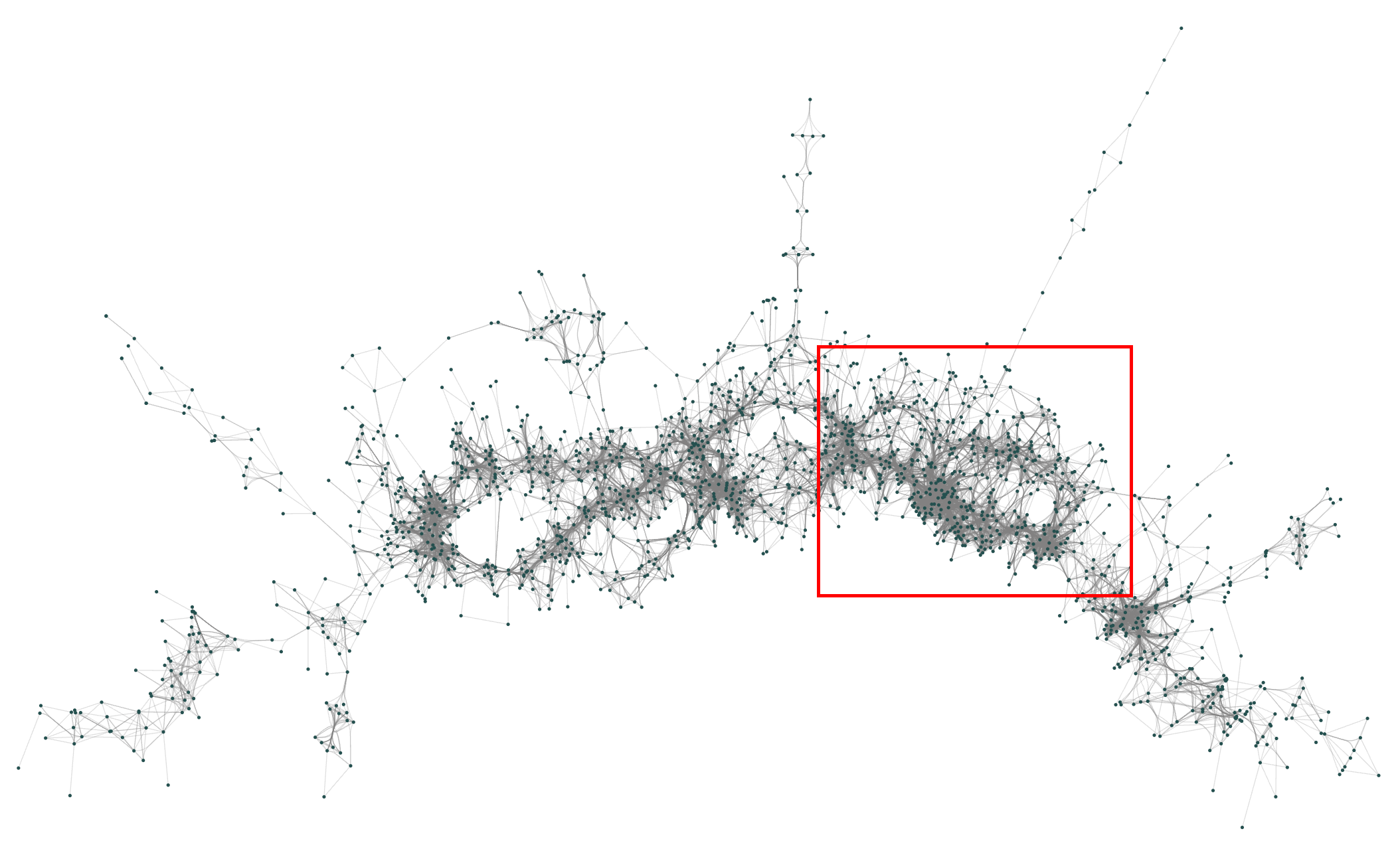} & \includegraphics[width=0.22\textwidth]{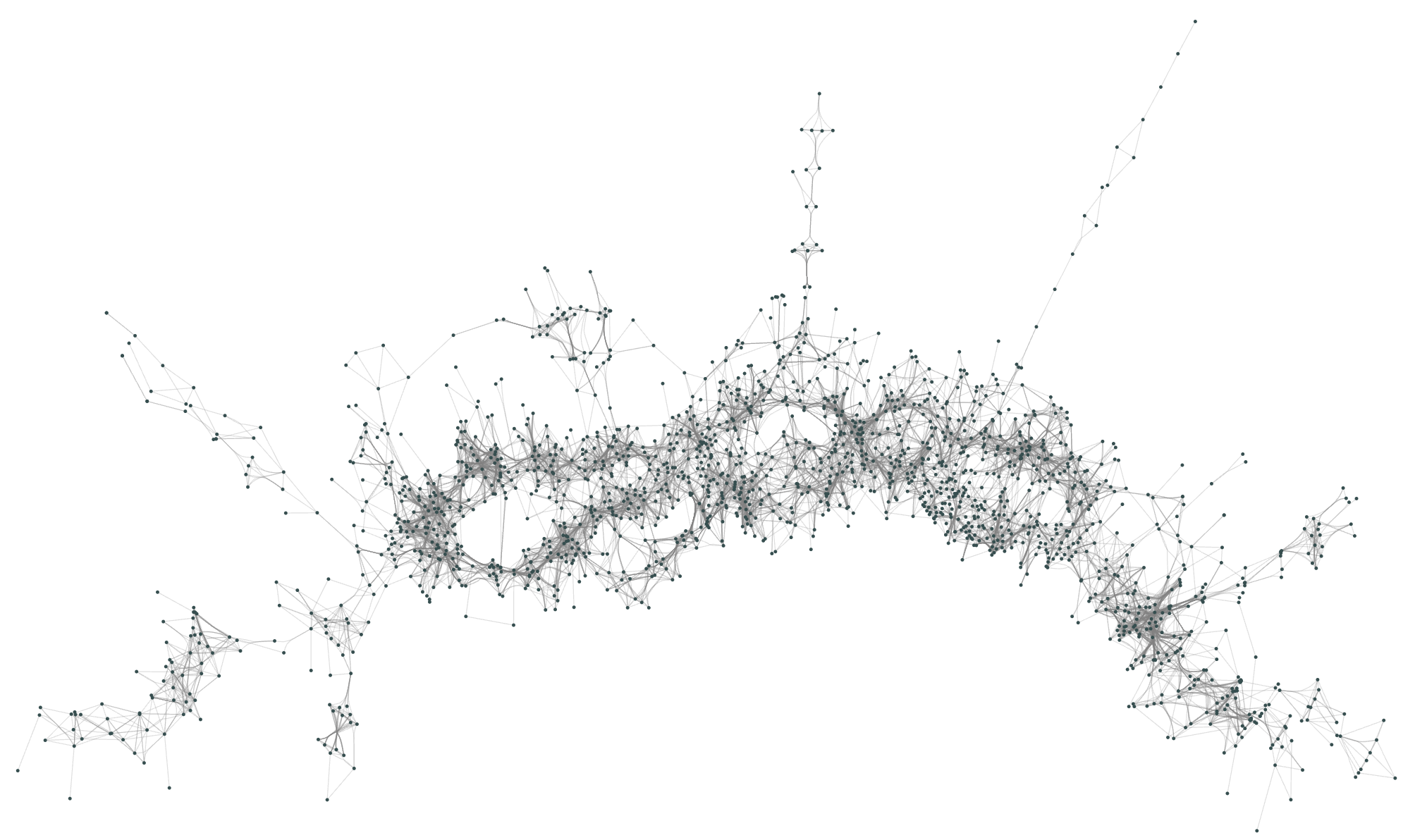} &  \includegraphics[width=0.22\textwidth]{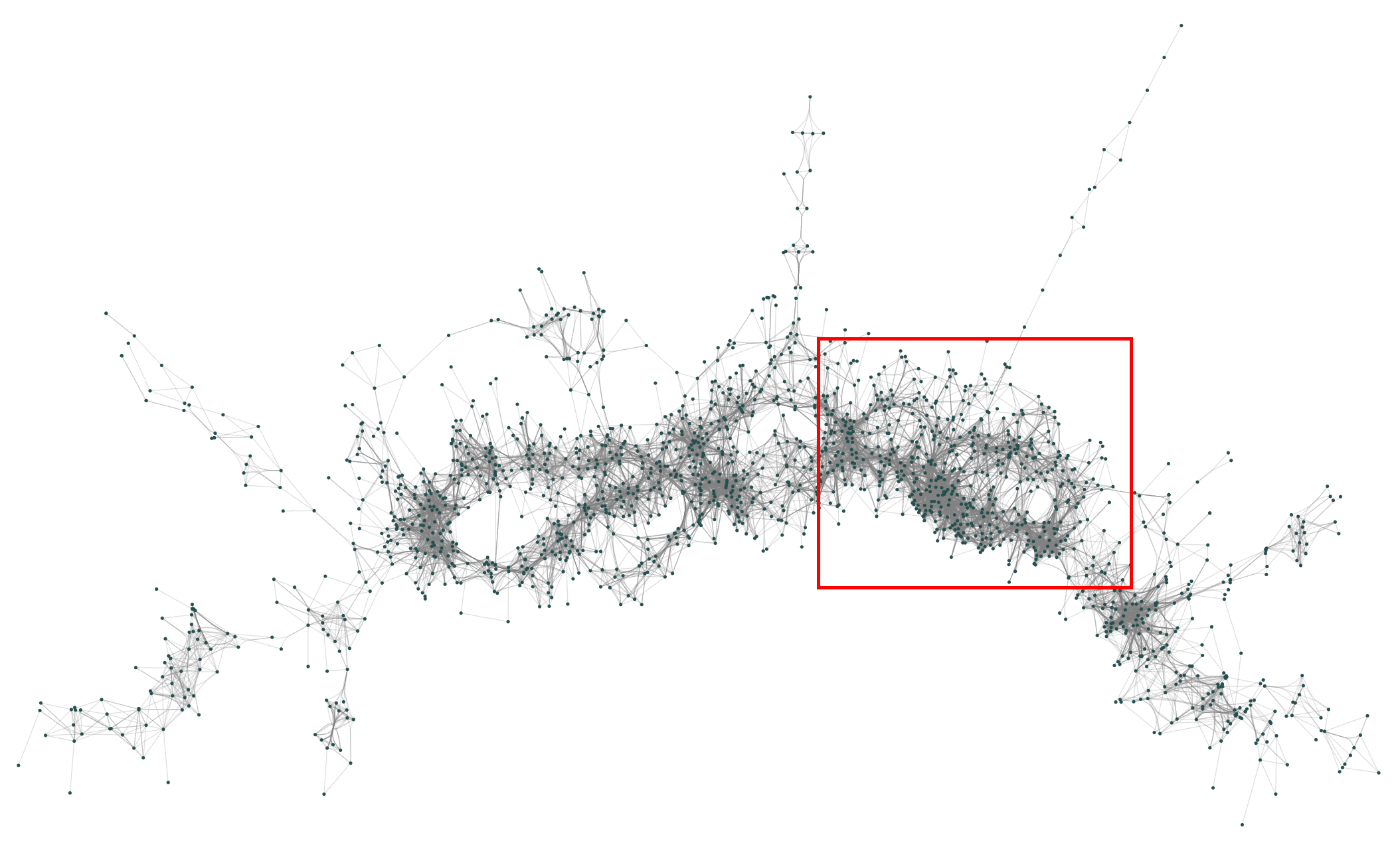} & \includegraphics[width=0.22\textwidth]{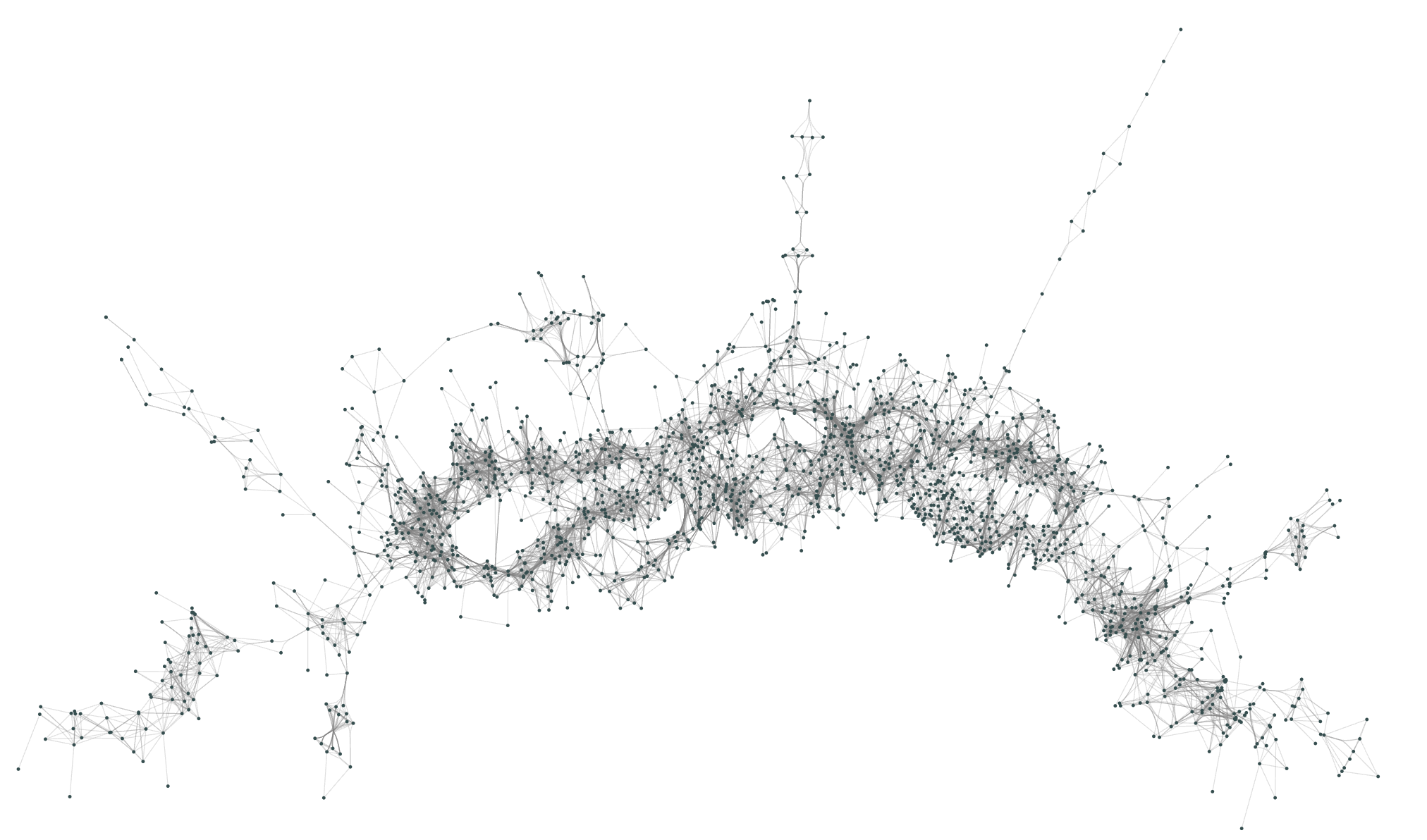} \\
        \hline
    \end{tabular}
    \caption{Visual comparison for the 6\_gion graph. SEB methods bundle the large dense cycles without disproportionally highlighting small cycles within, unlike EPB methods (see red insets). FEB methods generally maintain a similar structure as the original bundling methods.}
    \label{tab:gion-ss-compare}
\end{table*}

\begin{table*}[h!]
    \centering
    \footnotesize
    \begin{tabular}{|c|c||c|c|}
        \hline $D$  & $D'$  & $D_B$(FDEB) & $D'_B$(FFDEB) \\
        \hline \includegraphics[width=0.19\textwidth,angle=90]{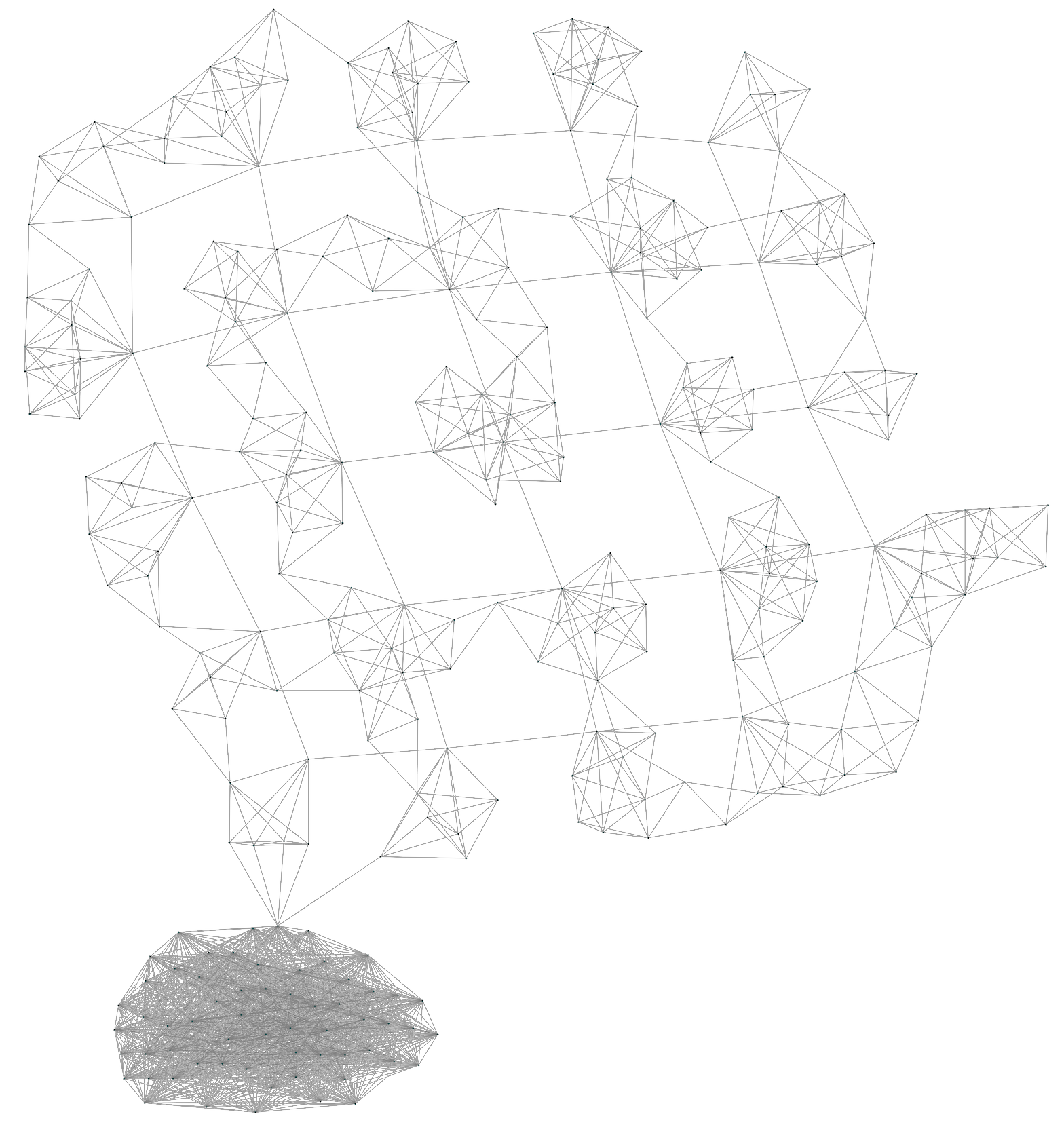}& \includegraphics[width=0.19\textwidth,angle=90]{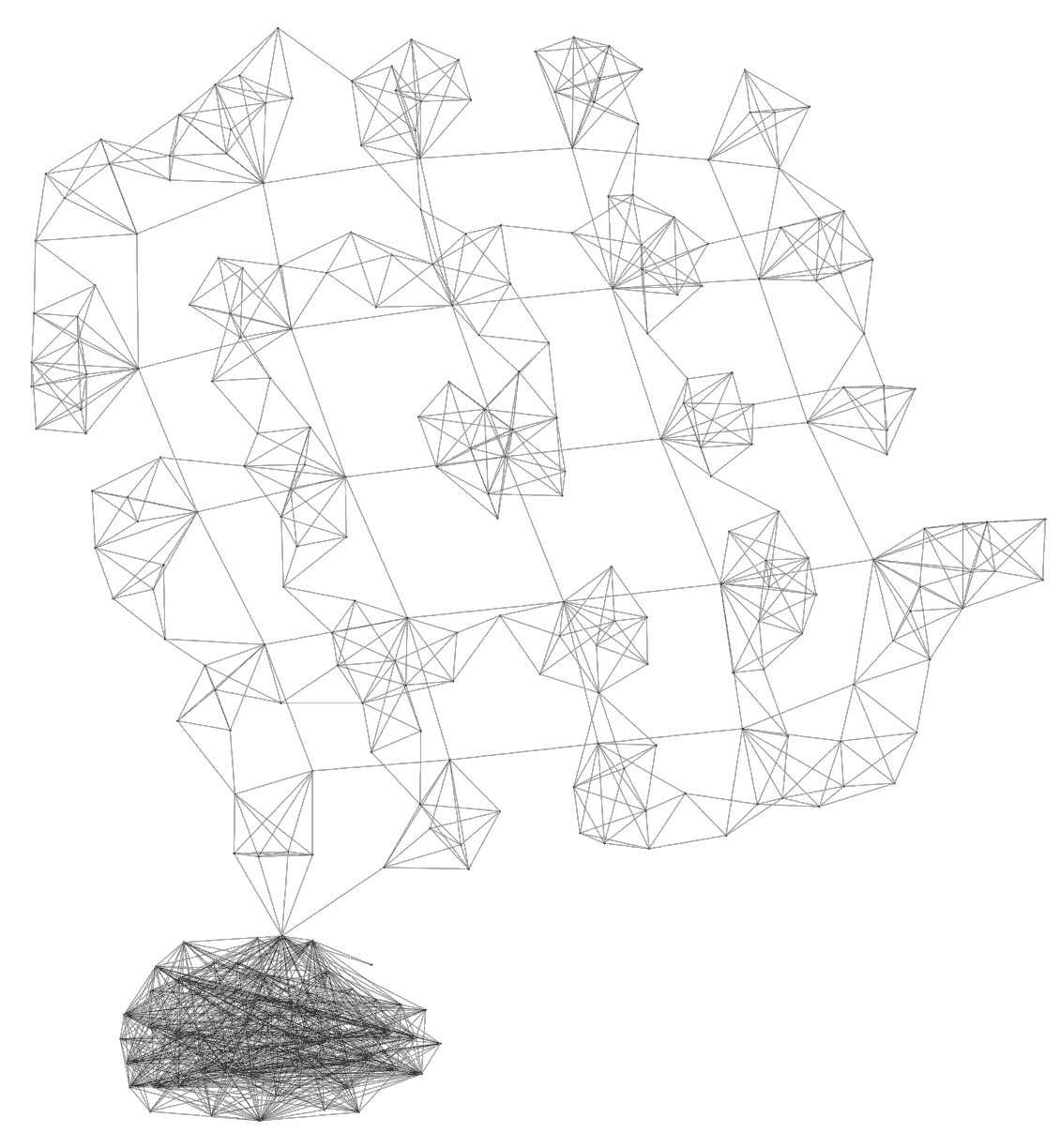} & \includegraphics[width=0.19\textwidth,angle=90]{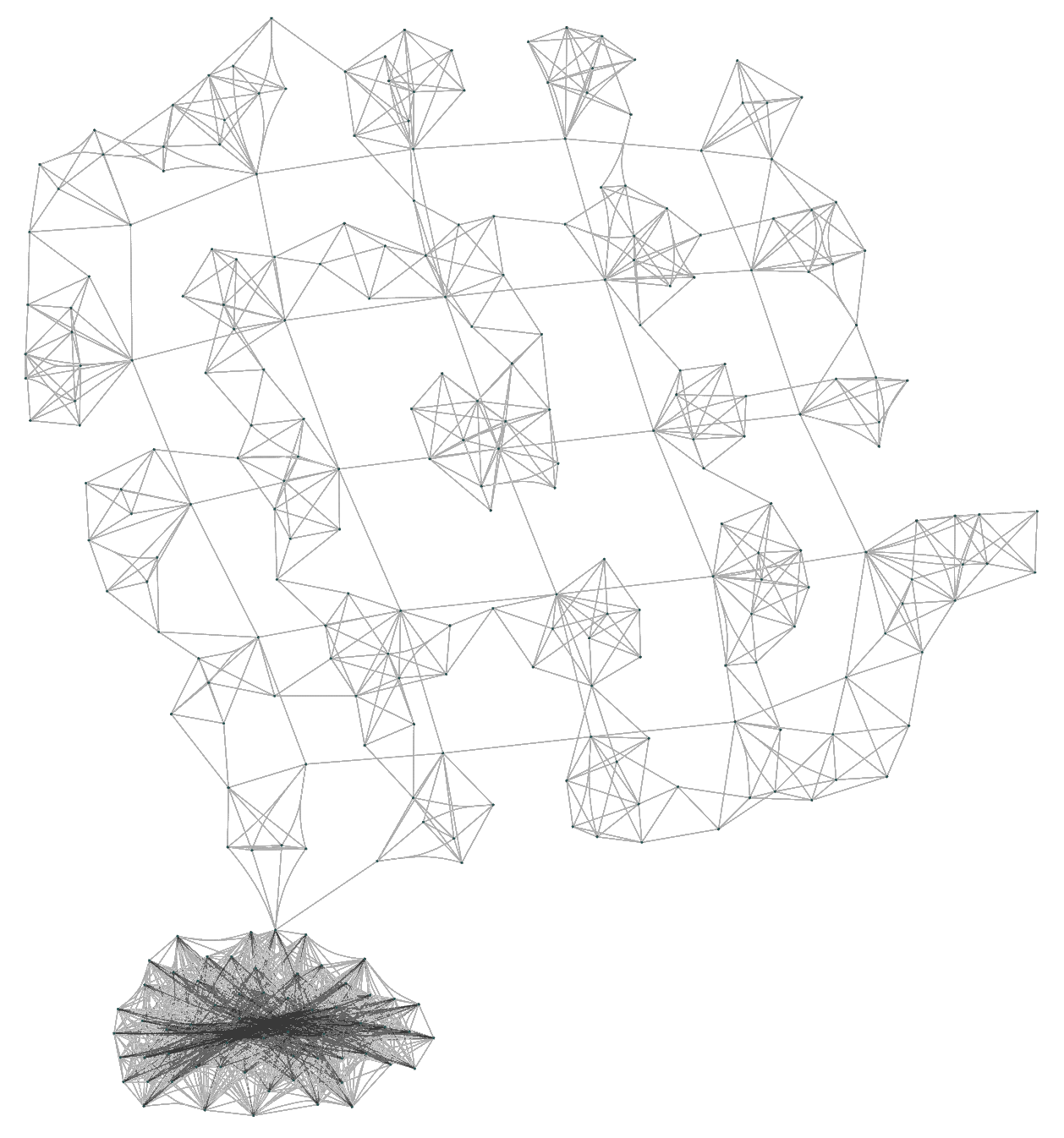} & \includegraphics[width=0.19\textwidth,angle=90]{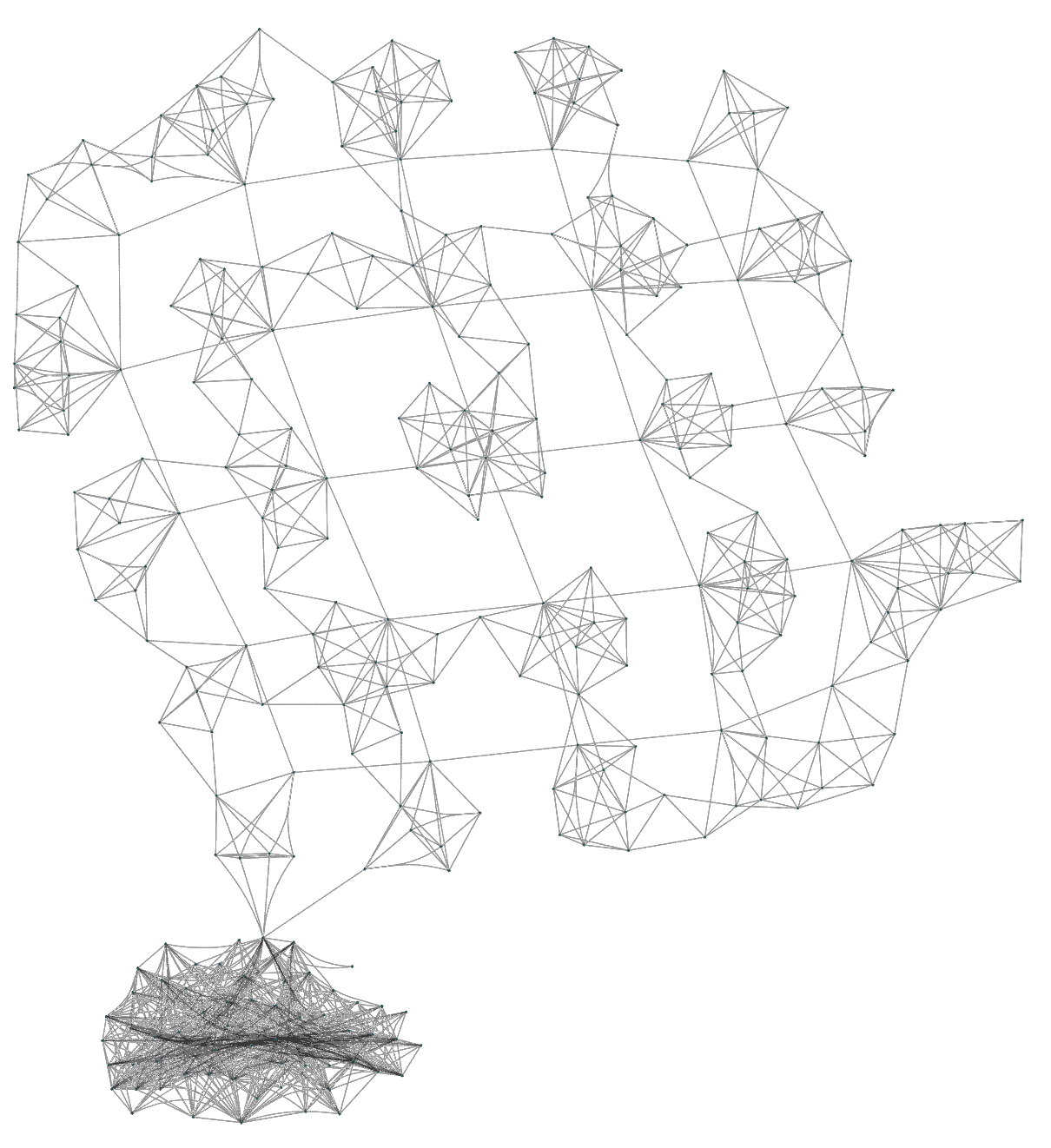}\\
        \hline $D_B$(EPB) & $D'_B$(FEPB)  & $D_B$(SEPB) & $D'_B$(FSEPB) \\
    \hline \includegraphics[width=0.19\textwidth,angle=90]{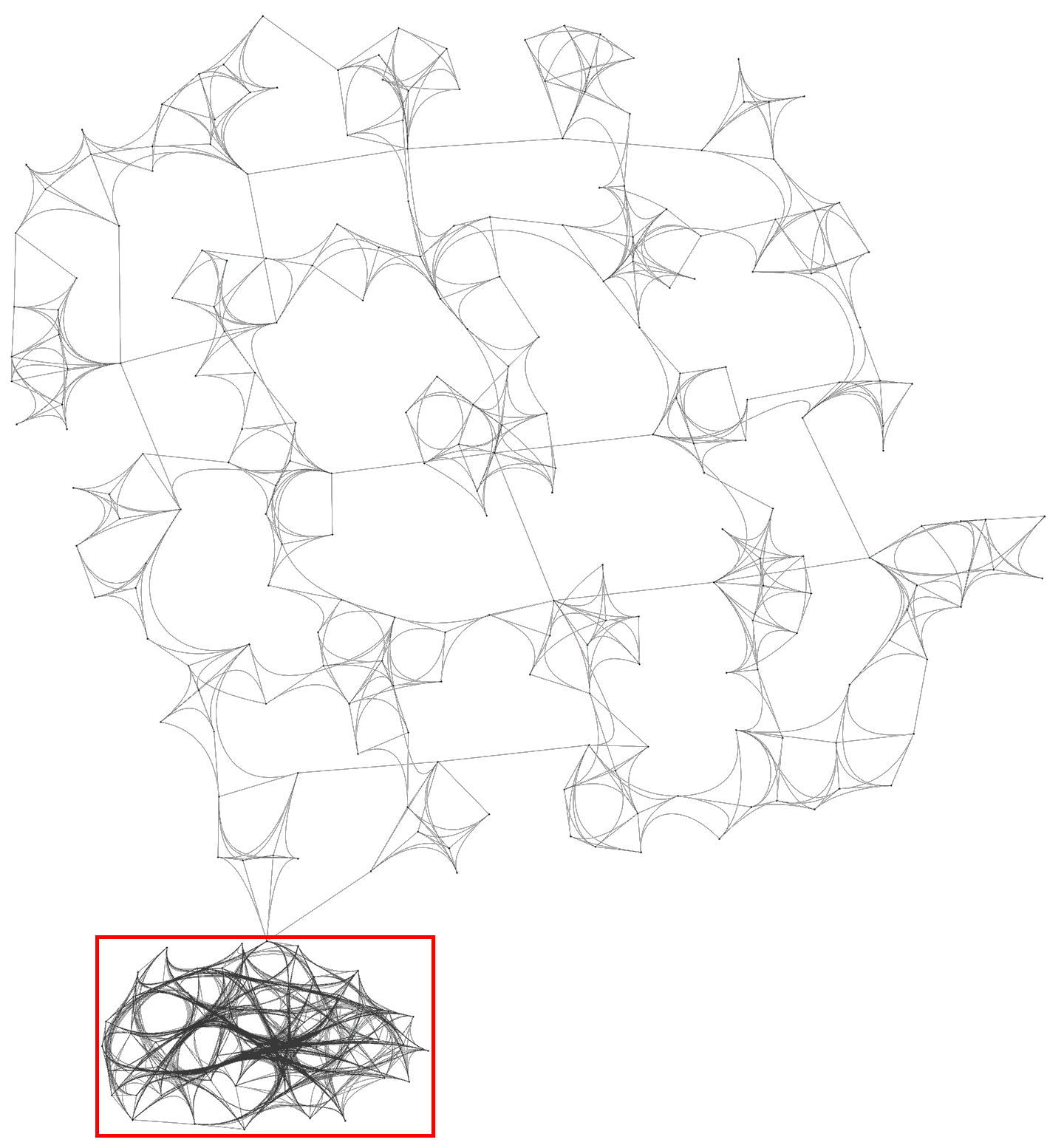} & \includegraphics[width=0.19\textwidth,angle=90]{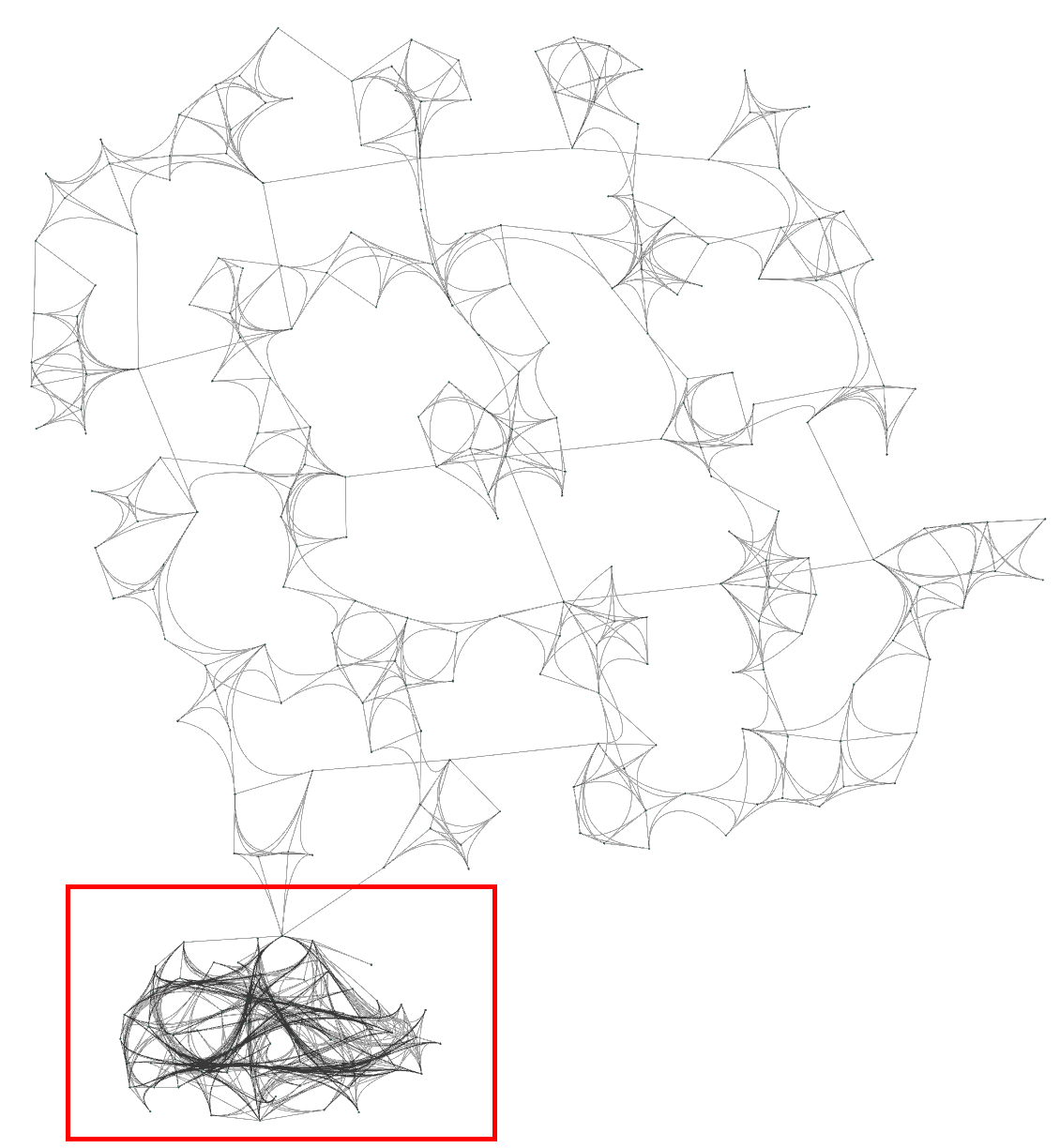} &  \includegraphics[width=0.19\textwidth,angle=90]{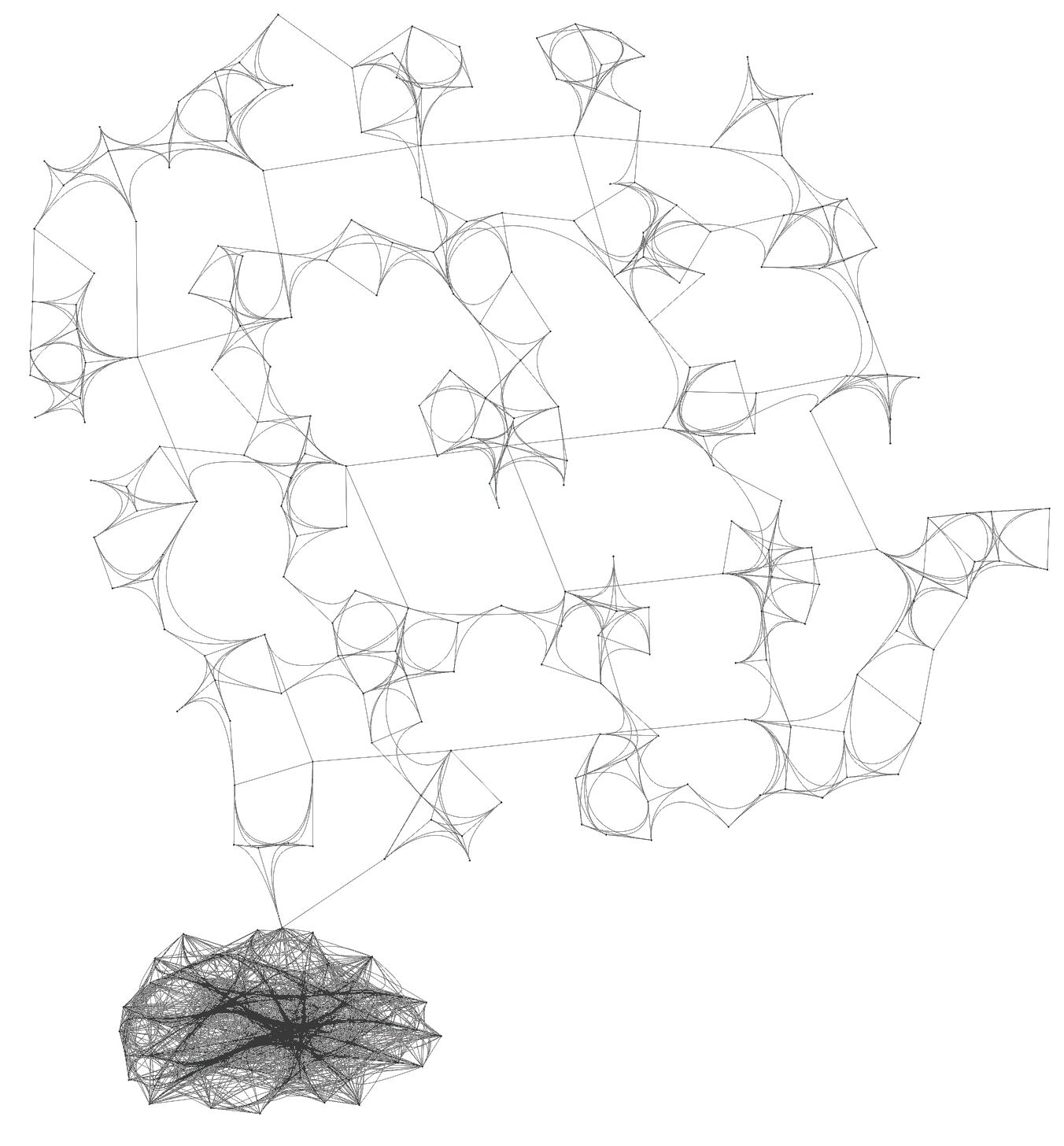} & \includegraphics[width=0.19\textwidth,angle=90]{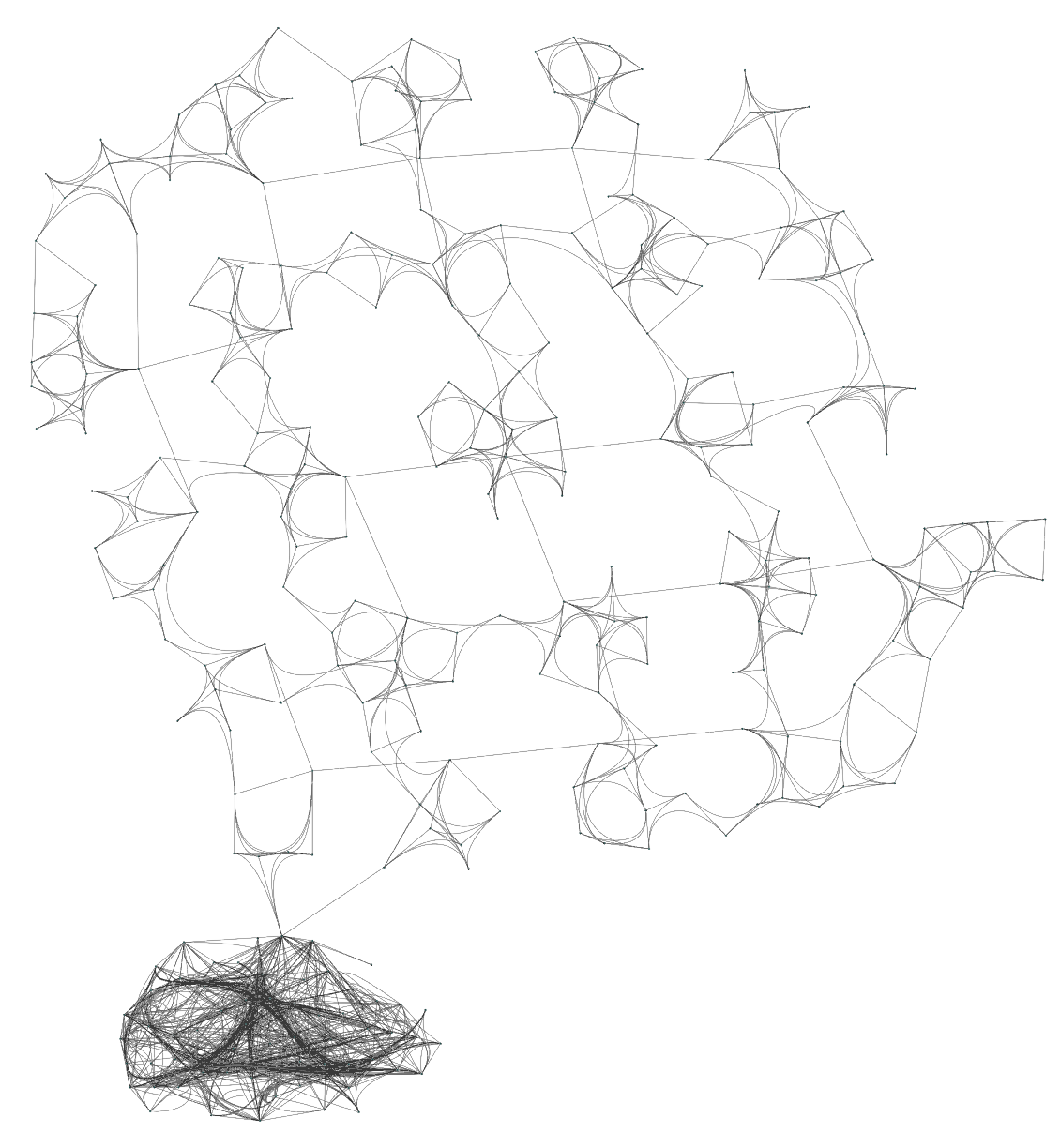}\\
    \hline $D_B$(SEB1) & $D'_B$(FSEB1)  & $D_B$(SEB2) & $D'_B$(FSEB2) \\
    \hline \includegraphics[width=0.19\textwidth,angle=90]{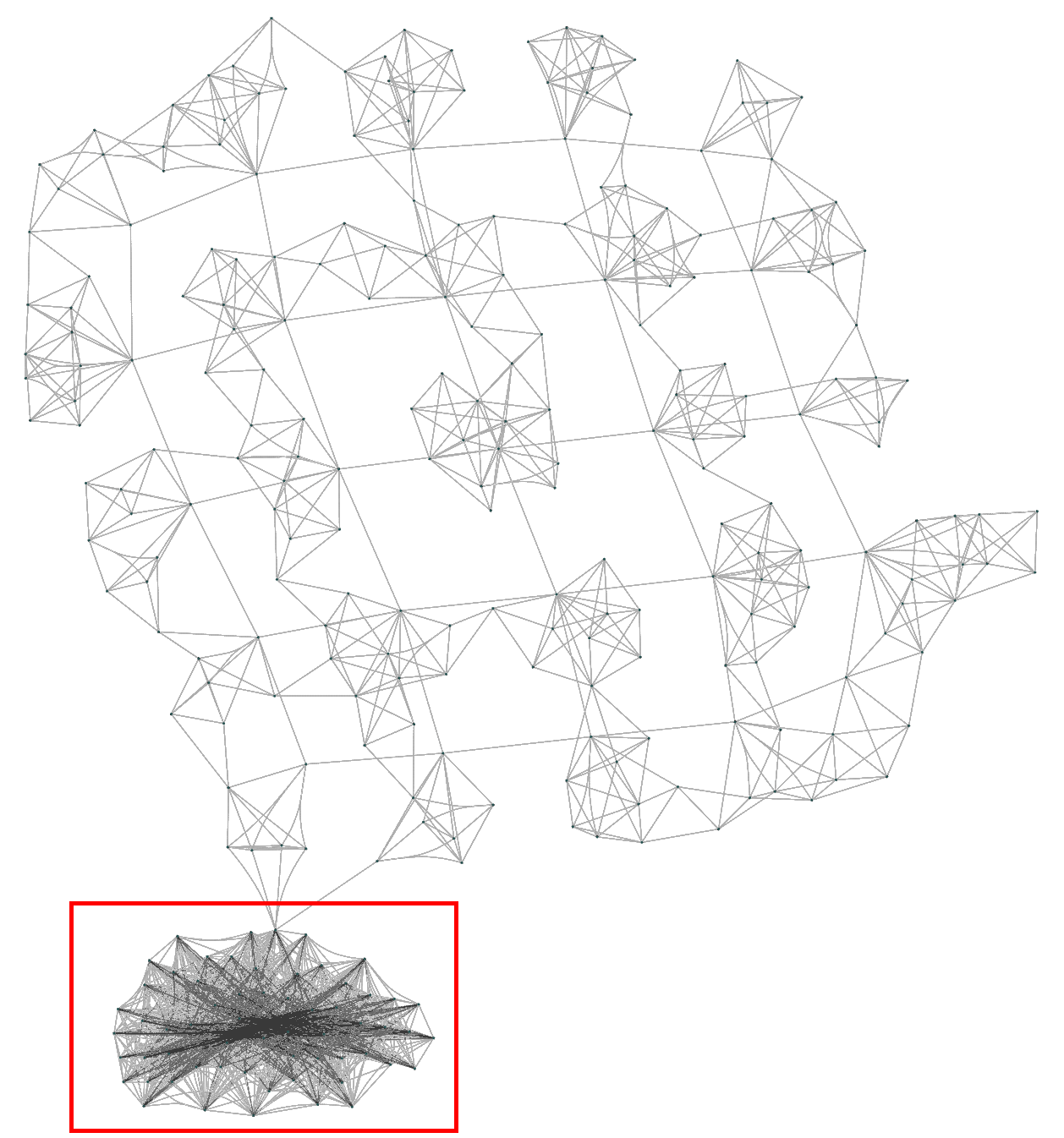} & \includegraphics[width=0.19\textwidth,angle=90]{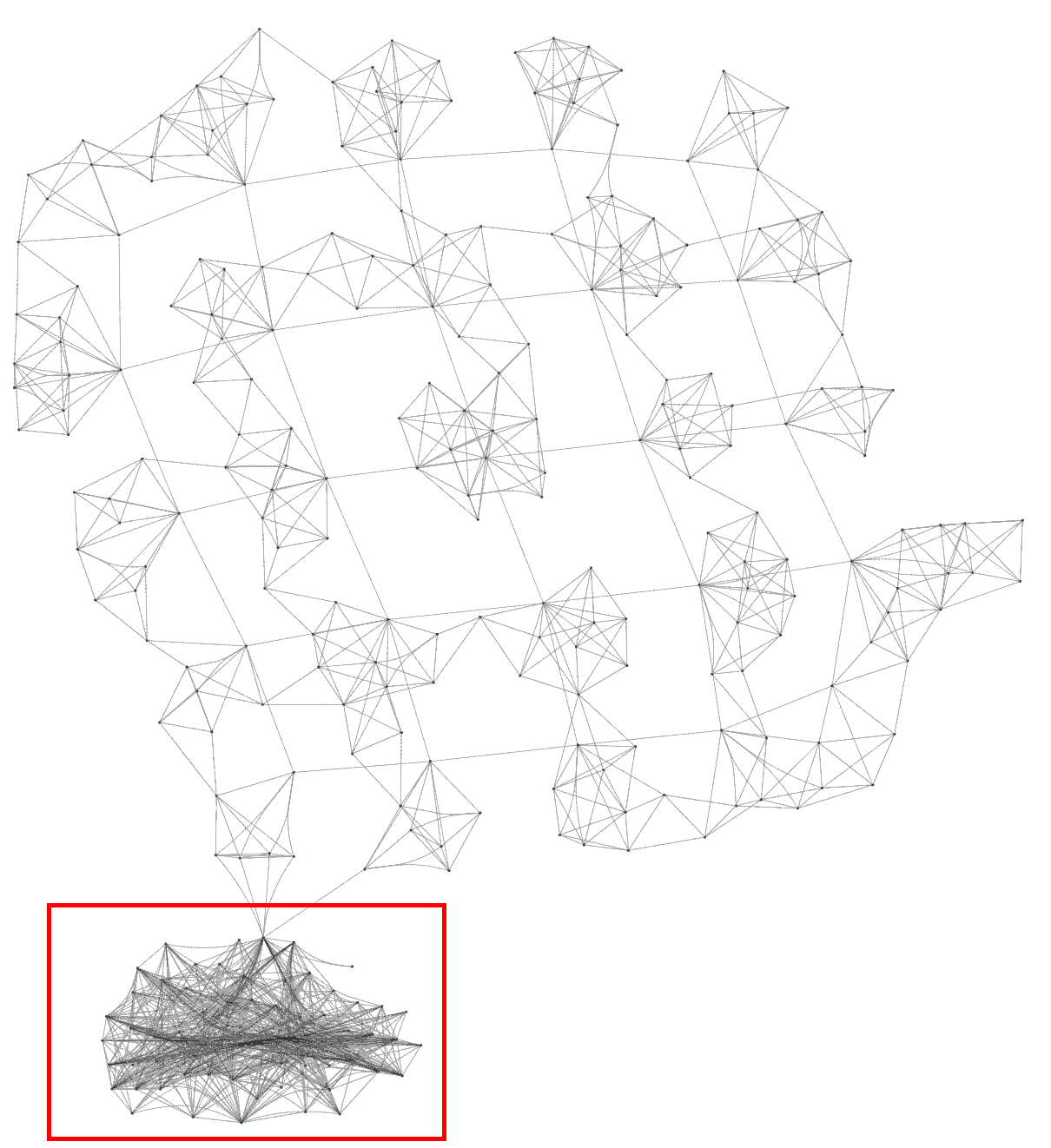} & \includegraphics[width=0.19\textwidth,angle=90]{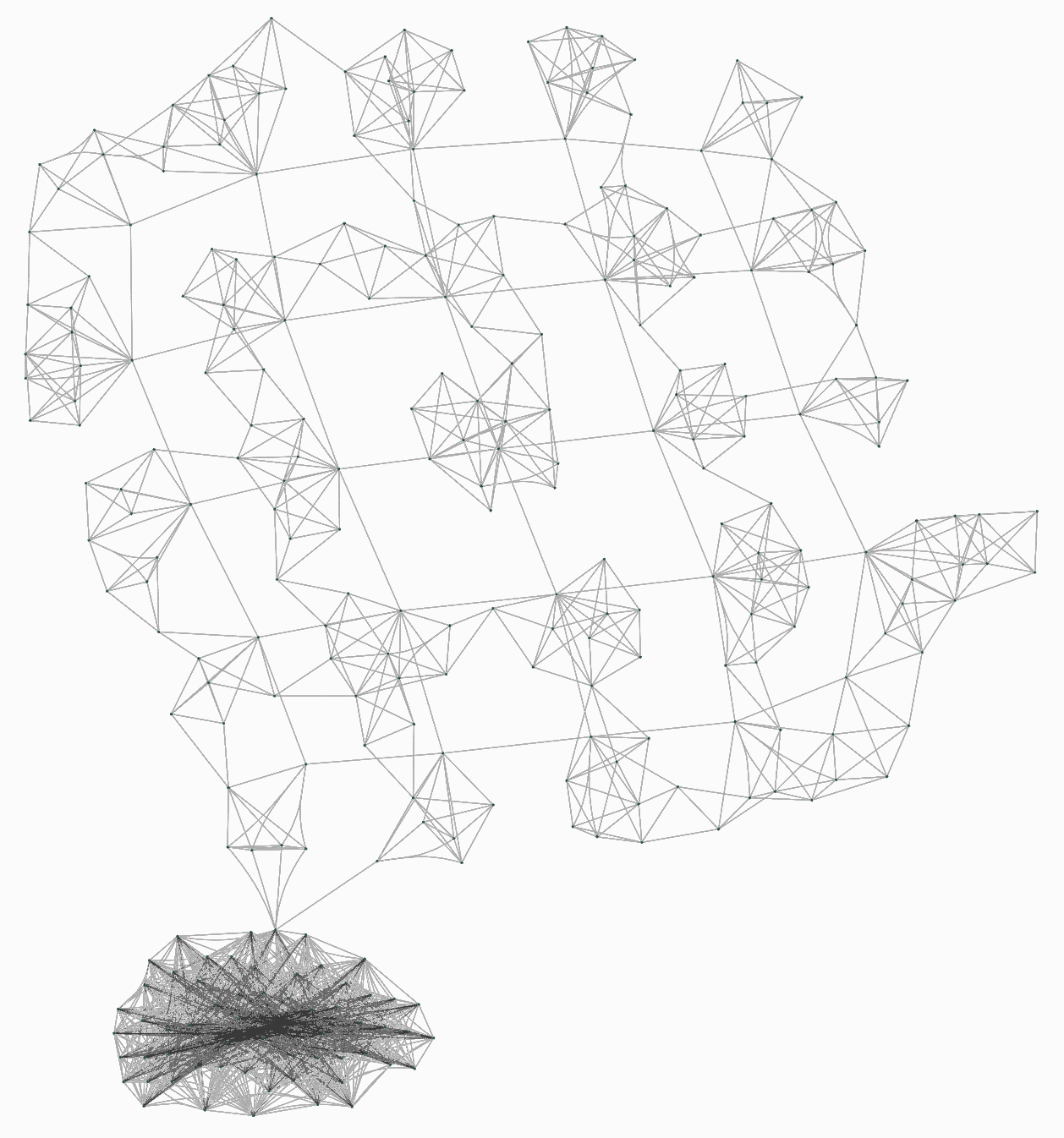} & \includegraphics[width=0.19\textwidth,angle=90]{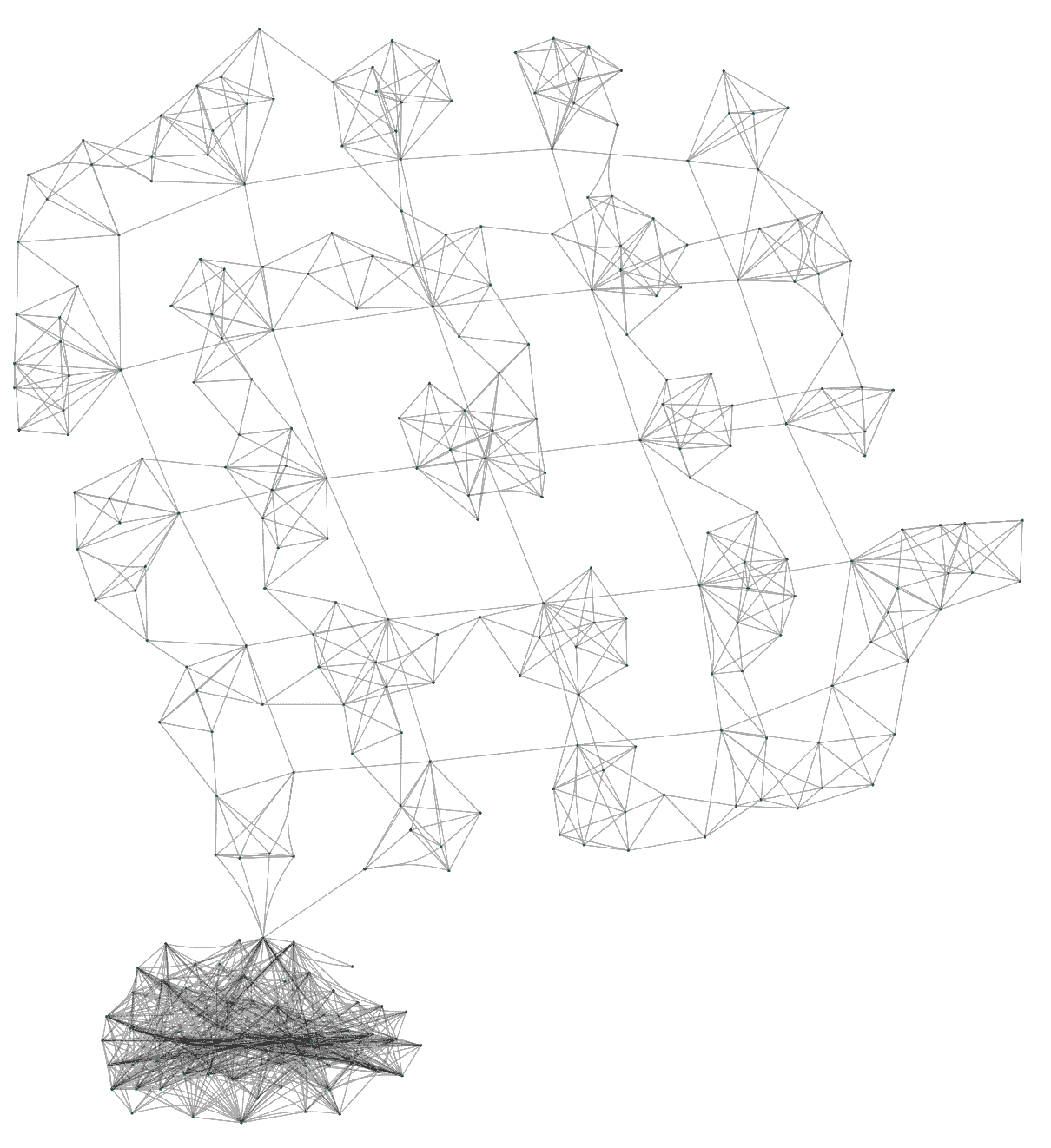}\\
        \hline
    \end{tabular}
    \caption{Visual comparison for the G443 graph. FEB generally manages to maintain a similar structure for the bundled drawing compared to the respective original bundling methods, although FEB with the EPB family creates a somewhat different bundling structure in the ``blob'' while FEB with the SEB methods maintains a similar structure in the blob (see red insets).}
    \label{tab:bh-ss-compare}
\end{table*}

\subsection{Visual Comparison}

%
%Regardless, despite the smaller number of edges, FEB may lead to a larger proportion of unbundled edges. For example, in FSEB2, the bundle between the top-left and top-middle cluster only includes a smaller proportion of edges compared to SEB2.

Table \ref{tab:facebook-ss-compare} shows visual comparisons between FEB and the original edge bundling methods on the scale-free Facebook graphs. 
For all FEB methods, the bundled drawing $D'_B$ by FEB is visually almost the same as the corresponding original bundled drawing $D_B$. 
Notably, FSEB2 can maintain the vertical edge bundle in SEB2, which is not present in the bundled drawing by other methods (see red insets).

Table \ref{tab:airtraffic-ss-compare} shows visual comparisons between FEB and the original edge bundling methods on a geographic graph. 
Overall, the bundled drawings $D'_B$ by FEB are visually almost the same as the corresponding original bundled drawing $D_B$, 
maintaining highly similar structures to $D_B$, with large bundles mostly preserved. 

Table \ref{tab:gion-ss-compare} shows visual comparisons on the GION graph 6\_gion. 
Clearly, FEB computes bundled drawings, which are visually highly similar to the original bundled drawings, maintaining the specific strength of each method.

Table \ref{tab:bh-ss-compare} shows visual comparisons on the black-hole graph G443, where FEB computes visually highly similar bundled drawings to the original bundled drawings.
Notably, FSEB methods show almost the same bundling structure as the original bundlings, while FEB with the EPB family (EPB, SEPB) bundles the dense ``blob'' with different details to the original bundlings (see red insets).

%Table \ref{tab:lastfm-ss-compare} shows the visual comparison between FEB and the original edge bundling methods on the scale-free Lastfm graph. 
%Similarly, for all FEB methods, the bundled drawing $D'_B$ by FEB is highly similar to the corresponding original bundled drawing $D_B$, especially on the FSEB drawings.  
%Meanwhile, the FEPB and FSEPB drawings show some differences in the bundling of edges compared to the original edge bundling methods, such as the bundles between the top-left clusters and the middle of the drawing not being as visible in the FEB drawings (see red insets). 
%See Appendix \ref{sec:feb_vis} for more visual comparisons.

\subsection{Discussion and Summary}

Our extensive experiments demonstrate the efficiency (i.e., 61\% runtime improvement on average) and effectiveness (i.e., highly similar edge bundling, 74\% by $FBQ_{SQ}$, to the original edge bundling) of the FEB framework.

\begin{figure}[h!]
    \centering
    \subfloat[Geo.]{
        \includegraphics[width=0.22\columnwidth]{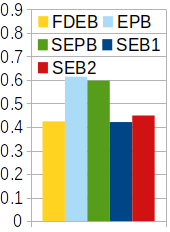}
        \label{fig:feb_geo_avg.png}
    }
    \subfloat[Scale-free]{
        \includegraphics[width=0.22\columnwidth]{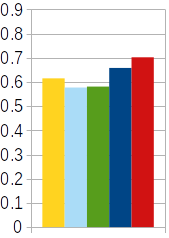}
        \label{fig:feb_sf_avg.png}
    }
    \subfloat[GION]{
        \includegraphics[width=0.22\columnwidth]{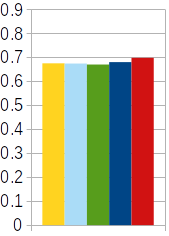}
        \label{fig:feb_gion_avg.png}
    }
    \subfloat[BH]{
        \includegraphics[width=0.22\columnwidth]{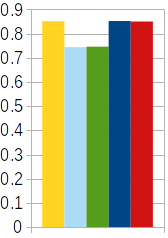}
        \label{fig:feb_bh_avg}
    }
    \caption{%XXXXX thin bars: one line XXXX 
    Average $FBQ_{JS}$ per data set. FEB with SEB methods perform better than FEB with EPB methods on black-hole graphs, while the opposite is true on geographic graphs.}
    \label{fig:feb_avg}
\end{figure}

Figure \ref{fig:feb_avg} shows $FBQ_{JS}$ per data set, 
%: for most data sets, $FBQ_{JS}$ is around 0.65 on average,
where FSEB methods perform 85\% similarity for black-hole graphs (see Figure \ref{fig:feb_bh_avg}).
The averages per data set show a similar pattern to the SEB experiment in Section \ref{sec:exp_comp}, i.e., FSEB methods obtain better metrics than FEPB methods on most non-geographic graphs (in particular, black-hole graphs), while the opposite holds on geographic graphs.

Visual comparisons support consistent results with the quality metrics: on the G443 graph in Table \ref{tab:bh-ss-compare}, FSEB methods obtain higher $FBQ_{JS}$ than FEPB methods, displaying more faithful, similar bundling of the blob to the original SEB bundlings. 
Meanwhile, on the Air Traffic graph in Table \ref{tab:airtraffic-ss-compare}, FEPB methods obtain higher $FBQ_{JS}$ than the FSEB methods, as they maintain bundles with similar sizes and densities as in the original EPB bundlings.

\emph{In summary, FEB methods achieve 61\% runtime improvement over the original edge bundling methods,  faithfully maintaining highly similar bundlings to the original bundlings by FBQ metrics (74\% by $FBQ_{SQ}$) and visual comparison. }

\section{SEB Experiments}
\label{sec:exp_comp}

This section presents experiments for evaluating SEB by extensive comparison with established edge bundling methods on various quality metrics. 
%By comparing SEB and SBQ to established edge bundling methods and metrics, we expect to be able to not only evaluate the potential strengths of SBQ compared to state-of-art edge bundling methods, but also examine the effectiveness of SBQ for evaluating the quality of edge bundled drawings compared to other metrics.
%
Since our methods are not graphics-based, we compare SEB to closely related edge bundling methods such as EPB, SEPB, and FDEB.

Specifically, we focus comparisons with  Edge-Path Bundling (EPB)~\cite{edge-path}, and Spanner-based Edge-Path Bundling (SEPB)~\cite{sepb}, as they are the state-of-art methods which have been shown to generally obtain lower ambiguity than other bundling methods, such as  Confluent drawings~\cite{dickerson2004confluent}, CUBu~\cite{van2016cubu}, and Winding Roads~\cite{lambert2010winding}, especially on higher orders of the ambiguity metric. 
We also compare SEB to Force-Directed Edge Bundling (FDEB)~\cite{forceDirBund},
as SEB is based on FDEB.
 %XXXXX on the ambiguity metric over other edge bundling methods. XXXXX REF XXXXX  

\begin{table*}[h]
    \centering
    \caption{Quality metrics comparison (lower values = better for all metrics) for each bundling method over all data sets. 
    SEB performs significantly better than EPB methods on distortion and ambiguity.
    SEB2 performs the best on distortion and ambiguity. EPB performs best on ink reduction.
    }
    \scriptsize
        \begin{tabular}{|l|l|l|l|l||l|l|l|l||l|l|l|l||l|l|l|l|}
    \hline
        & \multicolumn{4}{c||}{US Airlines (geographic)} & \multicolumn{4}{c|}{World Air Traffic (geographic)} & \multicolumn{4}{c||}{US Migration (geographic)} & \multicolumn{4}{c|}{yeastppi (scale-free)} \\ \hline
         & Ink & Dist & $\mathrm{Amb}^1$ & $\mathrm{Amb}^2$ & Ink & Dist & $\mathrm{Amb}^1$ & $\mathrm{Amb}^2$ & Ink & Dist & $\mathrm{Amb}^1$ & $\mathrm{Amb}^2$ & Ink & Dist & $\mathrm{Amb}^1$ & $\mathrm{Amb}^2$  \\ \hline
        FDEB & 0.811 & 0.016 & 0.697 & 0.018 & 0.73 & 0.108 & 0.697 & 0.111 & 0.757 & 0.027 & 0.804 & 0.416 & 0.741 & 0.025 & 0.894 & 0.78  \\ \hline
        EPB & \textbf{0.564} & 0.073 & 0.772 & 0.044 & \textbf{0.558} & 0.1 & \textbf{0.660} & 0.153 & \textbf{0.541} & 0.065 & \textbf{0.745} & 0.386 & 0.965 & 0.066 & 0.888 & 0.778  \\ \hline
        SEPB & 0.571 & 0.082 & 0.783 & 0.044 & 0.593 & 0.106 & 0.663 & 0.15 & 0.551 & 0.072 & 0.752 & \textbf{0.385} & 0.966 & 0.069 & 0.888 & 0.776  \\ \hline
        SEB1 & 0.832 & 0.01 & 0.687 & \textbf{0.016} & 0.777 & 0.092 & 0.692 & \textbf{0.101} & 0.761 & 0.023 & 0.792 & 0.412 & \textbf{0.733} & 0.015 & 0.88 & 0.764  \\ \hline
        SEB2 & 0.874 & \textbf{0.008} & \textbf{0.686} & 0.024 & 0.821 & \textbf{0.061} & 0.691 & 0.104 & 0.868 & \textbf{0.007} & 0.774 & 0.407 & 0.746 & \textbf{0.007} & \textbf{0.873} & \textbf{0.757}  \\ \hline
        & \multicolumn{4}{c||}{lastfm (scale-free)} & \multicolumn{4}{c|}{facebook (scale-free)} & \multicolumn{4}{c||}{6\_GION (GION)} & \multicolumn{4}{c|}{7\_GION (GION)}   \\ \hline
         & Ink & Dist & $\mathrm{Amb}^1$ & $\mathrm{Amb}^2$ & Ink & Dist & $\mathrm{Amb}^1$ & $\mathrm{Amb}^2$ & Ink & Dist & $\mathrm{Amb}^1$ & $\mathrm{Amb}^2$ & Ink & Dist & $\mathrm{Amb}^1$ & $\mathrm{Amb}^2$ \\ \hline
        FDEB & 1.067 & 0.039 & 0.946 & 0.808 & 0.917 & 0.084 & 0.839 & 0.032 & 1.156 & 0.104 & 0.551 & 0.082 & 0.964 & 0.146 & 0.114 & 0.012  \\ \hline
        EPB & 0.974 & 0.012 & 0.931 & 0.785 & \textbf{0.483} & 0.056 & 0.846 & 0.052 & \textbf{0.762} & 0.073 & 0.55 & \textbf{0.069} & 0.8 & \textbf{0.076} & 0.138 & 0.01  \\ \hline
        SEPB & \textbf{0.974} & 0.017 & 0.932 & 0.786 & 0.526 & 0.074 & 0.842 & 0.054 & 0.761 & 0.097 & 0.558 & 0.069 & \textbf{0.784} & 0.103 & 0.138 & \textbf{0.010}  \\ \hline
        SEB1 & 1.081 & 0.018 & 0.936 & 0.788 & 0.935 & 0.077 & 0.833 & 0.031 & 1 & 1.314 & 0.096 & 0.542 & 0.883 & 0.023 & 0.141 & \textbf{0.114} \\ \hline
        SEB2 & 1.086 & \textbf{0.006} & \textbf{0.930} & \textbf{0.777} & 1.005 & \textbf{0.033} & \textbf{0.814} & \textbf{0.027} & 1.369 & \textbf{0.048} & \textbf{0.524} & 0.072 & 1.015 & 0.098 & 0.117 & 0.011  \\ \hline
        & \multicolumn{4}{c||}{G443 (Black-hole)} & \multicolumn{4}{c|}{Cycle896 (Black-hole)} & \multicolumn{4}{c||}{Cycle907 (Black-hole)} & \multicolumn{4}{c|}{\textbf{Average}}  \\ \hline
         & Ink & Dist & $\mathrm{Amb}^1$ & $\mathrm{Amb}^2$ & Ink & Dist & $\mathrm{Amb}^1$ & $\mathrm{Amb}^2$ & Ink & Dist & $\mathrm{Amb}^1$ & $\mathrm{Amb}^2$ & Ink & Dist & $\mathrm{Amb}^1$ & $\mathrm{Amb}^2$ \\ \hline
        FDEB & 0.78 & 0.024 & 0.079 & 0.001 & 0.55 & 0.068 & 0.344 & 0.011 & \textbf{0.414} & 0.031 & 0.18 & 0.008 & 0.808 & 0.061 & 0.559 & 0.207 \\ \hline
        EPB & \textbf{0.745} & 0.061 & 0.213 & 0.002 & 0.784 & 0.048 & 0.4201 & 0.012 & 0.651 & 0.037 & 0.343 & 0.004 & \textbf{0.712} & 0.061 & 0.591 & 0.209  \\ \hline
        SEPB & 0.846 & 0.145 & 0.221 & 0.003 & 0.914 & 0.126 & 0.424 & 0.013 & 0.703 & 0.094 & 0.369 & 0.005 & 0.744 & 0.090 & 0.597 & 0.209 \\ \hline
        SEB1 & 0.781 & 0.023 & 0.079 & \textbf{0.001} & \textbf{0.550} & 0.068 & 0.341 & 0.009 & 0.416 & 0.03 & 0.174 & 0.003 & 0.833 & 0.053 & 0.552 & 0.201\\ \hline
        SEB2 & 0.791 & \textbf{0.016} & \textbf{0.078} & 0.001 & 0.561 & \textbf{0.047} & \textbf{0.340} & \textbf{0.008} & 0.421 & \textbf{0.025} & \textbf{0.172} & \textbf{0.003} & 0.869 & \textbf{0.032} & \textbf{0.545} & \textbf{0.199} \\ \hline
    \end{tabular}
    \label{tab:seb_metrics}
\end{table*}

We use various data sets from previous edge bundling and graph drawing studies: (i) real-world \emph{geographic} graphs with fixed positions assigned to each vertex~\cite{edge-path}; (ii) real-world \emph{scale-free} graphs with globally sparse, locally dense clusters, and short diameters~\cite{sparsematrices,networkrepository}; (iii) \emph{GION} graphs, biochemical networks with globally sparse, locally dense structures, and long  diameters~\cite{giondataset}; and (iv) \emph{black-hole} graphs with globally sparse (mesh- or cycle-like) and locally dense ``blobs'' structures~\cite{eades2018drawing}. 
Table \ref{tab:metric_datasets} shows the details. %including the number of vertices ($|V|$), the number of edges ($|E|$), density ($den$), and diameter ($dia$). 

\begin{table}[h]
    \centering
    \begin{tabular}{l| l | l | l | l | l}
        & type & $|V|$ & $|E|$ & $den$ & $dia$ \\
        \hline \hline
         airline & geographic & 235 & 1297 & 8.940 & 4 \\
         migration & geographic & 1702 & 6487 & 5.714 & 12 \\
         airtraffic & geographic & 1533 & 16480 & 10.759 & 6 \\
        yeastppi & scale-free & 2224 & 6609 & 2.972 & 11 \\
        facebook & scale-free & 4039 & 88234 & 21.846 & 8 \\
        lastfm & scale-free & 7624 & 27806 & 3.647 & 15 \\
        6\_gion & GION & 1785 & 20459 & 11.462 & 41 \\
        7\_gion & GION & 3010 & 41757 & 13.873 & 77 \\
        G443 & black-hole & 285 & 2009 & 7.049 & 12 \\
        Cycle896 & black-hole & 1031 & 22638 & 21.957 & 35 \\
        Cycle907 & black-hole & 823 & 14995 & 18.220 & 42 \\
    \end{tabular}
    \caption{Data sets including the number of vertices ($|V|$), the number of edges ($|E|$), density ($den$), and diameter ($dia$).}
    \label{tab:metric_datasets}
\end{table}

To compute drawings for data sets without given geometry (i.e., (ii)-(iv)), we use the \emph{Backbone} layout~\cite{backbone}, designed to untangle ``hairballs'' in drawings of large and complex graphs, shown to produce highly faithful drawings by shape-based and cluster-faithfulness metrics~\cite{hong2022dgg,meidiana2019quality}.

\begin{table*}[h]
    \centering
    \caption{Visual comparison for the scale-free Facebook graph. SEB2 preserves a bundle of edges in the top middle part that is combined into other bundles in the drawings by other methods, with FSEB2 preserving the same characteristic.}
    \scriptsize
    \begin{tabular}{|c|c||c|c|}
        %\hline $D$  & $D'$ & $D_B$  & $D'_B$ \\
        \hline $D$  & $D'$  & $D_B$(FDEB) & $D'_B$(FFDEB) \\
        \hline   \includegraphics[width=0.21\textwidth]{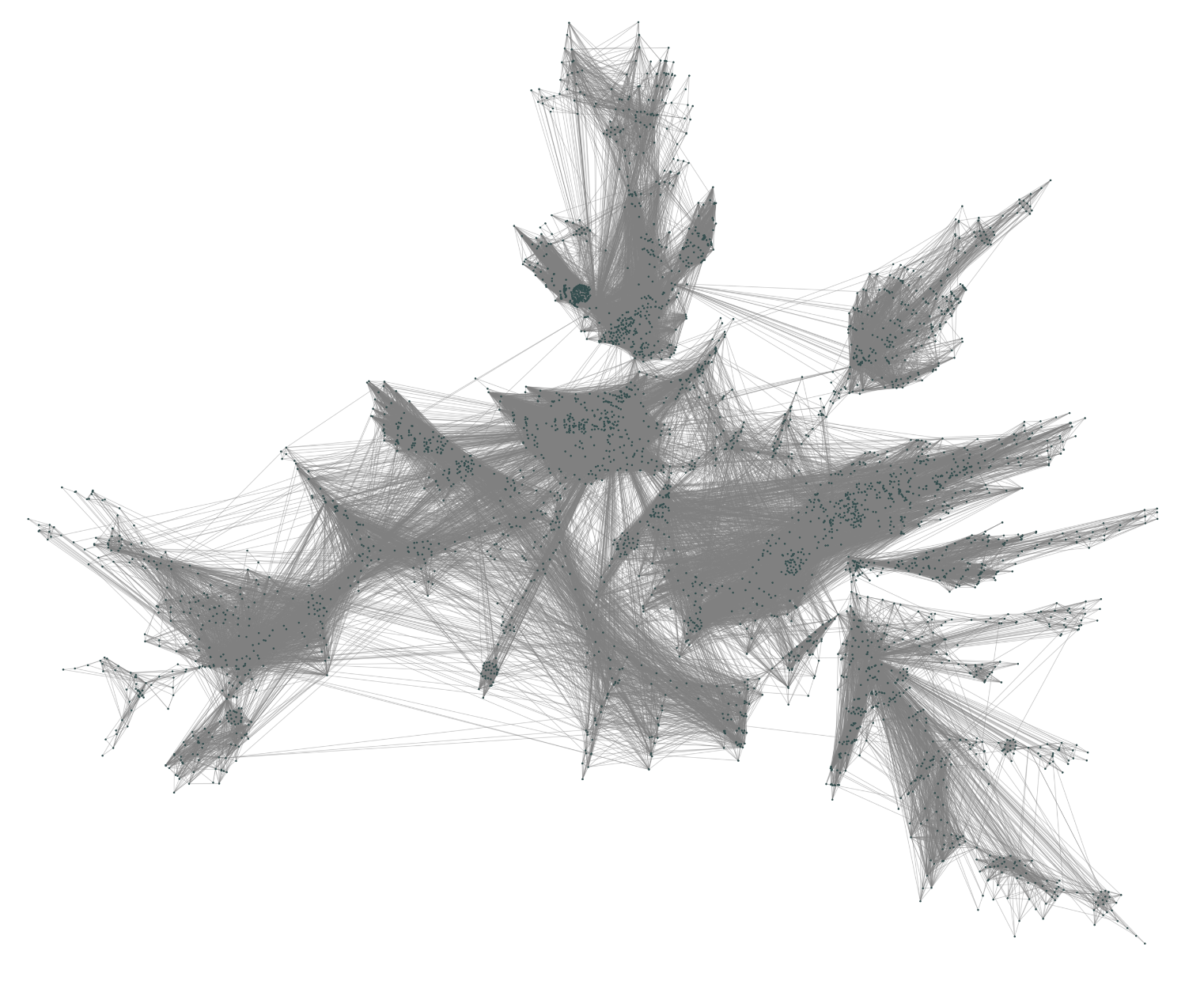}& \includegraphics[width=0.21\textwidth]{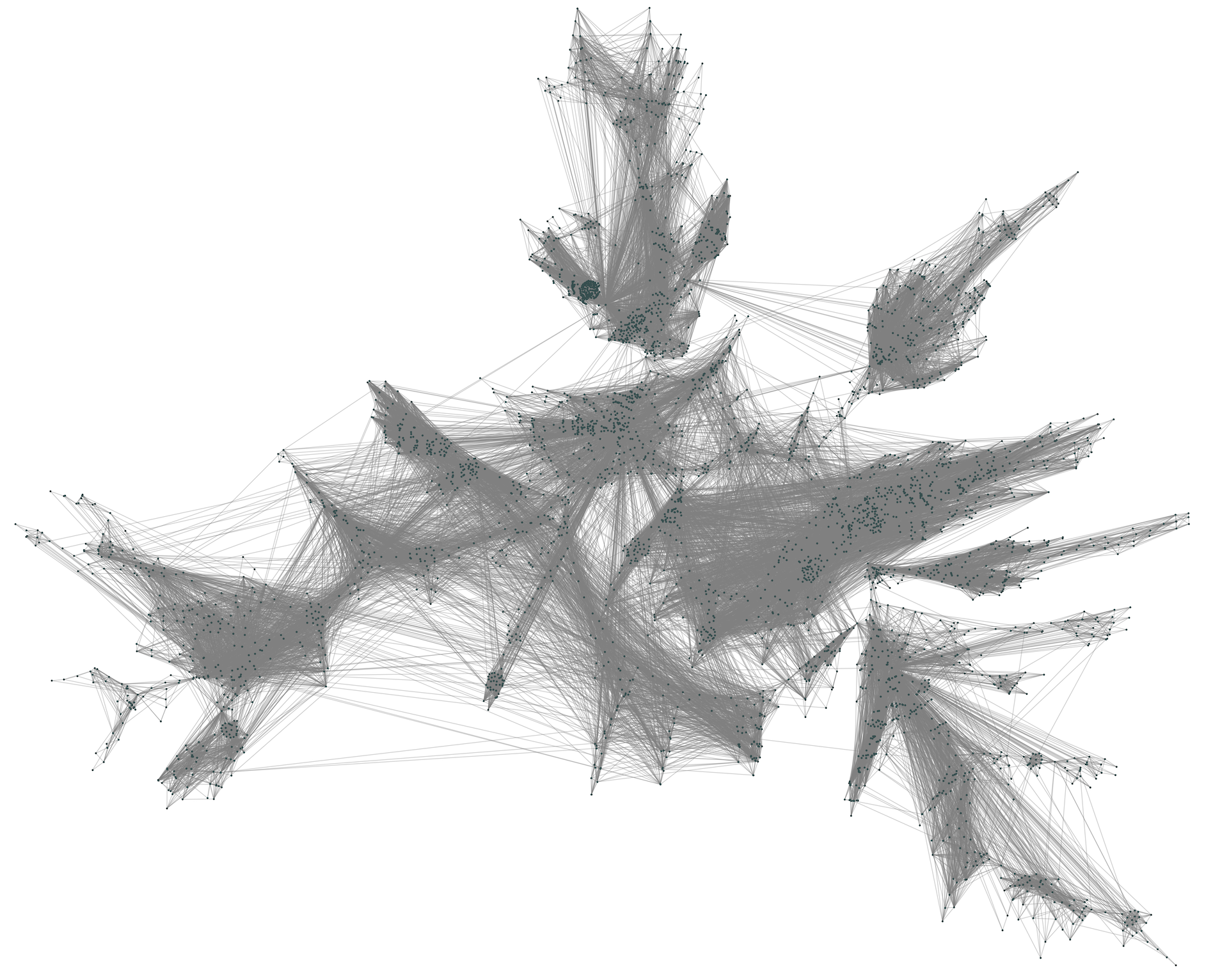} & \ \includegraphics[width=0.21\textwidth]{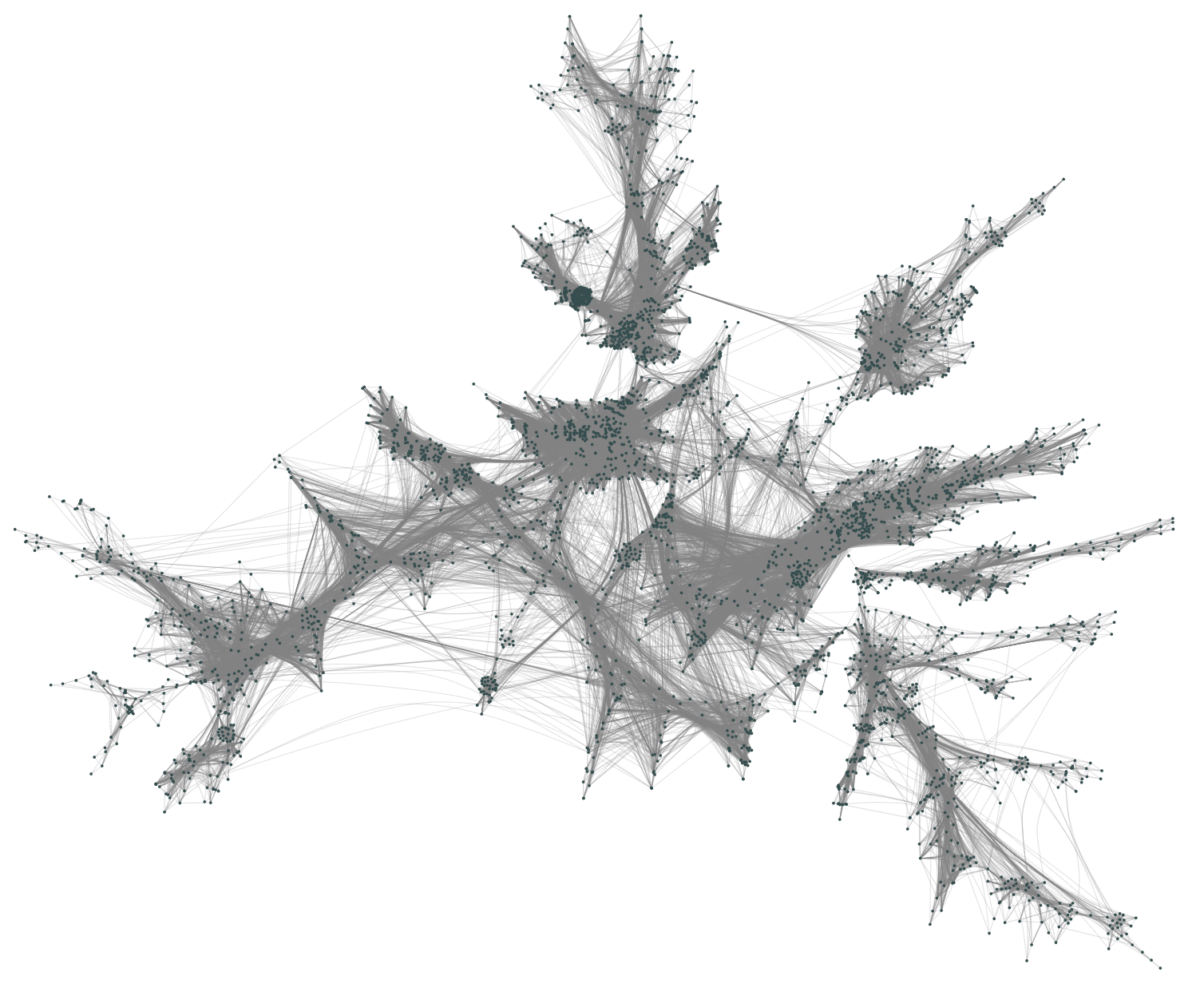}& \includegraphics[width=0.21\textwidth]{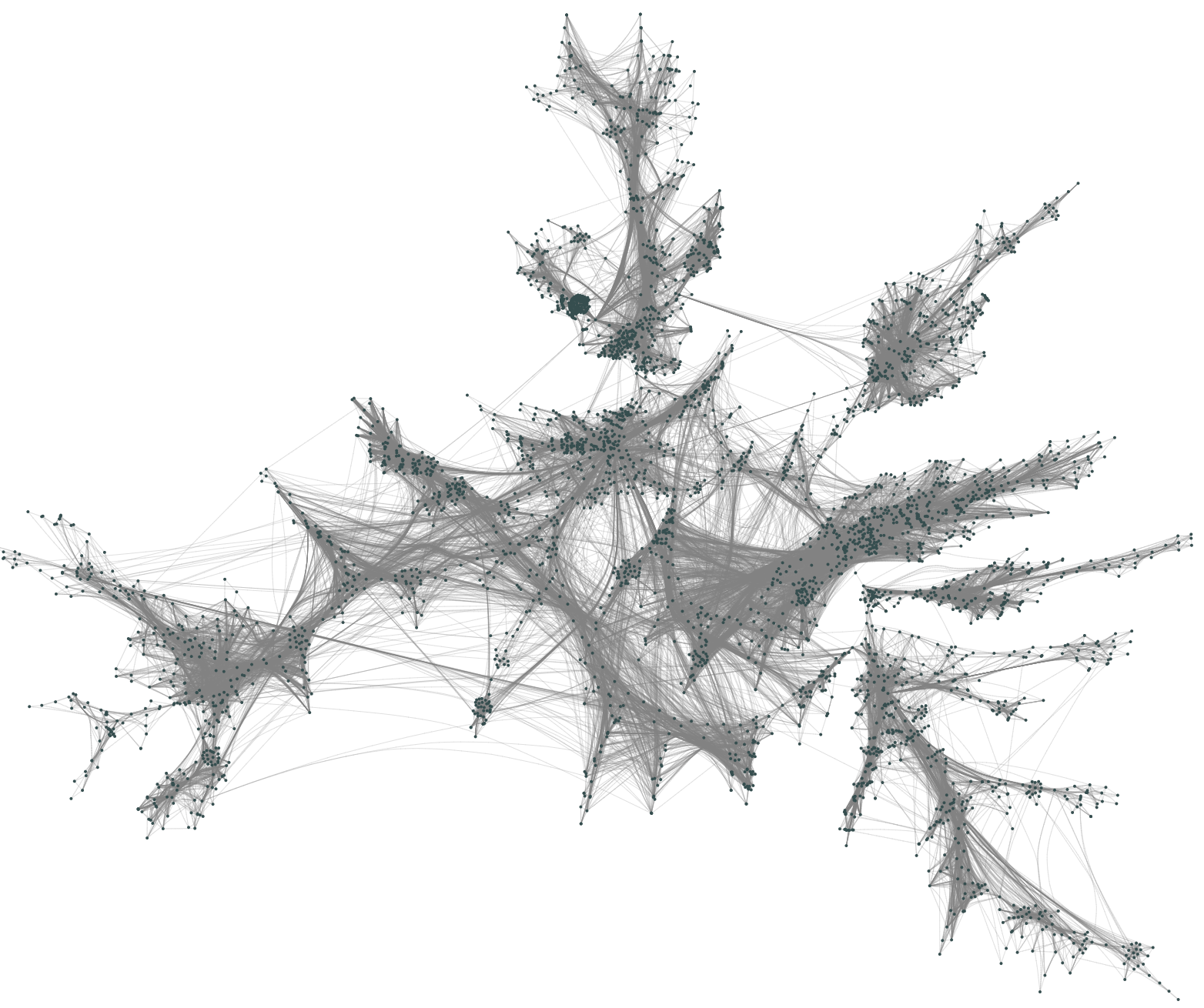}\\
        %\hline $D_B$  & $D'_B$ &  $D_B$  & $D'_B$ \\
        \hline $D_B$(EPB) & $D'_B$(FEPB) & $D_B$(SEPB) & $D'_B$(FSEPB) \\
    \hline \includegraphics[width=0.21\textwidth]{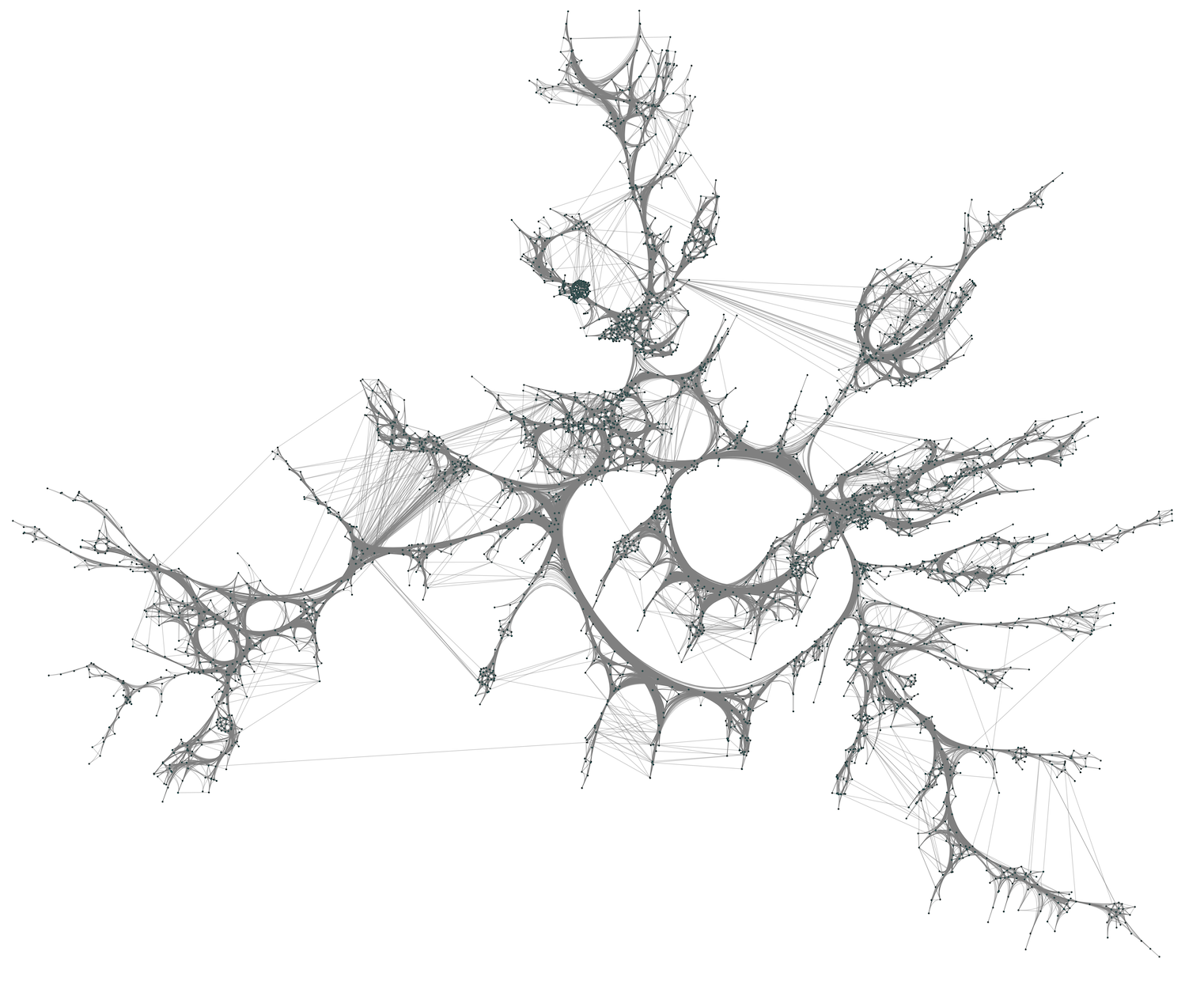} & \includegraphics[width=0.21\textwidth]{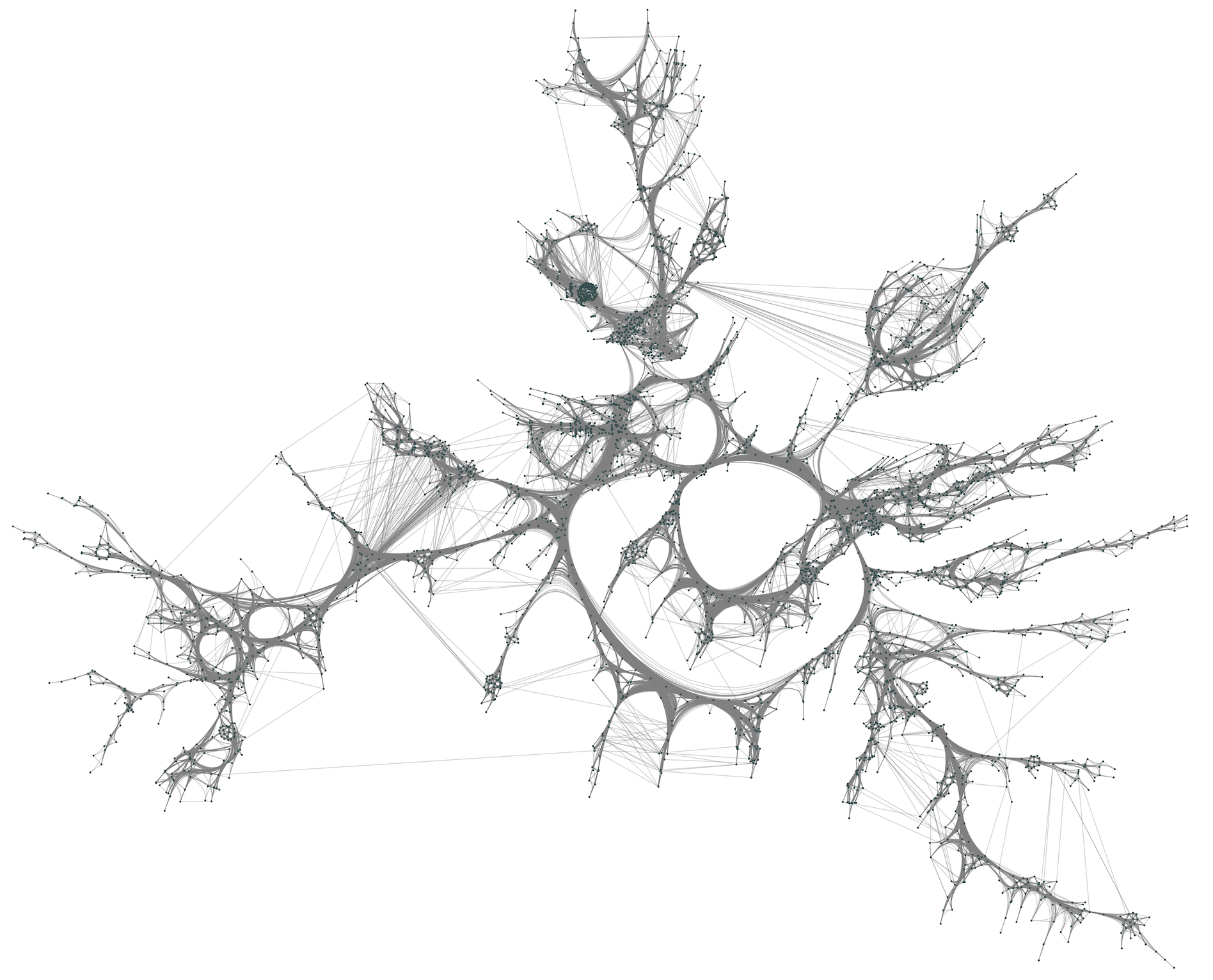} &  \includegraphics[width=0.21\textwidth]{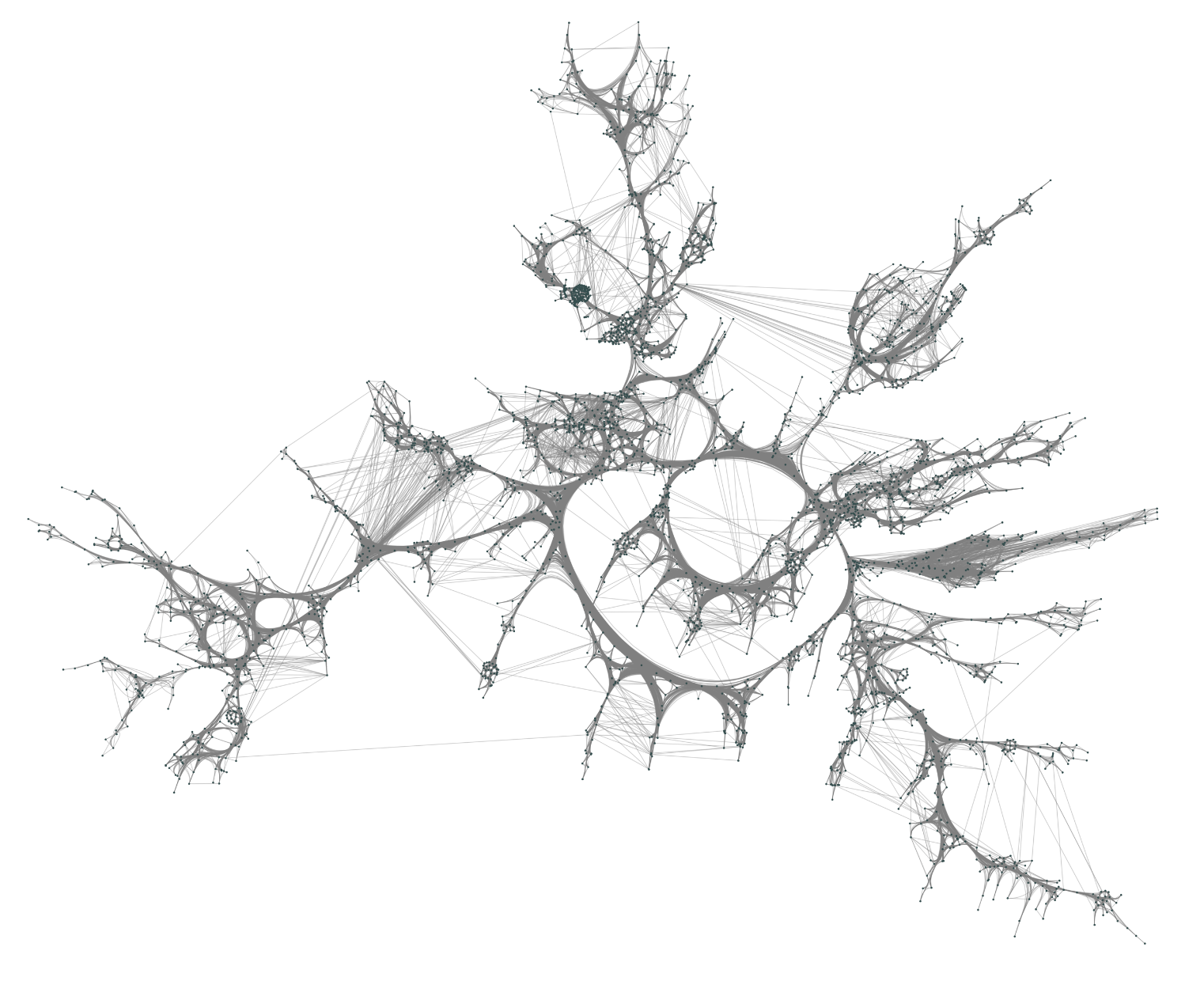} & \includegraphics[width=0.21\textwidth]{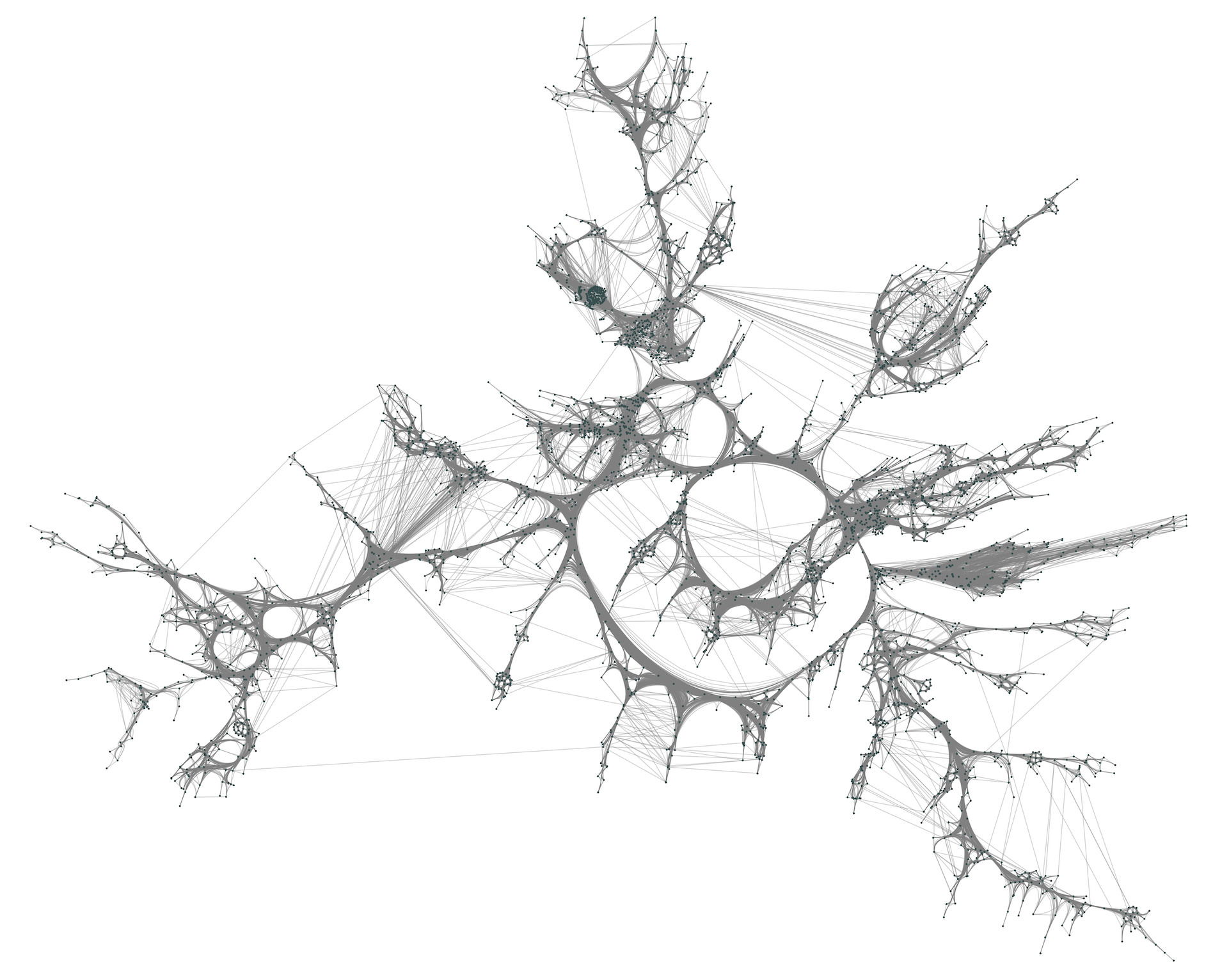}\\
        %\hline $D_B$  & $D'_B$ & $D_B$  & $D'_B$ \\
        \hline $D_B$(SEB1) & $D'_B$(FSEB) & $D_B$(SEB2) & $D'_B$(FSEB2) \\
    \hline\includegraphics[width=0.21\textwidth]{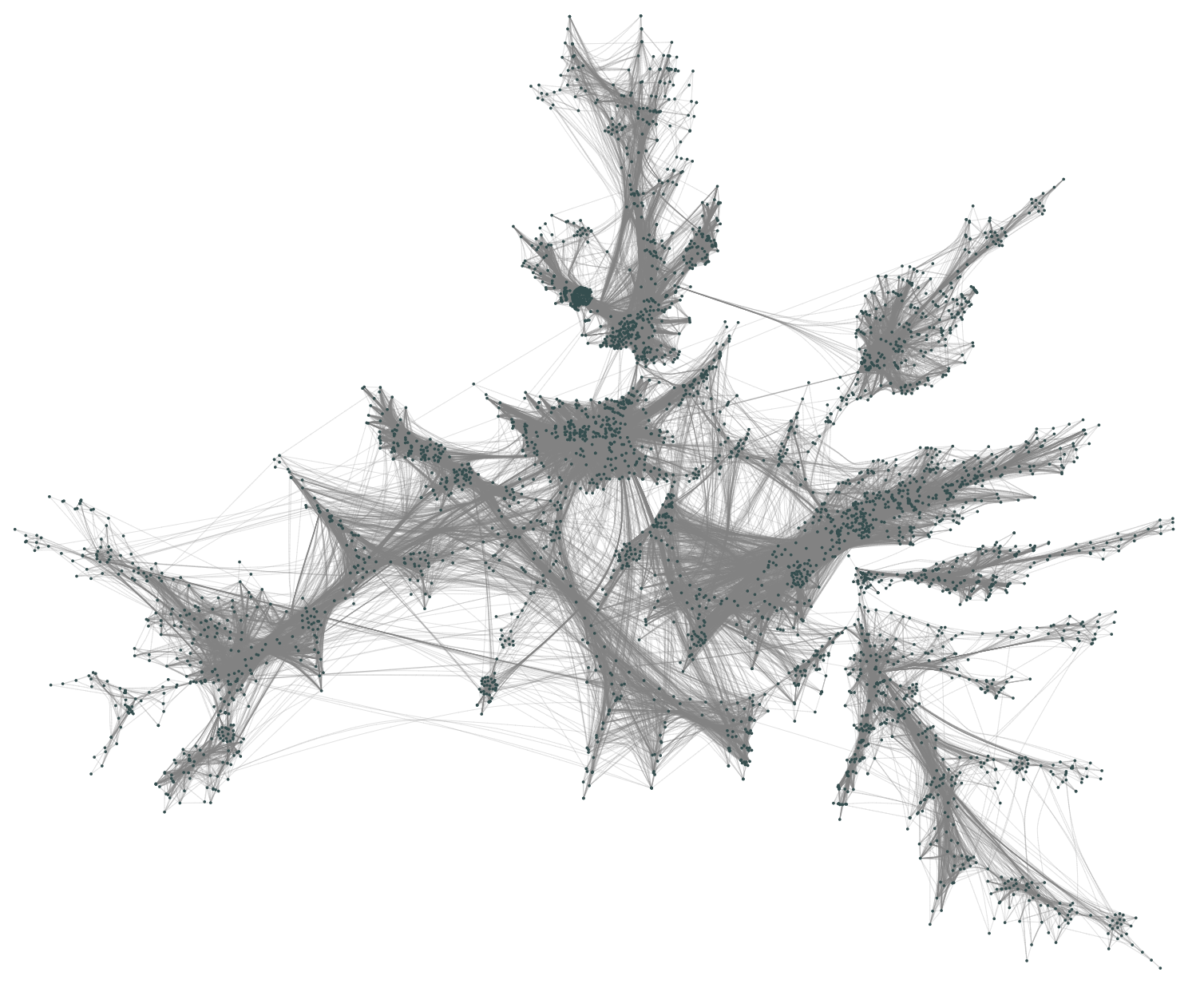} & \includegraphics[width=0.21\textwidth]{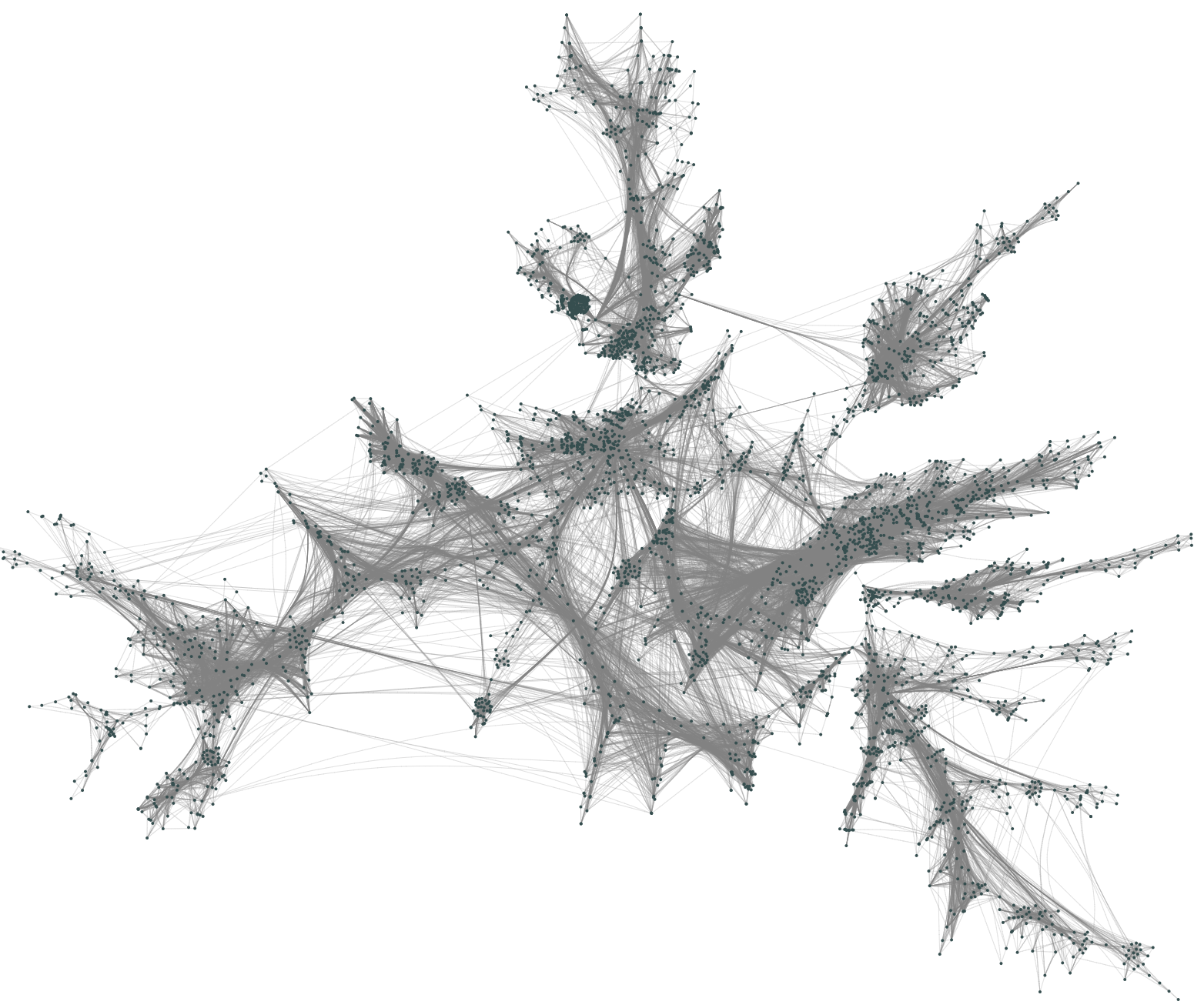} &   \includegraphics[width=0.21\textwidth]{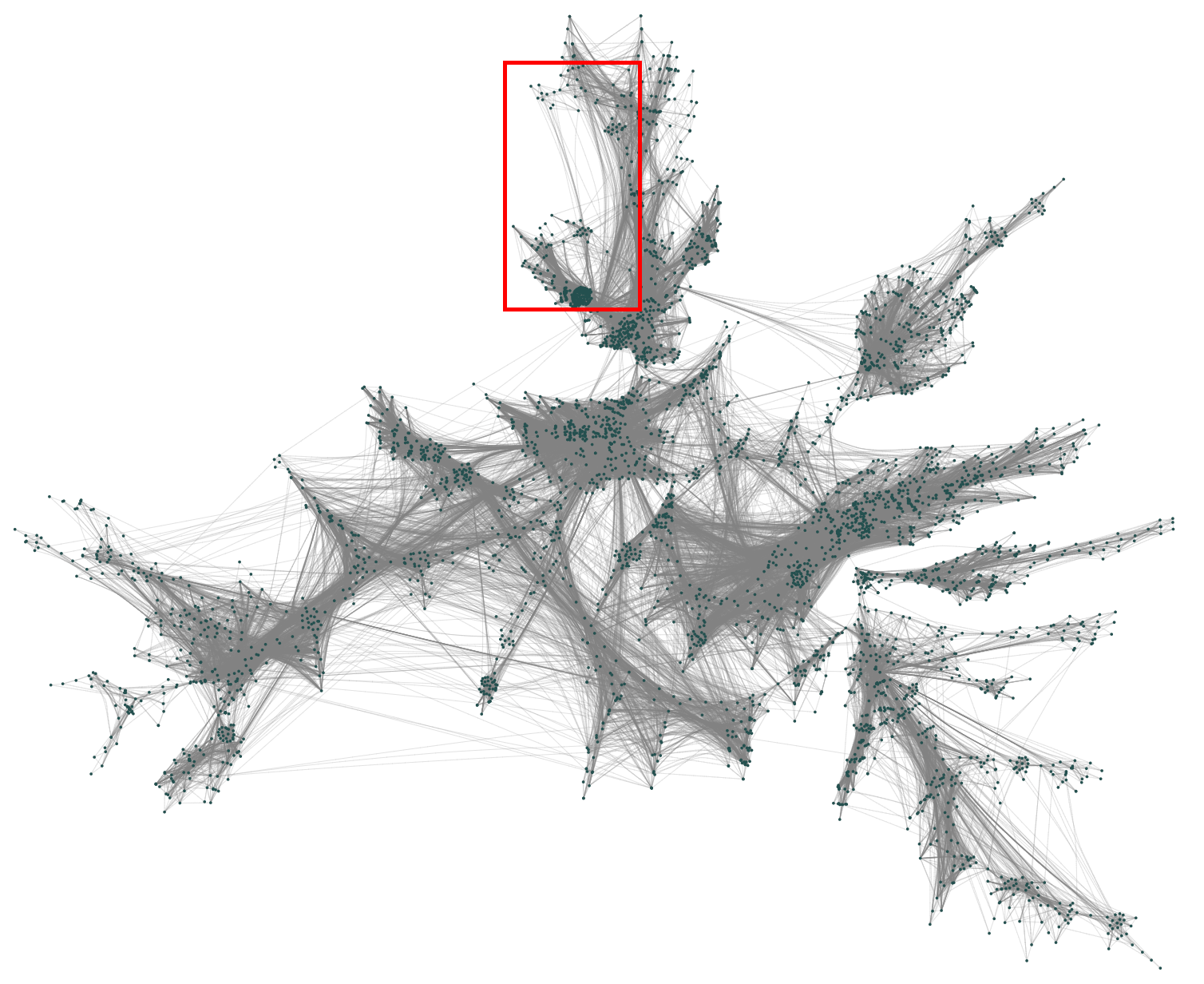} & \includegraphics[width=0.21\textwidth]{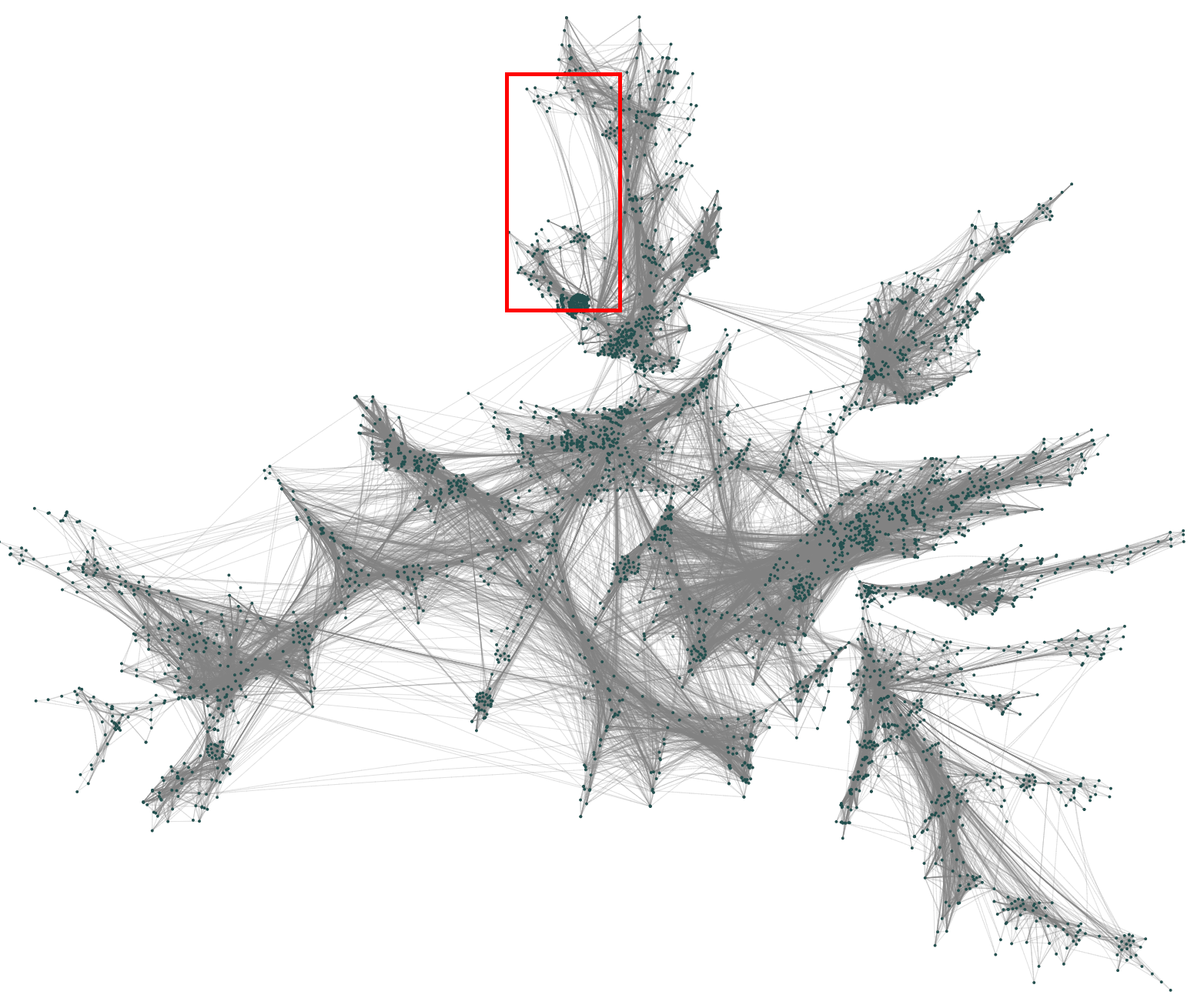}\\
        \hline
    \end{tabular}
    \label{tab:facebook-ss-compare}
\end{table*}

\subsection{Quality Metrics Comparison}

Table \ref{tab:seb_metrics} shows the details of all metrics (lower values mean better) for each bundling method over all data sets, where we denote SEB1 using $C_{ER_1}$ and SEB2 using $C_{ER_2}$.

The percentage improvement of our methods over the baseline methods are computed as follows: for example, to compute the improvement in $\mathrm{Amb}^1$ of SEB2 over EPB, we compute $\frac{\mathrm{Amb}^1(EPB)-\mathrm{Amb}^1(SEB2)}{\mathrm{Amb}^1(EPB)}$. 
We use the same formula for other metrics.

\begin{figure}[h!]
    \centering
    \subfloat[Ink.]{
        \includegraphics[width=0.22\columnwidth]{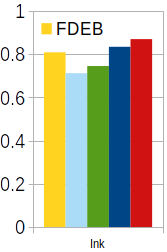}
    }
    \subfloat[Dist.]{
        \includegraphics[width=0.22\columnwidth]{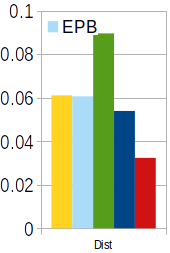}
    }
    \subfloat[$\mathrm{Amb}^1$]{
        \includegraphics[width=0.22\columnwidth]{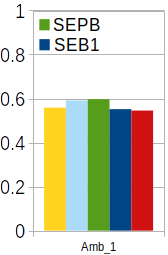}
    }
    \subfloat[$\mathrm{Amb}^2$]{
        \includegraphics[width=0.22\columnwidth]{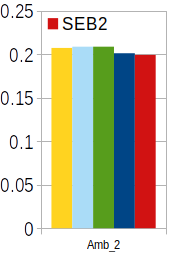}
    }
    %\begin{subfigure}[b]{0.19\textwidth}
    %    \centering
    %    \includegraphics[width=\textwidth]{seb_sbq_avg.png}
    %    \caption{SBQ}
    %\end{subfigure}
    \caption{%XXXXX thin bars to make one line XXXXX 
    Average metrics over all data sets (lower=better). 
    SEB methods perform significantly better than EPB methods on distortion and ambiguity, with average pairwise improvement of 46\% and 17\% respectively.}
    \label{fig:seb_avg}
\end{figure}

Overall, SEB methods perform better than EPB methods on distortion and ambiguity, especially for scale-free and black-hole graphs.
Figure \ref{fig:seb_avg} shows, on average, SEB2 performs the best on distortion and ambiguity, and EPB performs the best for ink reduction.

\begin{figure}[h!]
    \centering
    \subfloat[Ink.]{
        \includegraphics[width=0.22\columnwidth]{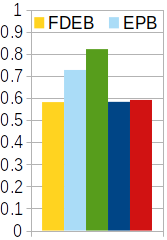}
    }
    \subfloat[Dist.]{
        \includegraphics[width=0.22\columnwidth]{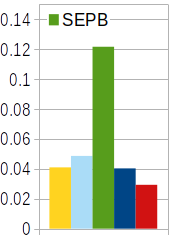}
    }
    \subfloat[$\mathrm{Amb}^1$]{
        \includegraphics[width=0.22\columnwidth]{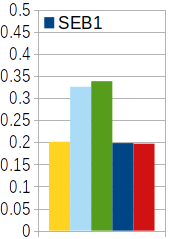}
    }
    \subfloat[$\mathrm{Amb}^2$]{
        \includegraphics[width=0.22\columnwidth]{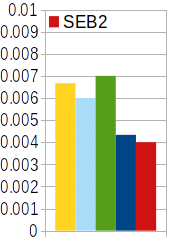}
    }
    %\begin{subfigure}[b]{0.19\textwidth}
    %    \centering
    %    \includegraphics[width=\textwidth]{seb_sbq_bh_avg.png}
    %    \caption{SBQ}
    %\end{subfigure}
    \caption{Average metrics for black-hole graphs. SEB methods outperform EPB methods on all metrics.}
    \label{fig:seb_bh}
\end{figure}

On {\em ink reduction}, EPB performs the best on most graphs. 
For some graphs, SEB methods obtain better ink reduction than EPB, such as the yeastPPI graph, which does not contain a clustering structure as other scale-free graphs, and the black-hole graphs with global cycle-like structures.

We use \emph{distortion} as the distortion value described in Section 2 subtracted by 1 (i.e., no distortion), to denote the divergence from the ideal level of distortion.
SEB2 outperforms the other methods on almost all graphs; for example, 46\% better than EPB on average.
Both SEB1 and SEB2 consistently perform better than FDEB.

For \emph{ambiguity}, SEB performs significantly better than EPB, where SEB2 performs the best overall.
For black-hole graphs, SEB2 performs best at 50\% better than EPB and a similar pattern is seen for scale-free graphs.
%to a lesser extent.
%
For GION graphs, SEB performs better than EPB on $\mathrm{Amb^1}$.
%, while the opposite is true on Amb2. 
%
For geographic graphs, SEB1 performs better on $\mathrm{Amb^2}$ for less dense graphs, while EPB performs better on $\mathrm{Amb^2}$ %XXX
for dense graph Air Traffic.

\subsection{Visual Comparison}

The first and third columns of Table \ref{tab:facebook-ss-compare} show visual comparison on the scale-free Facebook graph. 
SEB2 shows patterns of separating paths that were merged together in EPB and FDEB, such as the path going from the middle top cluster passing down through the small cluster (red inset), which is grouped together with the large middle bundle by other methods. 
EPB methods tend to highlight cycles in the graph, even to the detriment of faithfully preserving the clustering structure of the graphs, with dense clusters near the middle of the graph appearing as sparse ``branches'' splitting off the central cycle.

The first column of Table \ref{tab:airtraffic-ss-compare} shows a visual comparison on the geographic Air Traffic graph. 
SEB1 and SEB2 highlight long paths missed by EPB and FDEB: SEB1 separates the two long paths that start from the bottom left and pass through the middle of the drawing (red inset), while SEB2 separates out the long paths going through the lower half of the drawing which is missed by other bundling methods (red inset).

\begin{table*}[h]
    \centering
    \scriptsize
    \caption{Visual comparison for the Air Traffic graph, with original edge bundling methods on the left column and FEB on the right column. SEB methods separate out some paths (red insets) that are bundled together with other paths in FDEB. Meanwhile, the bundled drawing $D'_B$ by FEB is similar to the original bundled drawing $D_B$ for all bundling methods.}
    \begin{tabular}{|l|c|l|c|}
        \hline & $D$ &  & $D'$ \\
        \hline  & \includegraphics[width=0.33\textwidth]{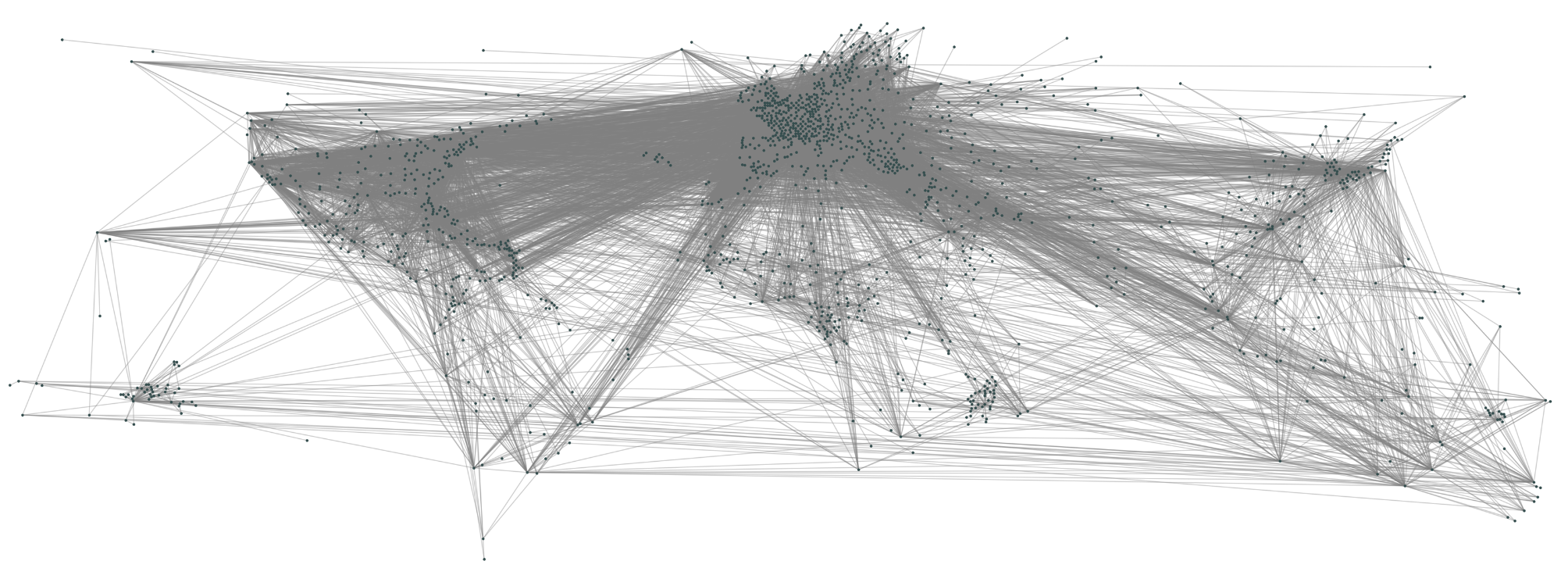}& & \includegraphics[width=0.33\textwidth]{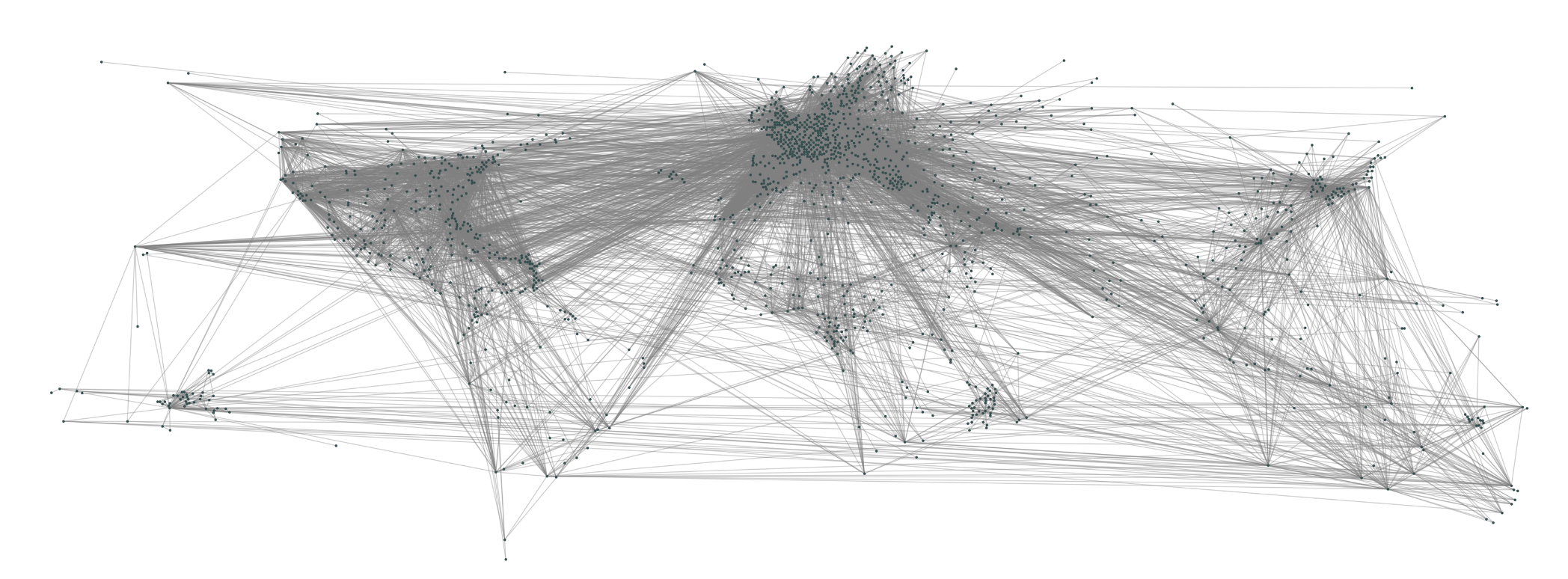}\\
        \hline & $D_B$ &  & $D'_B$ \\
        \hline \begin{sideways} FDEB \end{sideways} & \includegraphics[width=0.33\textwidth]{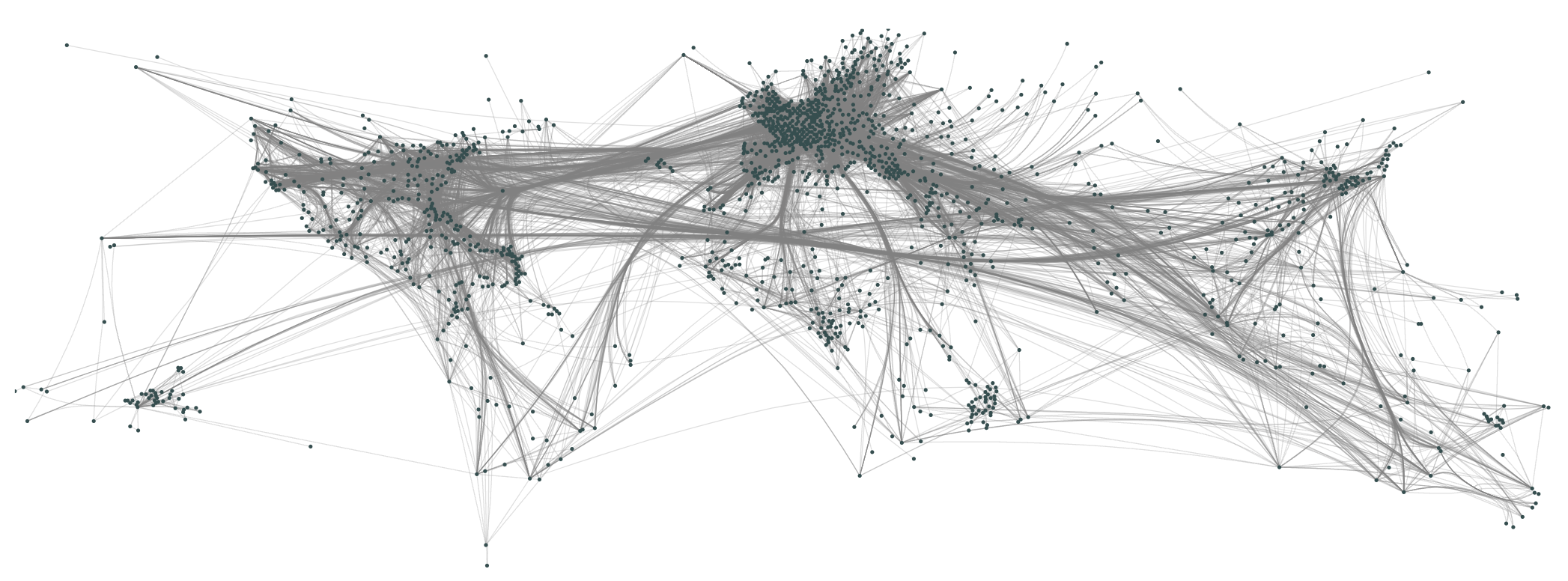}& \begin{sideways} FFDEB \end{sideways} & \includegraphics[width=0.33\textwidth]{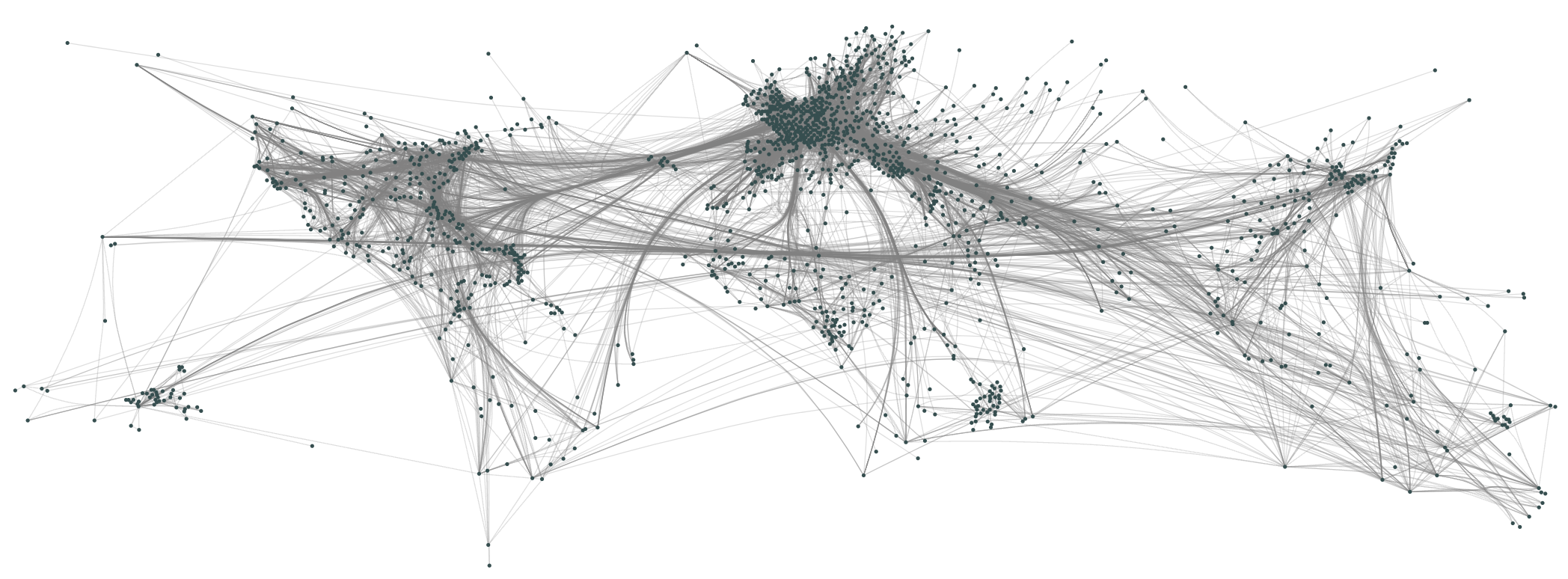}\\
        \hline \begin{sideways} EPB \end{sideways} & \includegraphics[width=0.33\textwidth]{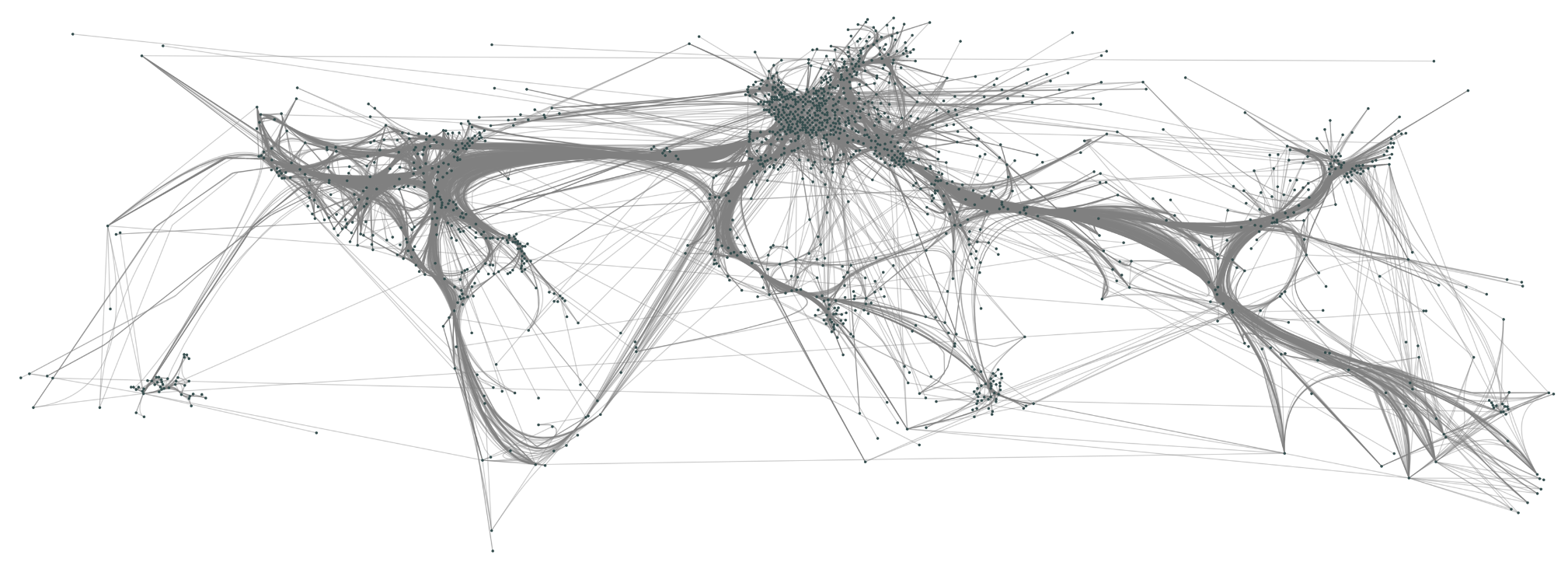} & \begin{sideways} FEPB \end{sideways} & \includegraphics[width=0.33\textwidth]{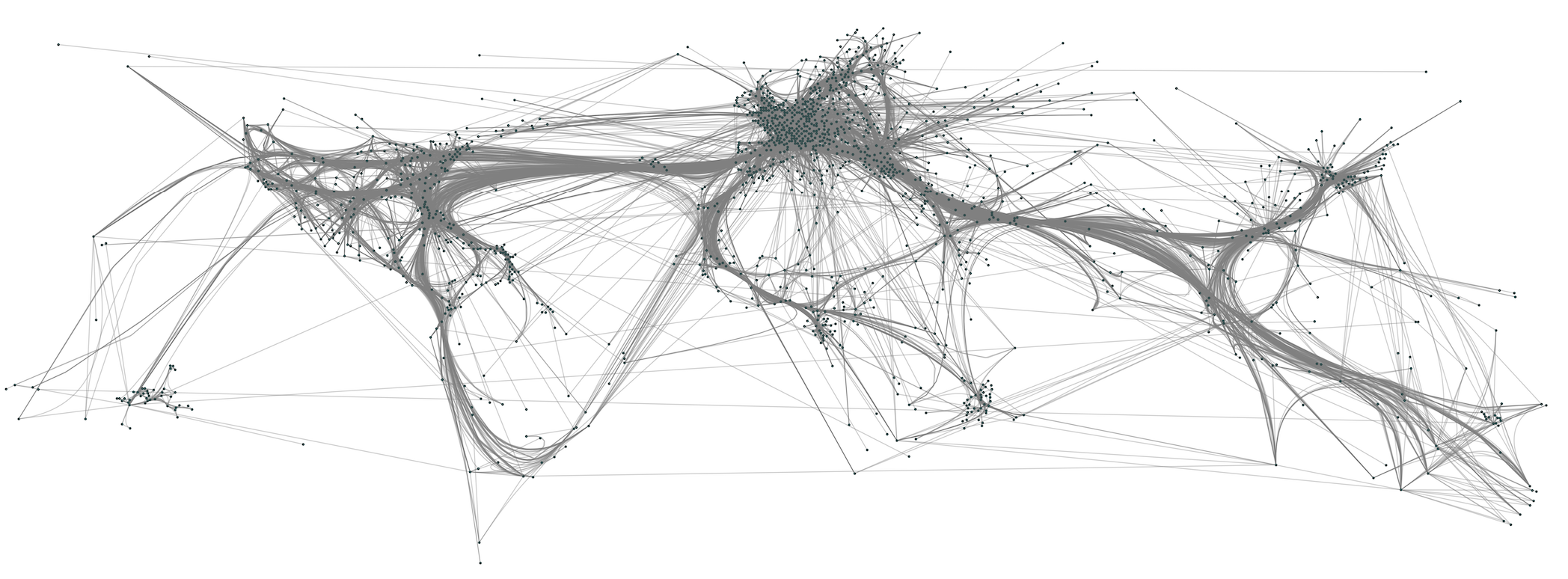}\\
        \hline \begin{sideways} SEPB \end{sideways} & \includegraphics[width=0.33\textwidth]{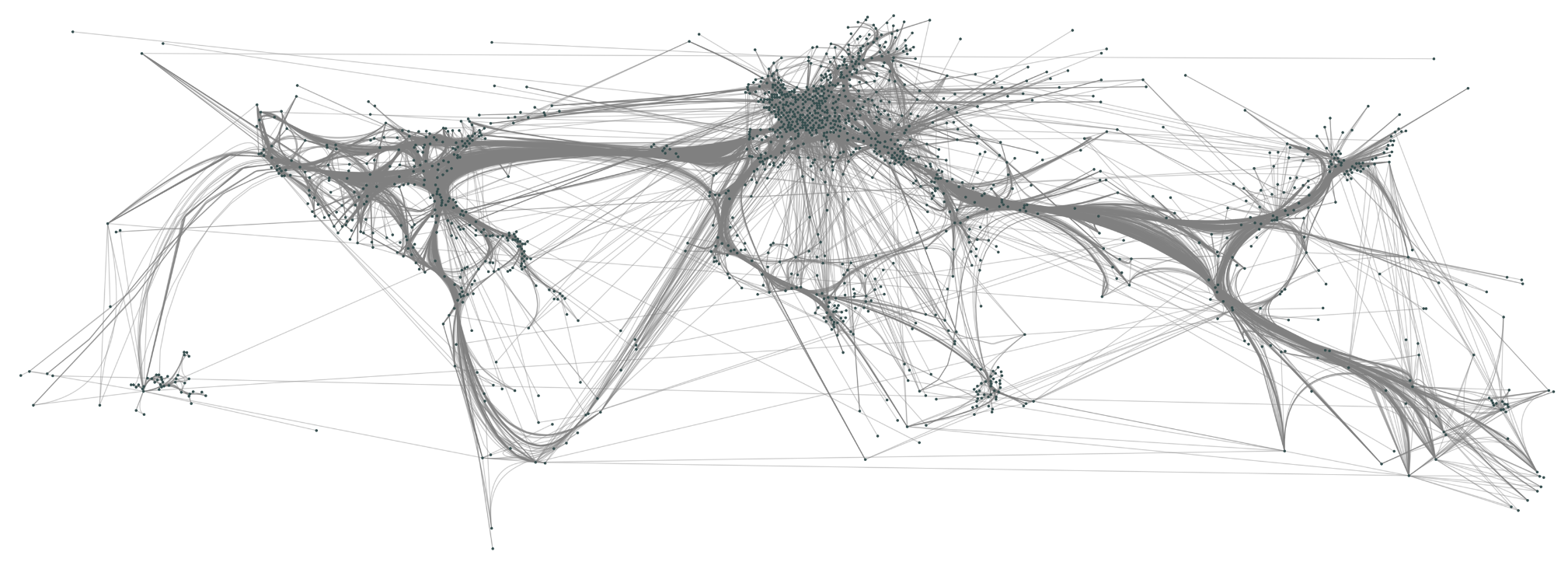} & \begin{sideways} FSEPB \end{sideways} & \includegraphics[width=0.33\textwidth]{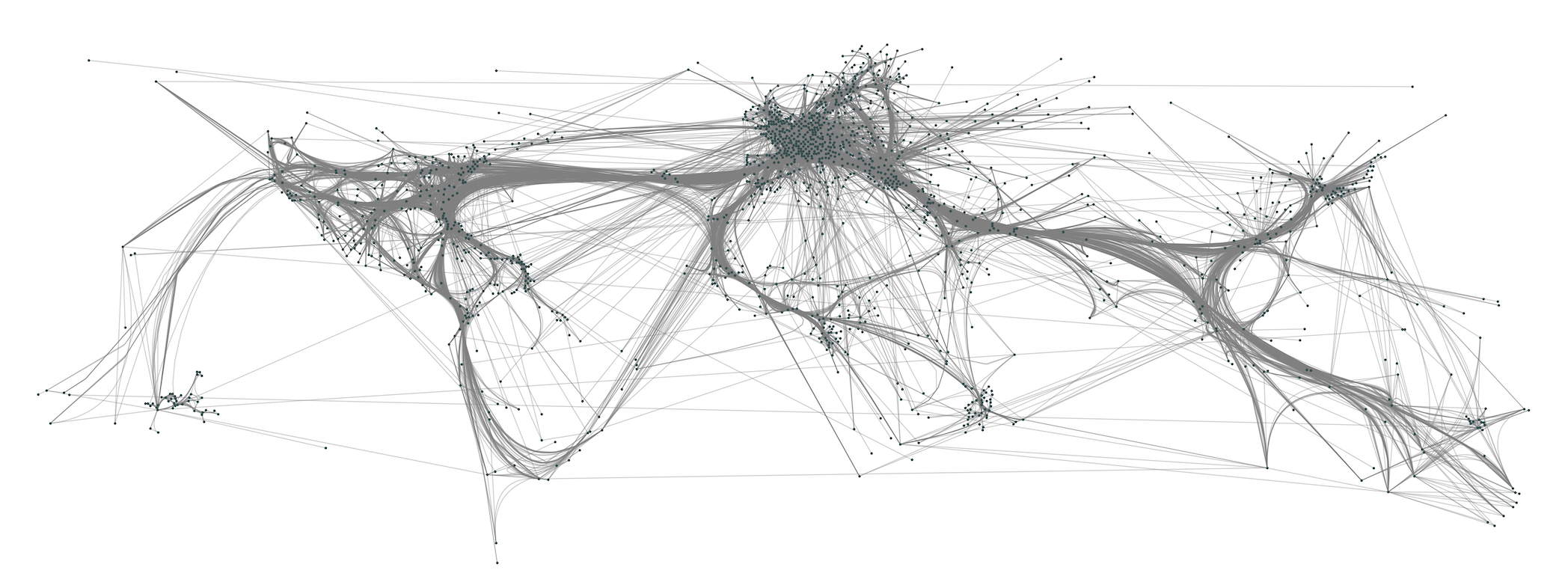}\\
        \hline \begin{sideways} SEB1 \end{sideways} & \includegraphics[width=0.33\textwidth]{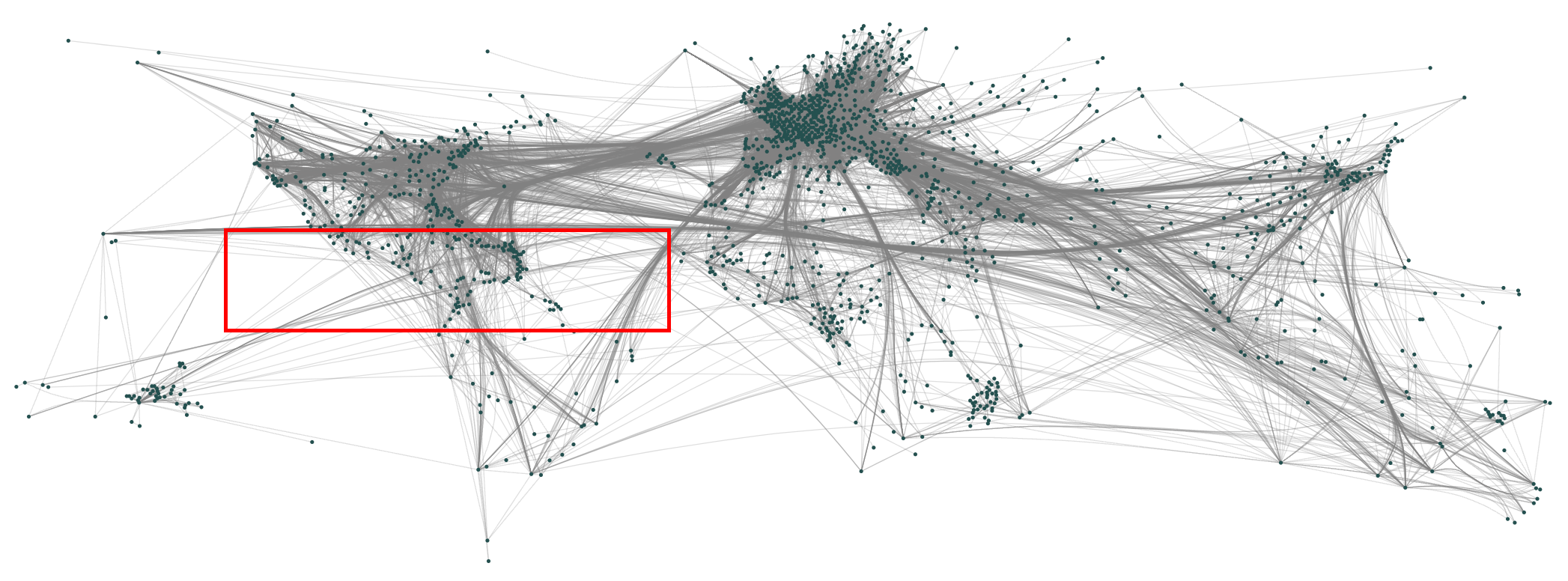} & \begin{sideways} FSEB1 \end{sideways} & \includegraphics[width=0.33\textwidth]{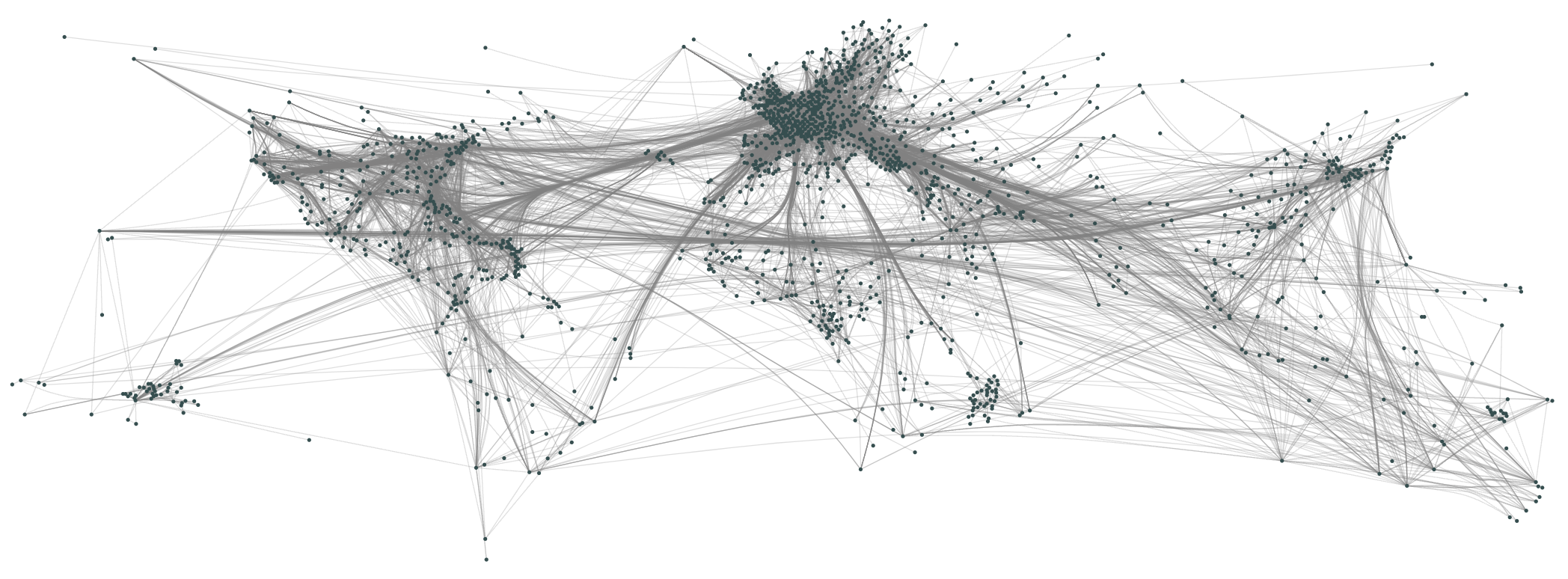}\\
        \hline \begin{sideways} SEB2 \end{sideways} & \includegraphics[width=0.33\textwidth]{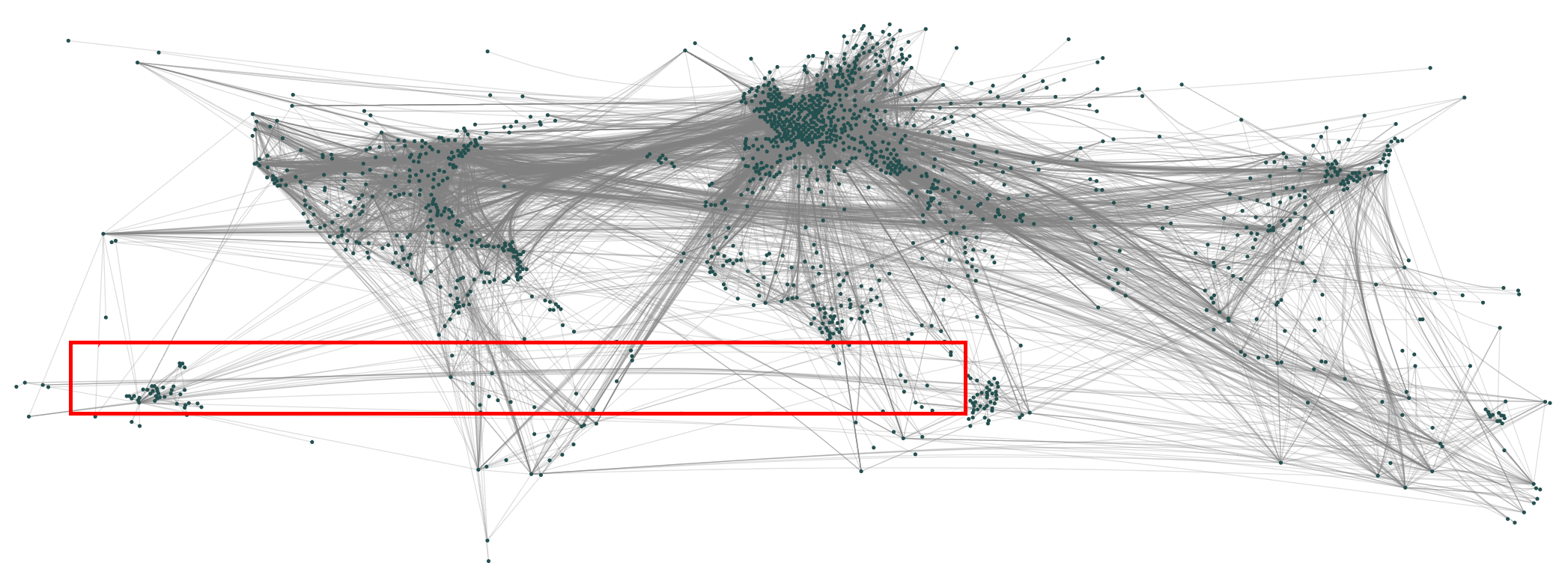} & \begin{sideways} FSEB2 \end{sideways} & \includegraphics[width=0.33\textwidth]{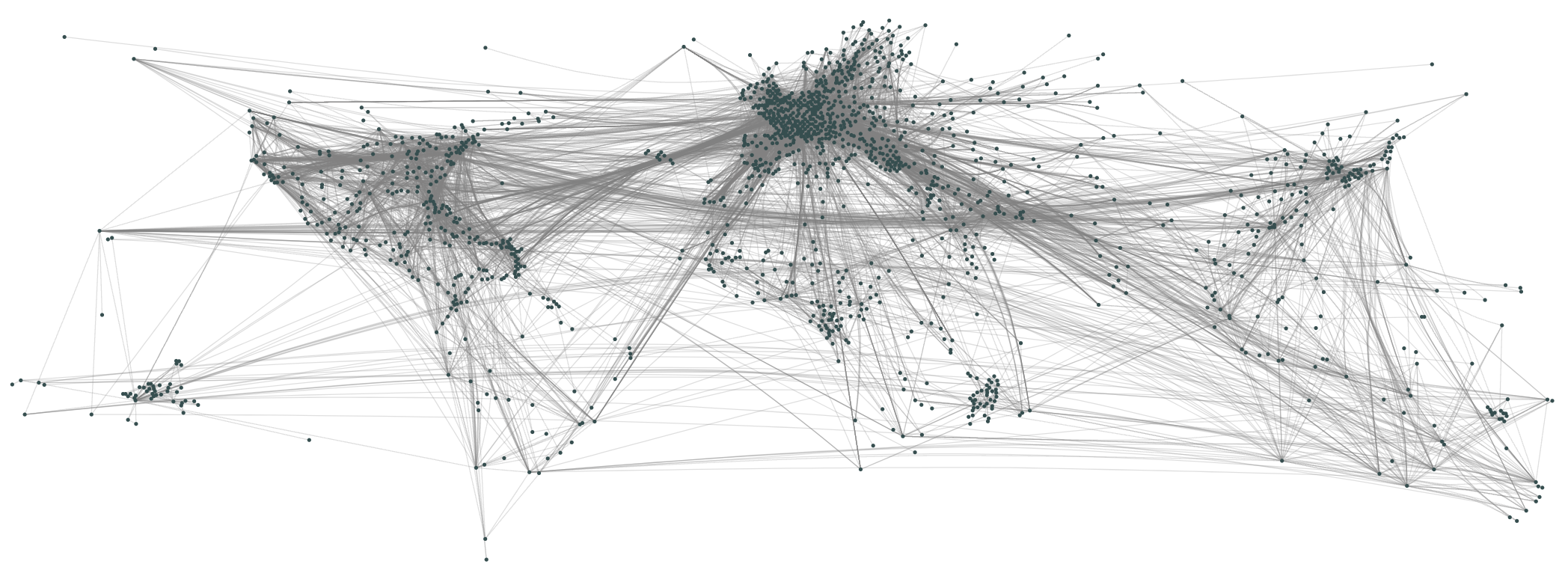}\\
        \hline
    \end{tabular}
    \label{tab:airtraffic-ss-compare}
\end{table*}

The first and third columns of Table \ref{tab:gion-ss-compare} show visual comparison on the 6\_gion graph. The SEB methods mainly bundle edges along the large, dense cycles, compared to the EPB methods which over-emphasizes smaller cycles compared to the original straight-line drawings (see red insets).

The first and third columns of Table \ref{tab:bh-ss-compare} show visual comparisons on the black-hole graph G443. While the EPB methods show more bundling than the SEB methods, this could show misleading structures, such as over-emphasizing certain cycles inside the blob (see red insets).

%The first and third columns of Table \ref{tab:lastfm-ss-compare} show a visual comparison on the scale-free graph Lastfm. In general, the SEB methods successfully create bundles highlighting the dense clusters of the graphs. Compared to EPB methods, SEb methods emphasize bundles around a few dense paths, while EPB emphasizes many less dense paths. This trend is most visible in the middle of the drawing (red insets), where SEB methods bundle edges along the dense ``bend'', while EPB divide the dense path into multiple smaller bundles.

\subsection{Discussion and Summary}

Extensive experiments demonstrate the effectiveness of SEB, outperforming state-of-art bundling methods: averaging the pairwise improvement, SEB2 obtains 46\% improvement on distortion and 17\% improvement on ambiguity over EPB.%, see Figure \ref{fig:seb_avg}.

Meanwhile, EPB methods perform better than SEB methods on ink reduction.
This indicates the negative correlation between quality metrics, where optimizing ink reduction may come at the expense of high distortion and ambiguity, consistent with the definitions of the metrics (i.e.,  reducing ink means bundling more edges, which introduces more distortion).

SEB methods consistently outperform EPB methods on ambiguity for most data sets, except geographic data sets, notably 50\% better for black-hole graphs (see Figure \ref{fig:seb_bh}). 
This may be due to the fact that the fixed geographic layouts often have very long edges, which are prioritized for bundling by EPB. 
Meanwhile, the drawings of non-geographic graphs are computed using graph layout algorithms, which often avoid long edges due to their optimization criteria. 
This may have caused EPB to be less effective for such drawings. 

%Another point is how, on some graphs, SEB outperforms EPB on $Amb^1$, but the opposite is true for $Amb^2$, likely due to EPB's tendency to bundle along long paths, which can reduce higher orders on ambiguity but may introduce ambiguity between direct neighbors.

%Figures \ref{fig:seb_geo}-\ref{fig:seb_gion} in Appendix \ref{sec:seb_perdata} show the averages for other data sets. 
%
On GION graphs, SEB performs better than EPB on $\mathrm{Amb^1}$, while EPB performs better on $\mathrm{Amb^2}$. 
SEB aims to preserve the spectrum of the graph in the drawing, which is related to the adjacency in the graphs, 
while EPB prioritizes bundling longer paths, especially relevant for the GION graphs with a long diameter; this may lead to SEB performing better on $\mathrm{Amb^1}$, and EPB performing better on $\mathrm{Amb^2}$.

\emph{In summary, SEB methods perform significantly better than EPB methods on distortion and ambiguity, on average 46\% and 17\% pairwise improvement respectively, indicating more spectrum faithful edge bundling.}
%, showing distinct bundles that are not visible in bundled drawings by EPB methods.}

\section{Conclusion}

We present new fast and faithful edge bundling methods, utilizing spectral sparsification. 
We first present the FEB framework for fast edge bundling based on spectral sparsification, and new faithfulness metrics FBQ for measuring how faithfully FEB preserves the ground truth structure of the original bundling, with two variants. 
Extensive experiments show that FEB runs 61\% faster than the original edge bundling methods while maintaining high similarity to the original bundling,  by the FBQ metrics (74\% by $FBQ_{SQ}$).

We then present SEB, introducing new effective resistance-based compatibility, which obtains 46\% improvement on distortion and 17\%  improvement on ambiguity than the state-of-the-art EPB methods.

Future work includes designing new faithfulness metrics for edge bundling, and new edge bundling methods to optimize the metrics. 

% if have a single appendix:
%\appendix[Proof of the Zonklar Equations]
% or
%\appendix  % for no appendix heading
% do not use \section anymore after \appendix, only \section*
% is possibly needed

% use appendices with more than one appendix
% then use \section to start each appendix
% you must declare a \section before using any
% \subsection or using \label (\appendices by itself
% starts a section numbered zero.)
%

% you can choose not to have a title for an appendix
% if you want by leaving the argument blank

% use section* for acknowledgment
\ifCLASSOPTIONcompsoc
  % The Computer Society usually uses the plural form
  \section*{Acknowledgments}
\else
  % regular IEEE prefers the singular form
  \section*{Acknowledgment}
\fi

This work was supported by an ARC (Australian Research Council) DP (Discovery Project) grant (\# DP190103301).

% Can use something like this to put references on a page
% by themselves when using endfloat and the captionsoff option.
\ifCLASSOPTIONcaptionsoff
  \newpage
\fi

% trigger a \newpage just before the given reference
% number - used to balance the columns on the last page
% adjust value as needed - may need to be readjusted if
% the document is modified later
%\IEEEtriggeratref{8}
% The "triggered" command can be changed if desired:
%\IEEEtriggercmd{\enlargethispage{-5in}}

% references section

% can use a bibliography generated by BibTeX as a .bbl file
% BibTeX documentation can be easily obtained at:
% http://mirror.ctan.org/biblio/bibtex/contrib/doc/
% The IEEEtran BibTeX style support page is at:
% http://www.michaelshell.org/tex/ieeetran/bibtex/
\bibliographystyle{IEEEtran}
% argument is your BibTeX string definitions and bibliography database(s)
\bibliography{IEEEabrv,template}
%
% <OR> manually copy in the resultant .bbl file
% set second argument of \begin to the number of references
% (used to reserve space for the reference number labels box)
%\begin{thebibliography}{1}

%\bibitem{IEEEhowto:kopka}
%H.~Kopka and P.~W. Daly, \emph{A Guide to \LaTeX}, 3rd~ed.\hskip 1em plus
%  0.5em minus 0.4em\relax Harlow, England: Addison-Wesley, 1999.

%\end{thebibliography}

% biography section
% 
% If you have an EPS/PDF photo (graphicx package needed) extra braces are
% needed around the contents of the optional argument to biography to prevent
% the LaTeX parser from getting confused when it sees the complicated
% \includegraphics command within an optional argument. (You could create
% your own custom macro containing the \includegraphics command to make things
% simpler here.)
%\begin{IEEEbiography}[{\includegraphics[width=1in,height=1.25in,clip,keepaspectratio]{mshell}}]{Michael Shell}
% or if you just want to reserve a space for a photo:

\begin{IEEEbiographynophoto}{Xingjue Jiang} is a student in the School of Computer Science, University of Sydney, Australia.
\end{IEEEbiographynophoto}

\begin{IEEEbiographynophoto}{Seok-Hee Hong} is a professor in the School of
Computer Science, University of Sydney,
Australia. Her research interests include graph drawing, computational geometry, information visualization, and visual analytics.
\end{IEEEbiographynophoto}

\begin{IEEEbiographynophoto}{Amyra Meidiana} is a researcher in the School of Computer Science, University of Sydney, Australia. Her research interests include graph drawing, information visualization, and visual analytics.
\end{IEEEbiographynophoto}

\begin{IEEEbiographynophoto}{Xianyuan Zeng} is a student in the School of Computer Science, University of Sydney, Australia.
\end{IEEEbiographynophoto}

% if you will not have a photo at all:
%\begin{IEEEbiographynophoto}{John Doe}
%Biography text here.
%\end{IEEEbiographynophoto}

% insert where needed to balance the two columns on the last page with
% biographies
%\newpage

%\begin{IEEEbiographynophoto}{Jane Doe}
%Biography text here.
%\end{IEEEbiographynophoto}

% You can push biographies down or up by placing
% a \vfill before or after them. The appropriate
% use of \vfill depends on what kind of text is
% on the last page and whether or not the columns
% are being equalized.

%\vfill

% Can be used to pull up biographies so that the bottom of the last one
% is flush with the other column.
%\enlargethispage{-5in}

% that's all folks
\end{document}